\pdfoutput=1
\documentclass[11pt,twoside,a4paper,cmspaper,final,collab]{cms-tdr}
\def\svnVersion{04b9023}\def\svnDate{2024/07/24}\def\cmsCernNoTag{CERN-EP-2023-304}\def\cmsCernDate{\today}\def\cmsMessage{Published in Physical Review D as \href{http://dx.doi.org/10.1103/PhysRevD.110.012013}{\doi{10.1103/PhysRevD.110.012013}.}}
\begin{document}\cmsNoteHeader{EXO-21-018}

\newcommand{\cmsTable}[1]{\ifthenelse{\boolean{cms@external}}{#1}{\resizebox{\textwidth}{!}{#1}}}
\newlength\cmsTabSkip\setlength{\cmsTabSkip}{1.2ex}
\newcommand{\MT}{\ensuremath{M_{\mathrm{T}}}\xspace}
\newcommand{\LT}{\ensuremath{L_{\mathrm{T}}}\xspace}
\newcommand{\ST}{\ensuremath{S_{\mathrm{T}}}\xspace}
\newcommand{\nossf}{\ensuremath{{\mathrm{OSSF}n}}\xspace}
\newcommand{\mossf}{\ensuremath{M_{\mathrm{OSSF}}}\xspace}  
\newcommand{\mmin}{\ensuremath{M_{\text{min}}}\xspace}
\newcommand{\ql}{\ensuremath{Q_{\Pell}}\xspace}
\newcommand{\ml}{\ensuremath{M_{\Pell}}\xspace}
\newcommand{\mee}{\ensuremath{M_{{\Pe\Pe}}}\xspace}
\newcommand{\mmm}{\ensuremath{M_{\PGm\PGm}}\xspace}
\newcommand{\meemin}{\ensuremath{\mee^{\text{min}}}\xspace}
\newcommand{\meemax}{\ensuremath{\mee^{\text{max}}}\xspace}
\newcommand{\mmmmin}{\ensuremath{\mmm^{\text{min}}}\xspace}
\newcommand{\mmmmax}{\ensuremath{\mmm^{\text{max}}}\xspace}
\newcommand{\mem}{\ensuremath{M_{{\Pe}\PGm}}\xspace}
\newcommand{\memmin}{\ensuremath{\mem^{\text{min}}}\xspace}
\newcommand{\mlt}{\ensuremath{M_{\Pell\PGt}}\xspace}
\newcommand{\mltmin}{\ensuremath{\mlt^{\text{min}}}\xspace}
\newcommand{\mtautau}{\ensuremath{M_{\PGt\PGt}}\xspace}
\newcommand{\mtautaumin}{\ensuremath{\mtautau^{\text{min}}}\xspace}
\newcommand{\mlttt}{\ensuremath{M_{\Pell\PGt/\PGt\PGt}}\xspace}
\newcommand{\mltttmin}{\ensuremath{\mlttt^{\text{min}}}\xspace}
\newcommand{\dR}{\ensuremath{\Delta{\mathrm{R}}}\xspace}
\newcommand{\dRmin}{\ensuremath{\Delta{\mathrm{R}}_{\text{min}}}\xspace}
\newcommand{\dz}{\ensuremath{\abs{d_{\mathrm{z}}}}\xspace}
\newcommand{\dxy}{\ensuremath{\abs{d_{\mathrm{xy}}}}\xspace}
\newcommand{\nj}{\ensuremath{N_{\mathrm{j}}}\xspace}
\newcommand{\nbj}{\ensuremath{N_{\PQb}}\xspace}
\newcommand{\ttphi}{\ensuremath{{\ttbar}{\phi}}\xspace}
\newcommand{\Wphi}{\ensuremath{{\PW}{\phi}}\xspace}
\newcommand{\Zphi}{\ensuremath{{\PZ}{\phi}}\xspace}
\newcommand{\Xphi}{\ensuremath{{\PX}{\phi}}\xspace}
\newcommand{\ttZ}{\ensuremath{{\ttbar}{\PZ}}\xspace}
\newcommand{\ttV}{\ensuremath{{\ttbar}\PV}\xspace}
\newcommand{\VVV}{\ensuremath{{\PV}{\PV}{\PV}}\xspace}
\newcommand{\WZ}{\ensuremath{{\PW}{\PZ}}\xspace}
\newcommand{\ZZ}{\ensuremath{{\PZ}{\PZ}}\xspace}
\newcommand{\ZG}{\ensuremath{{\PZ}{\PGg}}\xspace}
\newcommand{\ptlthree}{\ensuremath{p_{\mathrm{T3}}}\xspace}
\newcommand{\phiS}{\ensuremath{\phi_{\mathrm{S}}}\xspace}
\newcommand{\phiPS}{\ensuremath{\phi_{\mathrm{PS}}}\xspace}
\newcommand{\phiH}{\ensuremath{\phi_{\mathrm{H}}}\xspace}
\newcommand{\gpsiS}{\ensuremath{{\mathrm{g}}_{\psi {\mathrm{S}}}}\xspace}
\newcommand{\gpsiPS}{\ensuremath{{\mathrm{g}}_{\psi {\mathrm{PS}}}}\xspace}
\newcommand{\gtS}{\ensuremath{{\mathrm{g}}_{{\mathrm{t S}}}}\xspace}
\newcommand{\gtPS}{\ensuremath{{\mathrm{g}}_{{\mathrm{t PS}}}}\xspace}
\newcommand{\LambdaS}{\ensuremath{\Lambda_{\mathrm{S}}}\xspace}
\newcommand{\LambdaPS}{\ensuremath{\Lambda_{\mathrm{PS}}}\xspace}

\cmsNoteHeader{EXO-21-018}
\title{Search for a scalar or pseudoscalar dilepton resonance produced in association with a massive vector boson or top quark-antiquark pair in multilepton events at \texorpdfstring{$\sqrt{s} = 13\TeV$}{sqrt(s) = 13 TeV}}
 
\date{\today}

\abstract{
A search for beyond the standard model spin-0 bosons, $\phi$, that decay into pairs of electrons, muons, or tau leptons is presented. The search targets the associated production of such bosons with a \PW or \PZ gauge boson, or a top quark-antiquark pair, and uses events with three or four charged leptons, including hadronically decaying tau leptons. The proton-proton collision data set used in the analysis was collected at the LHC from 2016 to 2018 at a center-of-mass energy of 13\TeV, and corresponds to an integrated luminosity of 138\fbinv. The observations are consistent with the predictions from standard model processes. Upper limits are placed on the product of cross sections and branching fractions of such new particles over the mass range of 15 to 350\GeV with scalar, pseudoscalar, or Higgs-boson-like couplings, as well as on the product of coupling parameters and branching fractions. Several model-dependent exclusion limits are also presented. For a Higgs-boson-like $\phi$ model, limits are set on the mixing angle of the Higgs boson with the $\phi$ boson. For the associated production of a $\phi$ boson with a top quark-antiquark pair, limits are set on the coupling to top quarks. Finally, limits are set for the first time on a fermiophilic dilaton-like model with scalar couplings and a fermiophilic axion-like model with pseudoscalar couplings. 
}

\hypersetup{
pdfauthor={CMS Collaboration},
pdftitle={Search for a scalar or pseudoscalar dilepton resonance produced in association with a massive vector boson or top quark-antiquark pair in multilepton events at 13 TeV},
pdfsubject={CMS},
pdfkeywords={CMS, multilepton, dilepton, resonance, scalar, pseudoscalar}}

\maketitle 

\section{Introduction}\label{sec:Intro}

A search for a beyond-the-standard-model spin-0 boson, $\phi$, is presented, where the $\phi$ is produced in association with a \PW boson, a \PZ boson, or a top quark-antiquark pair ($\ttbar$), and decays into pairs of electrons, muons, or tau leptons.
The search uses events containing three or four charged leptons in the final state, selected from proton-proton ($\Pp\Pp$) collision data collected by the CMS experiment at the CERN LHC at $\sqrt{s} = 13\TeV$ from 2016--2018, and corresponding to an integrated luminosity of $138\fbinv$.
The analysis includes the leptonic and hadronic decays of tau leptons and targets events for which the two leptons from the $\phi$ decay form a localized excess in the dilepton mass spectrum.

This search targets possible new physics at the LHC that could underlie electroweak symmetry breaking.
Motivated by many extensions~\cite{Cacciapaglia:2019bqz,Ellwanger:2009dp,Maniatis:2009re,Buckley:2014fba} of the standard model (SM) that include additional spin-0 particles beyond the single Higgs boson ($\PH$), a minimal extension of the SM is considered that consists of a single neutral spin-0 boson, $\phi$~\cite{Casolino:2015cza,Chang:2017ynj}.

The effective Lagrangian relevant for this work is detailed in Ref.~\cite{Artoisenet:2013puc}. 
For the production of $\phi$ bosons with an SM \PW or \PZ boson, denoted $\Wphi$ and $\Zphi$, respectively, two coupling structures are considered. 
First, a scenario is considered in which the $\phi$ mixes with the SM \PH (H-like), yielding a coupling proportional to $\sin\theta$, where $\theta$ is the mixing angle, resulting in an interaction of the form: 

\begin{equation}
	{\mathcal{L}} \subset - 2 \, \sin \theta \, \, \frac{\phiH}{v} \, 
  \Big( \, m^2_{W} \, W^{+ \mu} W^-_{\mu} + \frac{1}{2} \, m^2_{\PZ}  \, Z^\mu Z_\mu \, \Big), 
\end{equation}
where $v \simeq 246\GeV$ is the Higgs field vacuum expectation value and $\phiH$ denotes the $\phi$ boson in the H-like scenario. 
Currently, there is an indirect bound on $\sin^2 \theta$ of approximately 0.1~\cite{Ghosh:2020ipy}.

Second, effective operators are considered that couple a scalar (S) or pseudoscalar (PS) $\phi$ boson to the $\mathrm{SU}(2)_{\mathrm{L}}$ field strength tensors $F^a_{\mu \nu}$ and $\widetilde{F}^a_{\mu \nu} = \frac{1}{2} \epsilon^{\mu \nu \rho \sigma} F^a_{\rho \sigma}$, where $a$ denotes the $\mathrm{SU}(2)_{\mathrm{L}}$ triplet index.
The operators take the form:

\begin{equation}
    {\mathcal{L}} \subset \frac{1}{\LambdaS} \,  \phiS  \,  F^{a \mu \nu}   \, F^a_{\mu \nu}
 + \frac{1}{\LambdaPS} \,  \phiPS  \,  F^{a \mu \nu}   \, \widetilde{F}^a_{\mu \nu}, 
\end{equation}
where $\LambdaS$ and $\LambdaPS$ are the mass scales of the effective interactions and $\phiS$ and $\phiPS$ denote the $\phi$ boson under the S or PS hypotheses, respectively.
The production of $\phi$ bosons in association with SM Higgs bosons or photons is not considered within the scope of this work. 

For the coupling of $\phi$ to fermions, flavor-conserving S or PS interactions of the following form are considered:

\begin{equation}
\label{eqn:ttphi}
{\mathcal{L}} \subset - \frac{\gpsiS}{\sqrt{2} } \, \phiS \,   \overline{\psi} \psi 
- \frac{\gpsiPS}{\sqrt{2} } \, \phiPS \,   \overline{\psi} i \gamma_5 \psi,
\end{equation}
where $\gpsiS$ and $\gpsiPS$ are dimensionless couplings to a given fermion field, $\psi$. 
These terms describe the associated production of the $\phi$ with $\ttbar$ ($\ttphi$). 
Probing the $\ttphi$ production mode is particularly motivated if the couplings are proportional to the fermion masses~\cite{Casolino:2015cza,Chang:2017ynj}. 
This is the case in the H-like coupling scenario where a scalar coupling between the $\phiH$ and fermions would be $\gpsiS = \sqrt{2}  \sin{\theta}   m_\psi/v$.

\begin{figure*}[hbt!]
\centering
\includegraphics[width=1.00\textwidth]{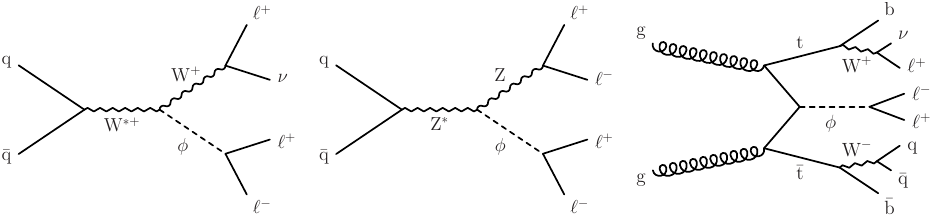}
\caption{\label{fig:XPhi_diagram} Example production and decay processes of $\Wphi$, $\Zphi$, and $\ttphi$ signals with multilepton final states, where $\Pell$ stands for electron, muon or tau lepton. Only leptonic decays of \PW and \PZ bosons are considered for $\Wphi$ and $\Zphi$ signals, while for the $\ttphi$ signal, \PW bosons from top quark decay can also decay hadronically.}
\end{figure*}

The three production modes, $\Wphi$, $\Zphi$, and $\ttphi$, populate a large, complementary multilepton signature space, and are collectively referred to as $\Xphi$. 
For all production modes, only decays of $\phi$ bosons to electron, muon, or tau lepton pairs are considered, as described by the couplings in Eq.~(\ref{eqn:ttphi}). 
The $\phi$ decays are assumed to conserve lepton number and electric charge, $\phi\to\Pell^+\Pell^-$, where $\Pell = \Pe$, $\PGm$, or $\PGt$, and are taken to occur at the production vertex.
As this search targets multilepton signatures, the \PW and \PZ bosons in $\Wphi$ and $\Zphi$ signal models are required to decay leptonically, whereas one of the \PW bosons in the $\ttphi$ signal model may decay hadronically.
The $\Wphi$ signal provides a signature with three leptons in the final state, primarily populating phase space regions with significant momentum imbalance due to the undetected neutrino from the W decay, low hadronic activity, and no heavy-flavored jets.
The $\ttphi$ signal can yield three- or four-lepton signatures, depending on the decay mode of the $\ttbar$ system, and populates the phase space with similarly large momentum imbalance, but with $\PQb$ jets and high hadronic activity.
The $\Zphi$ signal produces a four-lepton signature with low hadronic activity and no heavy-flavored jets, as well as no neutrinos in the final state.
A complete decay chain example is $\ttphi \to (\PQb\PWp)(\PAQb\PWm)(\Pell^+\Pell^-) \to (\PQb\Pell^+\PGn)(\PAQb\PQq\PAQq)(\Pell^+\Pell^-)$, where $\PQb$ is the SM bottom quark, $\PQq$ represents quarks of the first and second generations, and $\Pell$ ($\PGn$) represents the SM charged leptons (neutrinos).

Figure~\ref{fig:XPhi_diagram} illustrates possible production and decay processes of $\Wphi$, $\Zphi$, and $\ttphi$ that result in three- or four-lepton final states. 
This search targets $\phi$ masses between 15--350\GeV, excluding masses between 75--108\GeV for the $\phi\to\Pe\Pe/\PGm\PGm$ signal scenarios because of high SM \PZ boson background contributions. 
Decays of $\phi$ to gauge boson (top quark) pairs are no longer negligible when the $\phi$ boson mass is above twice that of the associated gauge bosons (top quarks) in each of the $\Xphi$ signals.
To facilitate comparisons of results across the signal phase space under consideration, an upper $\phi$ mass value of 350\GeV, slightly higher than twice the top quark mass, is chosen uniformly for all three $\Xphi$ signals.
For $\phi$ masses below the associated gauge boson (top quark) mass, the $\phi$ boson may also arise from a three-body decay of an on-shell gauge boson (top quark). 
The width of the $\phi$ boson is taken to be negligible in all signal models.

The search results are presented in terms of model-independent exclusions along with several model-dependent interpretations. 
For the model-independent results, $(3+3+2)\,3=24$ separate production and decay modes are probed separately in terms of the product of the production cross section and leptonic branching fraction $\mathcal{B}(\phi\to\Pell\Pell)$: three coupling structures (S, PS, and H) each for the $\Wphi$ and $\Zphi$ production modes, two coupling structures (S and PS) for the $\ttphi$  production mode, and three decay modes for the $\phi$  ($\Pe\Pe$, $\PGm\PGm$, and $\PGt\PGt$).
The model-dependent results are obtained from a subset of the experimental channels. 
First, direct bounds are set on models in which $\phi$ is a fermiophilic dilaton-like~\cite{Gildener:1976ih,Goldberger:2007zk,Ahmed:2019csf,Barger:2011nu} or a fermiophilic axion-like~\cite{Georgi:1986df,Mimasu:2014nea,Brivio:2017ije,Bauer:2018uxu} state, with couplings $\gpsiS$ and $\gpsiPS$ proportional to fermion masses. 
Only the $\ttphi$ production mode is considered in these cases, as fermiophilic particles do not couple to vector bosons.
Next, assuming an H-like $\phi$ production scenario, the $\Wphi$, $\Zphi$, or $\ttphi$ events are considered to be produced through mixing between the $\phi$ and SM Higgs boson, and constraints are set on the product of $\sin^2\theta$ and $\mathcal{B}(\phi\to\Pell\Pell)$. 
Under the additional assumption that the $\phi$ has the same branching fractions as the SM Higgs boson~\cite{Schabinger:2005ei,Barger:2007im,Chang:2017ynj}, constraints are set on $\sin^2\theta$ directly. 
All results presented in the paper, with accompanying material for reinterpretation, are provided in the \textsc{HEPData} record for this analysis~\cite{hepdata}. 
For the model-independent bounds, additional representation is provided in terms of the product of the production coupling constant squared and $\mathcal{B}(\phi\to\Pell\Pell)$, along with overlaid results from the $\Pe\Pe$, $\PGm\PGm$, and $\PGt\PGt$ decay channels for easier comparison. 

Spin-0 states produced in association with SM gauge bosons or top quark pairs and decaying into SM gauge bosons or fermions have been previously searched for by the LEP, Tevatron, and LHC experiments~\cite{ALEPH:1993sjl,L3:1996ome,LEPWG:2003ing,D0:2013cej,CDF:2013eju,CDF:2013kiv,CMS:2012bfw,ATLAS:2019vrd,CMS:2014wdm,ATLAS:2015xst,CMS:2014tll,ATLAS:2018alq,CMS:2020xwi,ATLAS:2020fzp,CMS:2018nsn,ATLAS:2018kot,ATLAS:2023ofo}. 
In comparison, this analysis constitutes a direct and model-independent search at the LHC for a dilepton resonance of any flavor produced in association with a W or a Z gauge boson, or a top quark pair.
In the context of the $\Xphi$ signals described above, the CMS Collaboration has previously probed the $\ttphi$ scenario with $\phi$ decays into dielectron or dimuon pairs with the combined 2016--2018 data set at $\sqrt{s}=13\TeV$~\cite{CMS:2019lwf}.
The current analysis also considers a $\ttphi\to\PGt\PGt$ signal and achieves increased sensitivity for the $\ttphi\to\Pe\Pe/\PGm\PGm$ signal over the entire mass range using improved event selection and analysis techniques.

This paper is organized as follows. 
The CMS detector is briefly described in Section~\ref{sec:CMSdetector}, followed by the summary of data and simulation samples in Section~\ref{sec:Samples}. 
Section~\ref{sec:objectSelections} covers event reconstruction, while Section~\ref{sec:eventselec} covers event selection and background estimation. 
Systematic uncertainties are discussed in Section~\ref{sec:Systematics} and results in Section~\ref{sec:Results}. 
The analysis is summarized in Section~\ref{sec:Summary}.

\section{The CMS detector} \label{sec:CMSdetector}

The central feature of the CMS apparatus is a superconducting solenoid of 6\unit{m} internal diameter, providing a magnetic field of 3.8\unit{T}. 
Within the solenoid volume are a silicon pixel and strip tracker, a lead tungstate crystal electromagnetic calorimeter (ECAL), and a brass and scintillator hadron calorimeter (HCAL), each composed of a barrel and two endcap sections. 
Forward calorimeters extend the pseudorapidity ($\eta$) coverage provided by the barrel and endcap detectors. 
Muons are measured in gas-ionization detectors embedded in the steel flux-return yoke outside the solenoid. 
A more detailed description of the CMS detector, together with a definition of the coordinate system used and the relevant kinematic variables, can be found in Ref.~\cite{CMS:2008xjf,CMS:2023gfb}.

Events of interest are selected using a two-tiered trigger system. 
The first level is composed of custom hardware processors, and uses information from the calorimeters and muon detectors to select events at a rate of around 100\unit{kHz} within a fixed latency of about 4\mus~\cite{CMS:2020cmk}. 
The second level, known as the high-level trigger, consists of a farm of processors running a version of the full event reconstruction software optimized for fast processing, and reduces the event rate to around 1\unit{kHz} before data storage~\cite{CMS:2016ngn}.

\section{Data samples and event simulation}\label{sec:Samples}

This analysis probes a data set that was collected in $\Pp\Pp$ collisions at $\sqrt{s} = 13\TeV$ and corresponds to an integrated luminosity of 138\fbinv, with 36.3, 41.5, and 59.8\fbinv recorded in the years 2016, 2017, and 2018, respectively~\cite{CMS:2021xjt, CMS:2018elu,CMS:2019jhq}.
The data presented here were collected using a combination of isolated single-electron (-muon) triggers with corresponding transverse momentum, \pt, thresholds of  27 (24), 32 (27), and 32 (24)\GeV in these three years. 

The rates of signal and SM background processes that give rise to isolated and prompt leptons are estimated from Monte Carlo (MC) simulations, which incorporate detailed detector and $\Pp\Pp$ collision properties. 
The \ZG, \WZ, \ttV, and triboson ($\VVV$) backgrounds, where \PV denotes a \PW or \PZ boson, are generated using \MGvATNLO~\cite{Alwall:2014hca} at next-to-leading order (NLO) accuracy in perturbative quantum chromodynamics (QCD). 
Version 2.2.2 (2.4.2) of \MGvATNLO is used for all background samples for 2016 (2017--2018) data.
The \ZG background includes all diagrams contributing to $\Pp\Pp\to \Pell\Pell\PGg$, with photons from both initial- and final-state radiation, and with a Lorentz-invariant mass requirement of $m(\Pell^+\Pell^-)>10\GeV$. 
The \ZZ background contribution produced from quark-antiquark annihilation is generated using \POWHEG 2.0~\cite{Nason:2004rx,Frixione:2007vw,Alioli:2010xd} at NLO, whereas the contribution from gluon-gluon fusion is generated at leading order (LO) using \MCFM 7.0.1~\cite{Campbell:2010ff}. 
The SM processes involving Higgs boson production are generated using \POWHEG 2.0, \MGvATNLO, and \textsc{JHUGen} 7.0.11~\cite{Gao:2010qx,Bolognesi:2012mm,Anderson:2013afp,Gritsan:2016hjl} at NLO, with a Higgs boson mass of 125\GeV. 
Processes with a single top quark and a \PZ boson or with four top quarks are simulated using \MGvATNLO at NLO in QCD.
Other small contributions from processes involving a single top quark and an electroweak boson or Higgs boson, two top quarks and two bosons, or three top quarks are simulated using \MGvATNLO at LO in QCD. 
Simulated event samples for the Drell--Yan (DY) and $\ttbar$ processes, which are used for systematic uncertainty studies, are generated at NLO with \MGvATNLO and \POWHEG 2.0, respectively.

The $\Xphi$ signal event generation and production cross section determination are performed at LO accuracy using \MGvATNLO 2.6.0 for $\ttphi$ and 2.6.5 for $\Wphi$ and $\Zphi$. 
The $\ttphi$ signal samples are generated with inclusive decays of the $\ttbar$ system, while the $\Wphi$ and $\Zphi$ signal samples are generated with leptonically decaying \PW and \PZ bosons, $\PW\to\ell\nu$ and $\PZ\to\ell\ell$, respectively. 
The samples satisfy an invariant mass requirement of $m(\Pell^+\Pell^-)>5\GeV$ and an angular separation requirement $\dR(\Pell^+\Pell^-)>0.05$ for the \PZ decay products, where $\dR \equiv [(\Delta\eta)^2+(\Delta\varphi)^2]^{1/2}$, $\eta$ is the pseudorapidity and $\varphi$ is the azimuthal angle~\cite{CMS:2008xjf}.

The  NNPDF3.0 LO or NLO parton distribution function (PDF) sets~\cite{Ball:2014uwa} are used for all background and signal samples for 2016 data, with a perturbative order matching that of the matrix element calculations. 
The NNPDF3.1 next-to-NLO PDF set~\cite{Ball:2017nwa} is used for all 2017--2018 samples.
To perform the parton showering, fragmentation, and hadronization of the matrix-element-level events, \PYTHIA~\cite{Sjostrand:2014zea} is used in all samples, with the event tune CUETP8M1~\cite{Khachatryan:2015pea} (CP5~\cite{CMS:2019csb}) in 2016 (2017--2018). \PYTHIA version 8.226 (8.230) is used for all background and signal samples for 2016 (2017--2018) data.
The MLM~\cite{Alwall:2007fs} or FxFx~\cite{Frederix:2012ps} jet matching schemes are used for \MGvATNLO samples at LO or NLO, respectively. 
The simulation of the response of the CMS detector to incoming particles is performed using the $\GEANTfour$ toolkit~\cite{Agostinelli:2002hh}. 
Additional $\Pp\Pp$ interactions from the same or nearby bunch crossings (pileup) are simulated with \PYTHIA and incorporated in the MC samples.

\section{Event reconstruction} \label{sec:objectSelections}

In each event, the primary vertex (PV) is taken to be the vertex corresponding to the hardest scattering in the bunch crossing, evaluated using tracking information alone, as described in Section 9.4.1 of Ref.~\cite{CMS-TDR-15-02}.
The full event information is then used by a particle-flow (PF) algorithm~\cite{Sirunyan:2017ulk}, which aims to reconstruct and identify each individual particle (photon, electron, muon, charged hadron, neutral hadron) with an optimized combination of information from the various elements of the CMS detector.

Electrons are reconstructed by geometrically matching charged-particle tracks from the tracking system with energy clusters deposited in the ECAL~\cite{CMS:2020uim}.
The energy of electrons is determined from a combination of the electron momentum at the primary interaction vertex as determined by the tracker, the energy of the corresponding ECAL cluster, and the energy sum of all bremsstrahlung photons spatially compatible with originating from the electron track.
The momentum resolution for electrons with $\pt \approx 45\GeV$ from $\PZ \to \Pe \Pe$ decays ranges from 1.6 to 5.0\%. 
It is generally better in the barrel region of the ECAL ($\abs{\eta}<1.479$) than in the endcaps, and also depends on the bremsstrahlung energy emitted by the electron as it traverses the material in front of the ECAL~\cite{CMS:2020uim,CMS-DP-2020-021}. 
To suppress undesired electrons originating from photon conversions in detector material and from the misidentification of hadrons, the electron candidates are required to satisfy shower shape and track quality requirements, using the medium cut-based criteria described in Ref.~\cite{CMS:2020uim}.
Electrons used in this analysis are required to also satisfy $\pt>10\GeV$ and $\abs{\eta}<2.4$.

Muons are reconstructed from compatible tracks in the inner tracker and the muon detectors~\cite{CMS:2018rym}. The energy of muons is obtained from the curvature of the corresponding track.
Additional track fit and matching quality criteria suppress the misidentification of hadronic showers that penetrate the calorimeters and reach the muon system.
The matching of muon system tracks to those measured in the silicon tracker results in a relative \pt resolution of 1\% in the barrel and 3\% in the endcaps for muons with \pt up to 100\GeV, and of better than 7\% in the barrel for muons with \pt up to 1\TeV~\cite{CMS:2018rym}. 
Muons used in this analysis must lie within the muon system acceptance, $\abs{\eta}<2.4$, and are required to have $\pt>10\GeV$.

Hadronically decaying tau lepton (\tauh) candidates are reconstructed from jets, using the hadrons-plus-strips algorithm~\cite{CMS:2018jrd}, which combines one or three tracks with a strip of energy deposits in the calorimeter. 
The energy deposits capture photons from neutral pion decay and electrons, and vary in size in $\eta$ and $\varphi$ as a function of \pt of the photon or electron candidate.
Reconstructed \tauh candidates must satisfy $\abs{\eta}<2.3$ and $\pt>20\GeV$, where \pt refers to the visible momentum of the tau lepton.

Jets are clustered from PF candidates using the anti-\kt algorithm~\cite{Cacciari:2008gp} with a distance parameter of 0.4, as implemented in the \textsc{FastJet} package~\cite{Cacciari:2011ma}. 
The jet momentum is given by the vector sum of all particle momenta in the jet.
Pileup interactions can contribute extra tracks and calorimetric energy depositions, increasing the apparent jet momentum. 
To mitigate this effect, tracks identified to be originating from pileup vertices are discarded, and an offset correction is applied to correct for remaining contributions. 
Jet energy corrections are derived from simulation studies so that the average measured energy of jets becomes identical to that of particle-level jets. 
In situ measurements of the momentum balance in dijet, $\text{photon}$+jet, $\PZ$+jet, and multijet events are used to determine any residual differences between the jet energy scale in data and in simulation, and appropriate corrections are made~\cite{CMS:2016lmd}. 
Additional selection criteria are applied to each jet to remove jets potentially dominated by instrumental effects or reconstruction failures.
The minimum \pt threshold for the jets selected in this analysis is 30\GeV, and the central axis of the jet is also required to be inside the muon system acceptance, $\abs{\eta}<2.4$. 
The selected jets must lie outside a cone defined by $\dR = 0.4$ relative to a selected muon, electron, or \tauh candidate, as defined later in this section. 

The missing transverse momentum \ptvecmiss is defined as the negative vector \pt sum of all the PF candidates in an event, and its magnitude is denoted as \ptmiss~\cite{Sirunyan:2019kia}. 
The pileup per particle identification algorithm~\cite{Bertolini:2014bba} is applied to reduce the pileup dependence of the \ptvecmiss observable.
The \ptvecmiss is computed using the PF candidates weighted by their probability to originate from the PV, and is modified to account for corrections to the energy scale of the reconstructed jets in the event~\cite{Sirunyan:2020foa}. 

Leptons produced from the decays of \PH and massive vector bosons (either directly or via a leptonically decaying tau lepton) are referred to as prompt leptons, and are often indistinguishable in momentum and isolation from those produced in signal events. 
Thus, the SM processes giving rise to three or more isolated leptons, such as \WZ, \ZZ, $\ttbar\PV$, $\PV\PV\PV$, and \PH boson production, are referred to as the prompt backgrounds in this analysis.
On the other hand, reducible backgrounds are defined as those from SM processes in which the jets are misidentified as leptons, or where the leptons originate from heavy-quark decays. 
Leptons from such sources are referred to as misidentified leptons, and SM background processes with such leptons are collectively labeled as ``MisID'' backgrounds.
Some examples of such backgrounds are $\PZ$+jets or $\ttbar$+jets production, in which the prompt leptons are accompanied by leptons that are within or near jets, 
hadrons that traverse the HCAL and reach the muon detectors, or hadronic showers with large electromagnetic energy fractions. 

The reducible backgrounds are significantly suppressed by applying stringent requirements on the lepton isolation and displacement. 
For electron and muon candidates, the relative isolation is defined as the scalar \pt sum of photon and hadron PF objects within a cone of fixed $\dR$ around the lepton, divided by the lepton \pt.
For electrons, the relative isolation is required to be less than $0.0478+0.506\GeV/\pt$ in the barrel section of the ECAL ($\abs{\eta}<1.479$) and less than $0.0658+0.963\GeV/\pt$ in the endcap section ($\abs{\eta}>1.479$), with $\dR=0.3$~\cite{CMS:2020uim}.
The relative isolation for muons is required to be less than 0.15 with $\dR=0.4$~\cite{CMS:2018rym}.
The isolation quantities are also corrected for contributions from particles originating from pileup vertices~\cite{CMS:2020uim,CMS:2018rym}.  
In addition to the isolation requirement, electrons in the barrel must satisfy $\dz<0.1\cm$ and $\dxy<0.05\cm$,
and in the endcap $\dz<0.2\cm$ and $\dxy<0.1\cm$, 
where $d_{\mathrm{z}}$ and $d_{\mathrm{xy}}$ are the longitudinal and transverse impact parameters of electrons with respect to the PV, respectively. 
Similarly, muons must satisfy $\dz<0.1\cm$ and $\dxy<0.05\cm$.
For both electrons and muons, the three-dimensional impact parameter significance (SIP3D) is defined as the absolute value of the impact parameter divided by its uncertainty~\cite{Sirunyan:2017ezt}. 
It is tuned to account for changes in detector and pileup conditions, and must be less than 10, 12, and 9 in 2016, 2017, and 2018 data, respectively.
All selected electrons within a cone of $\dR<0.05$ centered on a selected muon are discarded in order to reduce the inclusion of non-prompt electrons originating from muon bremsstrahlung. 

For \tauh leptons, the \textsc{DeepTau}~\cite{CMS:2022prd} algorithm is used to distinguish genuine \tauh lepton decays from jets originating from the hadronization of quarks or gluons, as well as from electrons or muons. 
Information from all individual reconstructed particles near the \tauh candidate axis is combined with properties of the \tauh candidate and of the event. 
The very tight working point of the jet discriminator of the \textsc{DeepTau} algorithm is used to suppress misidentified contributions from jets, with an identification efficiency of about 50\% depending on the visible \pt and $\eta$ of the \tauh candidate.
Similarly, the loose working points of the \textsc{DeepTau} electron and muon discriminators are used to suppress such misidentified contributions, with identification efficiencies of about 95\% and 99.9\%, respectively.
These result in misidentification probabilities of about 0.5\% for jets and electrons, and 0.05\% for muons, as measured in events enriched in DY+jets and W+jets processes. 
In addition to this multivariate requirement, \tauh candidates are required to satisfy $\dz<0.2\cm$. 
All selected \tauh candidates within a cone of $\dR<0.5$ of a selected electron or muon are also discarded to suppress contributions from electrons or muons misidentified as taus.

Reconstructed jets originating from \PQb hadrons are identified using the medium working point of the \textsc{DeepCSV} {\PQb} tagging algorithm~\cite{Sirunyan:2017ezt}. 
To suppress contributions due to  misidentified leptons originating from heavy-flavor jet decays, a \PQb tag veto is applied, where lepton candidates are discarded if any \PQb-tagged jet with a less stringent selection ($\pt>10\GeV$, $\abs{\eta}<2.5$) is found within a cone of $\dR<0.4$ centered on the candidate.
Selection criteria including requirements on ${d_{\mathrm{xy}}}$, ${d_{\mathrm{z}}}$, SIP3D, and \PQb-tag veto are collectively referred to as the lepton displacement veto.

These reconstruction and selection requirements result in typical efficiencies of 40--85\% for electrons, 65--90\% for muons, and 30--50\% for \tauh leptons, depending on the lepton \pt and $\eta$, as evaluated for prompt leptons originating from \PW and \PZ boson decays.
Similarly, the {\PQb} tagging working point achieves an identification efficiency of 60--75\% for {\PQb} quark jets depending on jet \pt and $\eta$, and a misidentification probability of about 10 (1)\% for {\cPqc} quark (light-quark and gluon) jets, respectively, as evaluated in events enriched in ${\ttbar}$ and multijet processes.

\begin{table*}[hbt!]
\centering
\topcaption{
A summary of control regions for the SM processes \ZZ, \ZG, \WZ, and \ttZ, and for the misidentified lepton backgrounds (MisID $\Pe/\PGm$ and MisID \PGt ).
The \ptmiss, \MT, 3L minimum lepton transverse momentum $\ptlthree$, $\ml$, and $\ST$ quantities are given in units of \GeV.
The 3L OnZ CR is further split into 3L MisID $\Pe/\PGm$ CR, 3L \WZ CR, and 3L \ttZ CR. The terminology is described in Section~\ref{sec:eventselec}.
}\label{tab:CRdef}
\cmsTable{
\renewcommand{\arraystretch}{1.1}
\begin{scotch}{l c c c c c c c}
  CR name                                             & \nossf & \mossf     & \nbj   & $\,\ptmiss$ & \MT     & $\ptlthree$ & $\,$Other selections \\ \hline
4L \ZZ                                            & OSSF2  & Double-OnZ & 0      & $\,\NA$     & \NA     & \NA         & $\,\NA$ \\
3L \ZG                                            & OSSF1  & BelowZ     & 0      & $\,\NA$     & \NA     & \NA         & $\,76<\ml<106$ \\[\cmsTabSkip]
3L OnZ                                              & OSSF1  & OnZ        & \NA    & $\,<$125    & $<$150  & \NA         & $\,\NA$ \\
\multicolumn{1}{l}{\,\,\,\,\,\,3L \WZ}            & OSSF1  & OnZ        & 0      & $\,<$125    & 50--150 & $>$20       & $\,\NA$ \\
\multicolumn{1}{l}{\,\,\,\,\,\,3L \ttZ}           & OSSF1  & OnZ        & $\ge$1 & $\,<$125    & $<$150  & $>$20       & $\,\nj\ge3$, $\,\;\ST>350$ \\
\multicolumn{1}{l}{\,\,\,\,\,\,3L MisID $\Pe/\PGm$} & OSSF1  & OnZ        & 0      & $\,<$100    & $<$50   & \NA         & $\,\NA$ \\[\cmsTabSkip]
2L1T MisID \PGt                                     & OSSF1  & OnZ        & \NA    & $\,<$100    & \NA     & \NA         & $\,\NA$ \\
\end{scotch}
}
\end{table*}

\section{Event selection and background estimation} \label{sec:eventselec}

In this analysis, events with three or more leptons satisfying the selection criteria given in Section~\ref{sec:objectSelections} are considered. 
Among these leptons, events must contain at least one muon with $\pt>26\;(29)\,\GeV$ in 2016 and 2018 (2017) or at least one electron with $\pt>30\;(35)\,\GeV$ in 2016 (2017--2018).
The analysis thresholds are set 3\GeV (2\GeV) higher than electron (muon) trigger thresholds. 

All events are categorized in seven distinct final states (channels) based on the number of light leptons ($\mathrm{L}= \Pe, \PGm$) and \tauh (T) candidates. 
These seven channels are mutually exclusive, and are defined as (i) 1L2T, with exactly one light lepton and exactly two \tauh candidates; (ii) 1L3T, with exactly one light lepton and three or more \tauh candidates; (iii) 2L1T, with exactly two light leptons and exactly one \tauh candidates; (iv) 2L2T, with exactly two light leptons and two or more \tauh candidates; (v) 3L, with exactly three light leptons and no \tauh candidates; (vi) 3L1T, with exactly three light leptons and one or more \tauh candidates; and (vii) 4L, with four or more light leptons and any number of \tauh candidates.

In the 4L channel, only the leading four light leptons in \pt are used in the subsequent analysis. 
Likewise, in the 3L1T, 2L2T, and 1L3T channels, only the leading one, two, and three \tauh candidates are used, respectively.

The main SM backgrounds are constrained using dedicated control regions (CRs), as described in detail in Sections~\ref{subsec:PromptBG} and \ref{subsec:MisID} and summarized in Table~\ref{tab:CRdef}.
Lepton and jet multiplicity categorization, transverse momenta, and invariant mass variables are used to define CRs enriched in specific SM background processes or MisID backgrounds. 
The CRs are then excluded from the analysis phase space. 
The remaining events are grouped in different signal regions (SR) to increase signal sensitivity across the probed dilepton mass spectra. 
The SRs are described in detail in Section~\ref{subsec:SignalRegions}. 

\begin{figure*}[hbt!]
\includegraphics[width=1.0\textwidth]{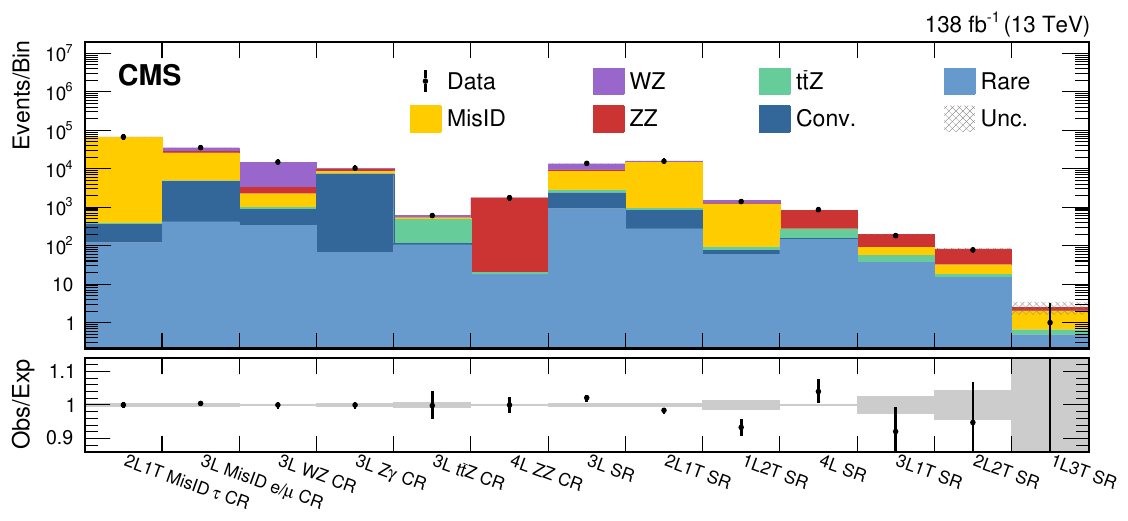}
\caption{\label{fig:CRSR3L4Lbins} 
Binned representation of the control and signal regions for the combined multilepton event selection and the combined 2016--2018 data set.
The CR bins follow their definitions as given in Table~\ref{tab:CRdef}, and the SR bins correspond to the channels as defined by the lepton flavor composition. 
The normalizations of the background samples in the CRs are described in Sections~\ref{subsec:PromptBG} and \ref{subsec:MisID}.
All three (four) lepton events are required to have $\ql=1~(0)$, and those satisfying any of the CR requirements are removed from the SR bins. 
All subsequent selections given in Tables~\ref{table:XPhieemmSR} and~\ref{table:XPhitautauSR} are based on events given in the SR bins. 
The lower panel shows the ratio of observed events to the total expected SM background prediction (Obs/Exp), and the gray band represents the statistical uncertainties in the background prediction.
}
\end{figure*}

In order to suppress contributions due to low-mass quarkonium resonances, cascade decays of heavy-flavor hadrons, and low-$\dR$ final-state radiation, events are discarded if $\mmin<12\,\GeV$ or $\dRmin<0.2$, where $\mmin$ and $\dRmin$ are defined as the minimum Lorentz-invariant mass and minimum $\dR$ of all lepton pairs in the event, irrespective of lepton charge and flavor. 
This requirement is applied to all CR and SR events.
A binned representation of the CRs and SRs illustrating the expected background composition is given in Fig.~\ref{fig:CRSR3L4Lbins}, where the normalizations of the background samples in the CRs are as described in Sections~\ref{subsec:PromptBG} and \ref{subsec:MisID}.

The multiplicity of jets ($\PQb$ jets) satisfying the selection criteria stated in Section~\ref{sec:objectSelections} is denoted as $\nj$ ($\nbj$).
The $\LT$ and $\HT$ variables are defined as the scalar \pt sum of all charged leptons (\Pe, \PGm, \tauh) and jets in an event, respectively. 
The invariant mass of all leptons in a given event is defined as $\ml$, and the absolute value of the charge sum of all leptons as $\ql$.
For the $\LT$, $\ml$, and $\ql$ calculations, only the charged leptons used to define the channel are used.
The variable $\ST$ is defined as the scalar sum of $\LT$, $\HT$, and \ptmiss.

The number of distinct opposite-sign same-flavor lepton pairs in an event is denoted as $\nossf$, where $n$ is the number of OSSF pairs. 
Some examples of the OSSF1 events are $\Pep\Pem\Pep$ in 3L,  $\PGmp\PGmm\tauh^+$ in 2L1T, $\Pep\Pem\Pep\PGmm$ in 4L. 
Only 4L and 2L2T events can be categorized as OSSF2, for example $\Pep\Pem\PGmp\PGmm$ in 4L and $\PGmp\PGmm\tauh^+\tauh^-$ in 2L2T. 
OSSF0 events are all those with no opposite-sign same-flavor pairs, as examples $\Pep\Pep\PGmp$ or $\Pep\Pep\Pep$ in 3L. 

\begin{figure*}[hbt!]
\includegraphics[width=1.0\textwidth]{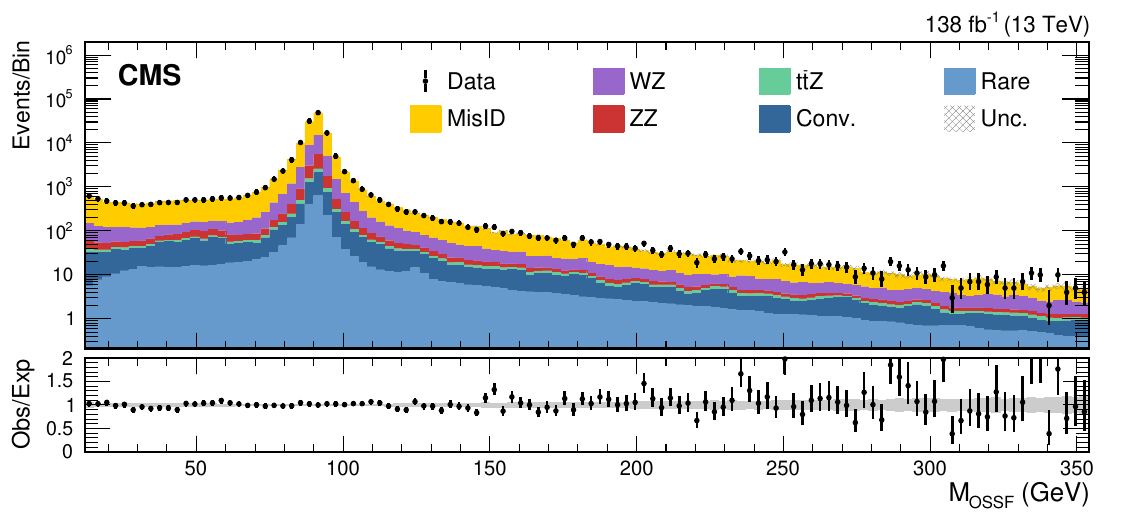}
\caption{\label{fig:CRSR3L4Lmass} 
The $\mossf$ spectrum for the combined 2L1T, 2L2T, 3L, 3L1T, and 4L event selection (excluding the \ZG CR) and the combined 2016--2018 data set.
All three (four) lepton events are required to have $\ql=1\,(0)$.
The lower panel shows the ratio of observed events to the total expected SM background prediction (Obs/Exp), and the gray band represents the statistical uncertainties in the background prediction.
}
\end{figure*}

To identify an event likely to contain a \PZ boson, $\mossf$ is defined to be the invariant mass of the OSSF dielectron or dimuon pair that is closest to 91\GeV, and an event with $76<\mossf <106\GeV$ is labeled as OnZ.
The OSSF1 or OSSF2 events that are not OnZ are labeled as OffZ, and if an OffZ event has $\mossf<76\GeV$, it is classified as BelowZ. 
The minimum lepton \pt in 3-lepton events is referred to as $\ptlthree$. 
The transverse mass for a single lepton $i$ is defined as $\MT^i = (2\ptmiss \pt^{i}[1-\cos(\ptvecmiss,\ptvec^{\; i})])^{1/2}$, where $\pt^{i}$ is the \pt of lepton $i$.
In 3L OSSF1 events, the lepton that is not used in the $\mossf$ pair is used in the calculation of $\MT$.
In the 3L events with two non-distinct OSSF pairs (such as in $\Pep\Pem\Pep$ or $\PGmp\PGmm\PGmp$), the events are classified as OnZ if either pair satisfies $76<\mossf<106\GeV$. 
Also in such events, the assignment of which leptons form the OSSF pair and which is used in the $\MT$ calculation is made simultaneously so that the event is OnZ, and $\MT$ is in the range 50--150\GeV. 
If such a choice is not kinematically possible, building an OnZ candidate is prioritized. 
Similarly, in 4L OSSF2 events with four electrons or muons, $\mossf$ is chosen to give the maximum number of distinct OSSF OnZ pairs, and such events are labeled as Single- or Double-OnZ, depending on whether they have one or two distinct OnZ OSSF pairs.
The $\mossf$ spectrum for the combined 2L1T, 2L2T, 3L, 3L1T, and 4L event selection (excluding the \ZG CR) is illustrated in Fig.~\ref{fig:CRSR3L4Lmass}.

\subsection{Prompt-lepton backgrounds}  \label{subsec:PromptBG}

The prompt-lepton backgrounds arise from processes in which all reconstructed leptons originate from decays of SM bosons. 
These backgrounds are irreducible, and their contributions are estimated with simulated event samples, which have been normalized and validated using data in dedicated CRs for each of the major \WZ, \ZZ, \ttZ, and \ZG processes, as summarized in Table~\ref{tab:CRdef}.
The normalizations for these processes are typically found to be within 20\% of the NLO theoretical cross sections, and are applied together with the associated uncertainties to the corresponding background estimates in the SRs.
These uncertainties include both statistical and systematic contributions, and take into account the contributions of events from other processes.
The measurements for the diboson processes are largely independent of one another because of the high purity of the corresponding CRs.
Since these backgrounds make significant contributions to the \ttZ-enriched CR, the normalization for this process is measured after the corresponding corrections have been obtained for the other backgrounds.

The $\ZZ\to 4\Pell$ and $\WZ\to 3\Pell\PGn$ processes are the primary prompt-background processes in the channels with four and three leptons, respectively. 
The $\PQq\PQq\to \ZZ$ and $\Pg\Pg\to \ZZ$ processes are considered collectively as the \ZZ background.
The 4L \ZZ CR require two OSSF lepton pairs with invariant masses consistent with the $\PZ$ boson mass. While the 3L \WZ CR require one OnZ lepton pair and $\MT$ consistent with the \PW mass. 
The 4L \ZZ CR has a purity greater than 99\%, whereas that of the 3L \WZ CR is greater than 75\%.
In both CRs, events with \PQb-tagged jets are vetoed, and in the 3L \WZ CR, the minimum lepton \pt cut is raised to $20\GeV$ and \ptmiss is required to be less than 125\GeV to suppress contributions from other background processes.
Relative uncertainties of 3--5\% are observed in the normalizations across the three data-taking periods for these processes.
The \ZZ and \WZ simulation samples are reweighted as functions of the jet multiplicity as well as the visible diboson \pt to match the simulated distributions to those of the data in these CRs, where the visible diboson \pt is defined as the vector \pt sum of the charged leptons in the event. 
This reweighting accounts for missing higher-order QCD and electroweak corrections, and yields an improved description of leptonic and hadronic quantities of interest in this analysis.

Production of \ttZ is a major prompt SM background process for all channels with $\nbj\geq1$.
A \ttZ enriched CR is created by selecting 3L events similarly to the 3L \WZ CR, but with inverted requirement of at least one \PQb-tagged jet, and with additional requirements of at least 3 jets and $\ST$ greater than 350\GeV.
The purity of the  3L \ttZ CR selection is about 60\%, and relative uncertainties of 15--25\% are measured in the normalizations across the three data-taking periods.

A smaller background contribution arises from initial- or final-state radiation photons that convert asymmetrically such that only one of the resultant leptons is reconstructed in the detector. 
The DY process with an additional photon is the dominant source of such backgrounds, collectively referred to as the conversion background.
The cross section of this process is normalized in a dedicated 3L \ZG CR, where the mass of the three-lepton system is required to be within the Z mass window, $(91\pm15)\GeV$, and events with \PQb-tagged jets are vetoed.
This CR targets $\PZ\to\Pell\Pell+\PGg$ events, where, for example, the photon converts in the detector and the energy of one of the four leptons is too low to satisfy the lepton selection criteria. 
Relative uncertainties of about 10\% are obtained in the normalizations across the three data-taking periods, where the quoted value also includes a lepton-flavor-dependent component because the fractions of internal and external conversions vary as a function of the electron multiplicity in the events.

Other SM processes that are not normalized in a dedicated CR in data are estimated from simulation samples and normalized to their theoretical cross sections. 
These processes consist of triboson, Higgs boson, and other rare SM contributions, and are collectively referred to as ``rare'' backgrounds. 

\subsection{Misidentified-lepton backgrounds} \label{subsec:MisID}

The misidentified-lepton backgrounds are estimated from data using three- or four-dimensional implementations of a matrix method~\cite{CMS:2015nep}, where the dimensionality corresponds to the lepton multiplicity in the targeted SRs.
For a given SR, the matrix method involves extrapolating the misidentified-lepton contributions from a set of sideband regions into the SR event-by-event using lepton ``misidentification'' and ``prompt'' rates. 

The sideband regions include all the lepton selection criteria described in Section~\ref{sec:objectSelections}, with the exception of the isolation requirements, which are relaxed.
We refer to the requirements in the SRs and sideband regions as ``tight'' and ``loose'', respectively.
Specifically, for the loose requirement, the relative isolation must be less than 1.0 for electrons and muons, and \tauh candidates are required to pass a looser working point of the $\textsc{DeepTau}$ algorithm jet discriminator.
Accounting for the possible combinations, for a given SR with 3 or 4 leptons, the matrix method uses 7 or 15 nonoverlapping sideband regions, respectively.
The sideband regions are mutually exclusive to the SRs by construction.

The misidentification rates are defined as the probabilities that misidentified leptons pass the tight selection, given that they satisfy the loose selection. 
Because of the isolation requirements used in the single-lepton triggers, background contributions with up to 2 (3) simultaneously misidentified leptons in 3 (4) lepton events can be predicted by this method. 
The fraction of signal events where all lepton candidates are misidentified leptons is found to be negligible in simulation. 
The DY+jets and $\ttbar$+jets processes are the dominant SM contributions to the total misidentified-lepton background in multilepton events.
Misidentification rates are derived in dedicated CRs in data and used in the matrix method in SRs.
Data-based misidentification rate measurements in DY+jets events are performed using a variant of the ``tag-and-probe'' method~\cite{CMS:2011aa} in three-lepton CR events.
In the 3L MisID and 2L1T MisID CRs, the OnZ leptons are taken as the tag leptons, and the additional lepton is taken as the misidentified-lepton probe, \eg, $\Pe\Pe\PGm$ and $\PGm\PGm\PGm$ events are used to measure the muon misidentification rates, while $\Pe\Pe\tauh$ and $\PGm\PGm\tauh$ events are used to measure the \tauh misidentification rates. 
In all data-based misidentification rate measurements, contributions due to prompt probe leptons are estimated and subtracted using MC simulation. 
Misidentification rates in $\ttbar$+jets events may differ by up to $50\%$ from those in DY+jets events for a given lepton flavor, because of different gluon, light quark, and heavy quark compositions, as well as different event kinematic properties.
As it is impractical to create a high purity $\ttbar$+jets enriched selection of events with well-defined misidentified-lepton probes in data, dedicated $\ttbar$+jets misidentification rates for all lepton flavors are obtained in simulated samples instead, using object and event selections compatible with the SR selections. 
These simulation-based rates are verified in dedicated data CRs enriched in $\ttbar$+jets contributions with a misidentified lepton, where one lepton is required to fail the three-dimensional impact parameter significance requirement or the {\PQb} tag veto described in Section~\ref{sec:objectSelections}. 

The lepton misidentification rates are measured as functions of various kinematic features of leptons and the hadronic properties of events that affect the lepton isolation.
All misidentification rates are parametrized as functions of the lepton \pt and $\abs{\eta}$.
For tau leptons, the misidentification rates are measured separately for one- and three-prong reconstructed \tauh candidates. 
The misidentification rate for each lepton flavor is corrected as a function of the recoil of the event, as well as the multiplicity of tracks originating from the PV and the jet multiplicity.
The recoil is defined as a projection along the lepton \pt axis of the vector sum of the \pt of all other leptons, jets, and \ptmiss in the event.
The associated corrections significantly improve the modeling of misidentified-lepton backgrounds in DY+jets events, in which the misidentified lepton often originates from a jet recoiling against the leptonically decaying \PZ boson system.
The final misidentification rates for all lepton flavors are obtained from a weighted average of the DY- and $\ttbar$-based measurements.
The weights are evaluated according to the expected DY-$\ttbar$ composition of the MisID background, as obtained from simulated samples in each SR category and for each \PQb-tagged jet multiplicity. 
Half of the difference between the rates derived from DY- and $\ttbar$-based measurements is assigned as a systematic uncertainty to account for inaccurate modeling of the expected background composition.
Typical electron and muon misidentification rates, relative to the loose selection, are in the range 5--30\%, whereas those of \tauh objects are found to be in the range 1--15\%.

Similarly, the prompt rates are defined as the probability for prompt and isolated leptons to pass the tight criteria given that they pass the loose criteria.
Within the matrix method, they constitute a correction, and account for the contribution to the lepton sidebands from prompt leptons that fail the tight selection criteria.
In data, the prompt rates for electrons, muons, and \tauh leptons are measured using the tag-and-probe method in DY-enriched $\Pe\Pe$, $\PGm\PGm$, and $\Pe\tauh/\PGm\tauh$ dilepton events, respectively, as functions of the lepton \pt and $\abs{\eta}$, and are found to be in the range 50--95\%.
In simulation, the prompt rates are similarly measured in DY and $\ttbar$ MC samples, using reconstructed leptons kinematically matched to generator-level prompt leptons ($\dR<0.2$).
The final prompt rates for all lepton flavors are taken from the DY-enriched data measurements.
The differences between the prompt rates derived from DY and $\ttbar$ MC samples are studied to assess their dependence on hadronic activity, and the impact of such systematic uncertainty contributions on the misidentified-lepton background estimate is found to be negligible. 

\begin{table*}[hbt!]
\centering
\topcaption{
Low- and high-mass signal region selections for $\Xphi\to\Pe\Pe/\PGm\PGm$ signals.
Events satisfying the control region requirements are vetoed throughout, and only those with a reconstructed $\phi$ candidate are retained using the specified dilepton mass variable.
The $\ST$, $\ptlthree$, and $\ml$ requirements are specified in units of \GeV.
The two entries in the labels, channels, and dilepton mass variables are provided for the $X\phi\to\Pe\Pe$ and $X\phi\to\PGm\PGm$ signal scenarios, as appropriate.
}\label{table:XPhieemmSR}
\cmsTable{
\renewcommand{\arraystretch}{1.1}
\begin{scotch}{  l c c c c c  c c c  c  c }
Label                             & Channels                     & \ql & \nossf  & $\mossf$       & $\nbj$  & $\ST$  & $\ptlthree$ & $\ml$    & Dilepton mass   \\ \hline 
$\Wphi(\Pe\Pe/\PGm\PGm)$ SR1Low   & 3L$(\Pe\Pe\PGm/\Pe\PGm\PGm)$ & 1   & 1       & OffZ           & 0       &\NA     &\NA      & $<$76,$\;>$106 & \mee/\mmm       \\
$\Wphi(\Pe\Pe/\PGm\PGm)$ SR2Low   & 3L$(\Pe\Pe\Pe/\PGm\PGm\PGm)$ & 1   & 1       & OffZ           & 0       &\NA     &\NA      & $<$76,$\;>$106 & \meemin/\mmmmin \\
$\Wphi(\Pe\Pe/\PGm\PGm)$ SR1High  & 3L$(\Pe\Pe\PGm/\Pe\PGm\PGm)$ & 1   & 1       & OffZ           & 0       & $>$200 & $>$15   & $>$150       & \mee/\mmm       \\
$\Wphi(\Pe\Pe/\PGm\PGm)$ SR2High  & 3L$(\Pe\Pe\Pe/\PGm\PGm\PGm)$ & 1   & 1       & OffZ           & 0       & $>$200 & $>$15   & $>$150       & \meemax/\mmmmax \\[\cmsTabSkip]
$\Zphi(\Pe\Pe/\PGm\PGm)$ SRLow    & 4L+3L1T+2L2T                 & 0   & $\geq$1 & Not double-OnZ & 0       &\NA     &\NA      &\NA           & \meemin/\mmmmin \\
$\Zphi(\Pe\Pe/\PGm\PGm)$ SRHigh   & 4L+3L1T+2L2T                 & 0   & $\geq$1 & Not double-OnZ & 0       & $>$200 &\NA      & $>$150       & \meemax/\mmmmax \\[\cmsTabSkip]
$\ttphi(\Pe\Pe/\PGm\PGm)$ SR1Low  & 3L$(\Pe\Pe\PGm/\Pe\PGm\PGm)$ & 1   & 1       & OffZ           & $\geq$1 & $>$350 &\NA      & $>$100       & \mee/\mmm       \\
$\ttphi(\Pe\Pe/\PGm\PGm)$ SR2Low  & 3L$(\Pe\Pe\Pe/\PGm\PGm\PGm)$ & 1   & 1       & OffZ           & $\geq$1 & $>$350 &\NA      & $>$100       & \meemin/\mmmmin \\
$\ttphi(\Pe\Pe/\PGm\PGm)$ SR1High & 3L$(\Pe\Pe\PGm/\Pe\PGm\PGm)$ & 1   & 1       & OffZ           & $\geq$1 & $>$400 & $>$15   & $>$100       & \mee/\mmm       \\
$\ttphi(\Pe\Pe/\PGm\PGm)$ SR2High & 3L$(\Pe\Pe\Pe/\PGm\PGm\PGm)$ & 1   & 1       & OffZ           & $\geq$1 & $>$400 & $>$15   & $>$100       & \meemax/\mmmmax \\
$\ttphi(\Pe\Pe/\PGm\PGm)$ SR3Low  & 4L+3L1T+2L2T                 & 0   & $\geq$1 & OffZ           &\NA      & $>$350 &\NA      &\NA           & \meemin/\mmmmin \\
$\ttphi(\Pe\Pe/\PGm\PGm)$ SR3High & 4L+3L1T+2L2T                 & 0   & $\geq$1 & OffZ           &\NA      & $>$400 &\NA      &\NA           & \meemax/\mmmmax \\
\end{scotch}
}
\end{table*}

\subsection{Signal regions} \label{subsec:SignalRegions}

The signals in the $\Xphi\to\Pe\Pe/\PGm\PGm$ scenarios are expected to produce narrow enhancements in reconstructed opposite-sign dielectron and dimuon mass ($\mee$ and $\mmm$) spectra, whereas wider enhancements are obtained in the $\Xphi\to\PGt\PGt$ scenarios because of undetected neutrinos originating from decays of tau leptons.
A set of reconstructed dilepton invariant mass variables with additional selection criteria discriminates the $\Xphi$ signal from the SM backgrounds, where the binning scheme and range of these variables are chosen to maximize sensitivity according to the expected signal shapes and SM backgrounds.
These are summarized in Tables~\ref{table:XPhieemmSR} and~\ref{table:XPhitautauSR} for the $\Xphi\to\Pe\Pe/\PGm\PGm$ narrow resonance search and the $\Xphi\to\PGt\PGt$ wide resonance search, respectively.

For the $\Xphi\to\Pe\Pe/\PGm\PGm$ signal scenarios, $\mee$ and $\mmm$ are utilized.
The SM-background-enriched mass window around the Z boson mass is excluded, dividing the mass spectra into two mass regions.
The SRs containing events with a $\phi$ mass below or above the $\PZ$ boson mass window are referred to as the low- or high-mass region, respectively. 
The low (high) dilepton mass range spans 12--76 (106--366)\GeV. 
In events with more than one opposite sign $\Pe\Pe$ pair, such as $\Pep\Pep\Pem$, masses are defined for two possible pairings of $\Pem$ and $\Pep$. The lower of the two $\phi$ masses is referred to as  $\meemin$ and the higher one  is referred to as $\meemax$. 
Dimuon decay scenarios are handled similarly, where $\mmmmin$ ($\mmmmax$) is the minimum (maximum) opposite-sign $\PGm\PGm$ invariant mass per event. 
On the other hand, for the $\Xphi\to\PGt\PGt$ signal, where tau can subsequently decay to $\Pe$ or $\PGm$, a different categorization based on three new mass variables is adopted because of the absence of a narrow resonance in DY background, where the full mass range above 12\GeV, including the $\PZ$ boson mass region, is used. 
These variables correspond to the minimum invariant mass among all $\tauh\tauh$ pairs ($\mtautaumin$), $\Pe\PGm$ pairs ($\memmin$), and $\Pe\tauh$ or $\PGm\tauh$ pairs ($\mltmin$), as well as the minimum of $\mtautaumin$ and $\mltmin$ ($\mltttmin$).
For all three decay modes of $\phi$, the targeted mass ranges for reconstructed dilepton masses are taken to be wider than the probed $\phi$ mass range (15--350\GeV) to minimize loss of signal acceptance due to detector resolution effects.

To increase signal sensitivity, further selections consisting of requirements on lepton and jet multiplicities, total lepton charge, minimum lepton \pt, and the combined invariant mass of all leptons in the event ($\ml$) are used. 
For the $\Xphi\to\Pe\Pe/\PGm\PGm$ samples with high mass, requiring larger $\ST$,  $\ptlthree$, and $\ml$ suppresses the MisID background and increases sensitivity. 
For the $\Xphi\to\PGt\PGt$ signals, because of significant contributions from misidentified tau lepton candidates, a minimum \pt threshold of 30\GeV is applied to the \tauh candidates in all 3-lepton channels. 
In various signal regions, minimum $\ST$ requirements are applied as an approximate measure of the effective mass of all particles produced in the targeted final state.
Signal events with $\ttbar$ pairs are distinguished from others based on $\nbj$. 
Specifically, the $\Zphi$ and $\Wphi$ SRs include only events with $\nbj=0$, while most $\ttphi$ SRs require $\nbj \geq1$. 
For the $\ttphi$ signal in the 4L channel, no requirement is imposed on $\nbj$ selection because of the high-$\ST$ requirement and the already low total background expectation.
Furthermore, all SRs are required to have a total absolute charge consistent with the probed signal scenario, i.e. $\ql=0\,(1)$  for all $4\,(3)$ lepton events.
In SRs with significant background contributions, events with a single $\phi$ candidate are considered separately from those  with more than one candidate, to mitigate the effects of dilepton mispairings in the resolution of the reconstructed $\phi$ mass.

In order to improve the $\phi$ selection efficiency in channels with multiple candidates, certain signal-specific kinematic features are used to help correctly identify the $\phi$ decay products.
In the $\Wphi(\Pe\Pe/\PGm\PGm)$ SR2 channels with ambiguity in OSSF pair construction, the lepton with $\MT$ closest to the \PW boson mass in the $(81\pm30)\GeV$ mass window is selected first, and the other two leptons are taken to reconstruct the $\phi$ mass. 
If no \PW candidate in the specified $\MT$ window is found, or if one is found but the remaining two leptons are not of opposite charge, then the dilepton pair with the minimum or maximum mass is used.
Similarly, in the $\ttphi(\Pe\Pe/\PGm\PGm)$ SR2 channels, the lepton with minimum $\MT$ is identified first, and the remaining two leptons are labeled as the $\phi$ decay products, provided they form an opposite-sign pair.
In the 4L channel of the  $\Zphi(\Pe\Pe/\PGm\PGm)$ SR with OSSF2, the $\PZ$ candidate is identified using the $\mossf$ variable and a mass window of $(91\pm15)\GeV$, and the other two leptons are taken as $\phi$ decay products. 
These $\phi$ selection algorithms were evaluated using simulations and found to correctly identify the $\phi$ decay products in more than 70\% of events with multiple $\phi$ candidates.

In each $\Xphi(\Pe\Pe/\PGm\PGm)$ SR, $1\,(5)\GeV$-wide dilepton mass bins are used to probe the targeted phase space for low (high) $\phi$ masses.
These bin widths are chosen to be consistent with the narrow-width assumption for the $\phi$ boson as well as the detector resolution effects on the $\mee$ and $\mmm$ spectra over the probed mass range.
The $\Xphi\to\Pe\Pe/\PGm\PGm$ signal mass hypotheses that are closer to the dilepton mass bin boundaries than to the bin centers are probed with a modified binning scheme, where the mass bin boundaries are shifted by half the value of the bin widths.
A smoothing procedure is separately applied to the prompt and MisID background contributions in these SRs, using a nonparametric kernel density estimation method with a Gaussian kernel~\cite{Cranmer:2000du}.
A fixed-width kernel is used except within 10\% from each edge in the mass spectra, where the kernel size is reduced significantly to prevent artificial shaping of the expected background distributions. 
The smoothing procedure mitigates the impact of statistical fluctuations in the expected background spectra and ensures a stable signal sensitivity across the probed mass bins. 
No additional uncertainties have been added for the smoothing procedure since different choices of kernel widths yield differences within the existing uncertainties.
No smoothing procedure is applied to $\Xphi(\PGt\PGt)$ SRs, as these mass spectra are binned using wider bins with variable widths ranging 5--60\GeV to achieve smoothly behaving expected background distributions.

\begin{table*}[hbt!]
\centering
\topcaption{
Signal selections for $\Xphi\to\PGt\PGt$ signals.
Events satisfying the control region requirements are vetoed throughout, and only those with a reconstructed $\phi$ candidate are retained using the specified dilepton mass variable.
The $\ST$, $\ptlthree$, and $\ml$ requirements are specified in units of \GeV.
}\label{table:XPhitautauSR}
\cmsTable{
\renewcommand{\arraystretch}{1.1}
\begin{scotch}{  l c c  c c c c  c c c  c  c }
Label                         & Channels       & \ql & \nossf  & $\mossf$      & $\nbj$  & $\ST$   & $\nj$ & $\ptlthree$ & $\ml$  & Dilepton mass \\ \hline 
$\Wphi(\PGt\PGt)$ SR1         & 3L             & 1   & 0       &\NA            & 0       & $>$200  &\NA    & $>$15       & $>$150 & $\memmin$ \\ 
$\Wphi(\PGt\PGt)$ SR2         & 2L1T+1L2T      & 1   & 0       &\NA            & 0       & $>$200  &\NA    & $>$30       & $>$150 & $\mltmin$ \\
$\Wphi(\PGt\PGt)$ SR3         & 1L2T           & 1   & 1       &\NA            & 0       & $>$200  &\NA    & $>$30       & $>$150 & $\mtautaumin$ \\[\cmsTabSkip]
$\Zphi(\PGt\PGt)$ SR1         & 4L+2L2T        & 0   & 1       &\NA            & 0       & $>$200  &\NA    &\NA          &\NA     & $\memmin$ \\
$\Zphi(\PGt\PGt)$ SR2         & 3L1T           & 0   & 1       &\NA            & 0       & $>$200  &\NA    &\NA          &\NA     & $\mltmin$ \\
$\Zphi(\PGt\PGt)$ SR2         & 2L2T           & 0   & 0       &\NA            & 0       & $>$200  &\NA    &\NA          &\NA     & $\mltmin$ \\
$\Zphi(\PGt\PGt)$ SR3         & 2L2T           & 0   & 2       &\NA            & 0       & $>$200  &\NA    &\NA          &\NA     & $\mtautaumin$ \\[\cmsTabSkip]
$\ttphi(\PGt\PGt)$ SR1        & 3L             & 1   & 0       &\NA            & 0       & $>$400  & $>$1  & $>$15       & $>$100 & $\memmin$ \\
$\ttphi(\PGt\PGt)$ SR2        & 2L1T+1L2T      & 1   & 0       &\NA            & 0       & $>$400  & $>$1  & $>$30       & $>$100 & $\mltmin$ \\
$\ttphi(\PGt\PGt)$ SR3        & 1L2T           & 1   & 1       &\NA            & 0       & $>$400  & $>$1  & $>$30       & $>$100 & $\mtautaumin$ \\
$\ttphi(\PGt\PGt)$ SR4        & 3L             & 1   & 1       & OffZ          & $>$0    & $>$400  & $>$1  & $>$15       & $>$100 & $\memmin$ \\
$\ttphi(\PGt\PGt)$ SR4        & 3L             & 1   & 0       &\NA            & $>$0    & $>$400  & $>$1  & $>$15       & $>$100 & $\memmin$ \\
$\ttphi(\PGt\PGt)$ SR5        & 2L1T+1L2T      & 1   & 0       &\NA            & $>$0    & $>$400  & $>$1  & $>$30       & $>$100 & $\mltmin$ \\
$\ttphi(\PGt\PGt)$ SR6        & 1L2T           & 1   & 1       &\NA            & $>$0    & $>$400  & $>$1  & $>$30       & $>$100 & $\mtautaumin$ \\
$\ttphi(\PGt\PGt)$ SR7        & 3L1T           & 0   & 1       & OffZ          &\NA      & $>$400  &\NA    &\NA          &\NA     & $\mltttmin$ \\
$\ttphi(\PGt\PGt)$ SR7        & 3L1T           & 0   & 0       &\NA            &\NA      & $>$400  &\NA    &\NA          &\NA     & $\mltttmin$ \\
$\ttphi(\PGt\PGt)$ SR7        & 2L2T           & 0   & 2       & OffZ          &\NA      & $>$400  &\NA    &\NA          &\NA     & $\mltttmin$ \\
$\ttphi(\PGt\PGt)$ SR7        & 2L2T           & 0   & $<$2    &\NA            &\NA      & $>$400  &\NA    &\NA          &\NA     & $\mltttmin$ \\
$\ttphi(\PGt\PGt)$ SR7        & 1L3T           & 0   & 1       &\NA            &\NA      & $>$400  &\NA    &\NA          &\NA     & $\mltttmin$ \\
\end{scotch}
}
\end{table*}

To summarize all mass spectra listed above, 6 low-mass and 6 high-mass $\mee$ and $\mmm$ spectra are probed for each of the $\Xphi\to\Pe\Pe/\PGm\PGm$ signals, and 13 dilepton mass spectra are probed for the $\Xphi\to\PGt\PGt$ signals. 
This results in a total of 37 mass spectra covering all dilepton decay modes.

\section{Systematic uncertainties}\label{sec:Systematics}

All background and signal estimates have uncertainties because of the finite number of events in simulated samples or data sidebands.
These statistical uncertainties are typically less important, but are nonetheless propagated to the results. 

Systematic uncertainties arise from the corrections applied to the background and signal simulations.
These include corrections for the efficiencies of the electron and muon triggers, electron charge misidentification probability, lepton reconstruction, identification and isolation, and lepton displacement veto selection, as well as the lepton energy scale and resolution modeling, $\PQb$ tagging efficiency, pileup modeling, and energy scale corrections for jets and $\ptmiss$.

Each of the uncertainty sources is studied for the main SM backgrounds (\WZ, \ZZ, \ttZ, and \ZG) and various signal samples covering all probed $\phi$ mass hypotheses in the different production modes. 
The impact of each source is evaluated by varying the corresponding correction factor up and down within one standard deviation of its associated uncertainty.
The resulting variations in the mass spectra are then used to define an envelope of the impact from each source of systematic uncertainty. 
The uncertainties that only affect the overall normalization of the expected backgrounds play a less important role in the resonant search, particularly in the  $\Xphi\to\Pe\Pe/\PGm\PGm$ signal scenarios. 
Such uncertainties are collectively labeled as ``flat'' in the discussion below.

Uncertainties in the lepton trigger and selection efficiencies are largely mass-independent, and are in the 1--15\% range, depending on the lepton flavor, \pt, and $\eta$.
Uncertainties affecting the lepton energies, which account for any mismodeling of the overall energy scale and resolution in simulated samples, are evaluated for the background and signal processes and are taken to be correlated. 
These uncertainty sources affect only the normalization of the background distributions. 
However, for the signal, they are the most important uncertainties, as they affect the mean and width of the reconstructed signal mass distributions. 
A maximum shift of $0.5\,(0.1)\%$ is observed in the mean of the reconstructed resonant $\phi$ mass distribution for dielectron (dimuon) decays. The width of the resonant signal changes by around $2\%$ for low $\phi$ masses and up to $6\%$ for the largest $\phi$ mass scenarios, for both electrons and muons.
The correction and uncertainty in the electron charge misidentification rate, obtained in a dedicated DY-enriched dielectron selection of data events, is found to have a negligible impact. 
Uncertainties in the signal acceptance, as well as the acceptance and cross section of the dominant SM backgrounds, due to the choices of factorization and renormalization scales~\cite{Cacciari:2003fi} and PDFs~\cite{Ball:2014uwa,Ball:2017nwa} are found to be negligible. 

The uncertainty in the integrated luminosity is partially correlated between data-taking years. 
The integrated luminosities of the 2016, 2017, and 2018 data-taking periods have uncorrelated uncertainties ranging from 1.2--2.5\%~\cite{CMS:2021xjt,CMS:2018elu,CMS:2019jhq}, and a correlated uncertainty of 1.6\%. 
The pileup modeling correction has an associated uncertainty of 3\% in the normalization of the dilepton mass distributions, evaluated by varying the total inelastic $\Pp\Pp$ cross section used in the correction procedure up and down by 5\%~\cite{CMS:2018mlc,ATLAS:2016ygv}.

Dedicated uncertainties are considered for the modeling of primary SM backgrounds, including \WZ, \ZZ, \ttZ, and \ZG  processes, which were normalized to data in dedicated CRs. 
The relative uncertainties in the normalizations for \WZ, \ZG, \ttZ, and \ZZ backgrounds are 3--5, 10, 15--25, and 4--5\%, respectively, in all three years of data collection.
The diboson \pt correction typically has a flat 1--5 (4--9)\% effect on the \WZ (\ZZ) background, while the jet multiplicity reweighting has an effect up to 10 (3--30)\%.
For the rare background processes, a $50\%$ systematic uncertainty is assigned to the theoretical cross section estimates to cover any higher-order effects and PDF uncertainties.

The uncertainty in the misidentified lepton background estimation, which is obtained from data via the matrix method, is dominated by the uncertainties in the lepton misidentification rates.
The relative statistical uncertainties in the measurement of the misidentification rates are the dominant source of uncertainty, and are typically in the 10--30\% range. 
As the \tauh (\Pe and \PGm) misidentification rates are extrapolated from low \pt to $\pt>80\,(50)\GeV$, these uncertainties are doubled, and a $60\%$ relative uncertainty for such high-\pt leptons is assigned.
In summary, the lepton misidentification rates have typical relative uncertainties of $10$, $30$, and $60\%$ in the low-\pt (10--20\GeV for light leptons and 10--30\GeV for \tauh), 
medium-\pt (20--50\GeV for light leptons and 30--80\GeV for \tauh), and high-\pt ($>50\GeV$ for light leptons and $>80\GeV$ for \tauh) regions, respectively.
These uncertainties in the total misidentified lepton background correspond to the contribution of 20--50\%.
The uncertainties are uncorrelated across lepton flavors, the three \pt regions, and the three data-taking periods.
In addition, process-dependent uncertainties in the lepton misidentification rates are considered as a separate source of uncertainty.
These are estimated by comparing the misidentification rates observed in the DY- and $\ttbar$-enriched samples, and are typically in the range 5--25$\%$, correlated across the data-taking periods.

The uncertainty sources, the affected processes, the resulting uncertainties in the yields of those processes, and the correlations across the data-taking periods are summarized in Table~\ref{tab:systematics}.
The overall uncertainties in the total expected backgrounds are largely dominated by those affecting the \WZ, \ZZ, \ttZ, and MisID processes. 

\begin{table*}[hbt!]
\centering
\topcaption{
Sources, magnitudes, impacts, and correlation properties of systematic uncertainties in the signal regions.
Magnitude refers to the relative change in the underlying uncertainty source, whereas impact quantifies the resultant relative change in the signal and background yields passing the event selection. 
Uncertainty sources marked as ``Yes'' under the Correlation column are correlated across the 3 years of data collection, and those marked with an asterisk in the Impact column are mass-dependent.
} \label{tab:systematics}
\cmsTable{
\renewcommand{\arraystretch}{1.1}
\begin{scotch}{l c c c c c}
Uncertainty source                          & Magnitude     & Type         & Processes               & Impact    & Correlation \\ \hline
Statistical                                 & 1--100\%      & Per event    & All MC samples          & 1--100\%     & No  \\ 
Integrated luminosity                       & 1.2--2.5\%    & Per event    & Conversion/Rare/Signal  & 1.2--2.5\%   & Yes \\ 
Pileup                                      & 5\%           & Per event    & All MC samples          & $<$5\%       & Yes \\
Trigger efficiency                          & 1--4\%        & Per lepton   & All MC samples          & $<$2\%       & No  \\
Electron reco., ID and iso. efficiency      & 1--5\%        & Per lepton   & All MC samples          & 1--3\%       & No  \\
Muon reco., ID and iso. efficiency          & 1--5\%        & Per lepton   & All MC samples          & 1--3\%       & No  \\
Tau lepton reco., ID and iso. efficiency    & 5--15\%       & Per lepton   & All MC samples          & 5--25\%*     & No  \\
Electron energy scale and resolution        & $<$2\%        & Per lepton   & All MC samples          & $<$10\%*     & Yes \\ 
Muon energy scale and resolution            & 2\%           & Per lepton   & All MC samples          & $<$10\%*     & No  \\
Tau lepton energy scale                     & $<$10\%       & Per lepton   & All MC samples          & $<$5\%*      & No  \\ 
Lepton displacement veto efficiency         & 1--2\%        & Per lepton   & All MC samples          & 3--5\%       & No  \\
$\PQb$ tagging efficiency                   & 1--10\%       & Per jet      & All MC samples          & 1--5\%       & No  \\ 
Jet energy scale                            & 1--10\%       & Per jet      & All MC samples          & $<$10\%      & No  \\
Unclustered energy scale                    & 1--25\%       & Per event    & All MC samples          & $<$3\%       & No  \\
Electron charge misidentification           & 30\%          & Per lepton   & All MC samples          & $<$1\%       & No  \\ 
\WZ normalization                           & 3--5\%        & Per event    & \WZ                     & 3--5\%       & No  \\
\ZZ normalization                           & 4--5\%        & Per event    & \ZZ                     & 4--5\%       & No  \\
\ttZ normalization                          & 15--25\%      & Per event    & \ttZ                    & 15--25\%     & No  \\
Conversion normalization                    & 10--50\%      & Per event    & \ZG/Conversion          & 10--50\%     & No  \\
Rare normalization                          & 50\%          & Per event    & Rare                    & 50\%         & No  \\
Prompt and misidentification rates          & 20--60\%      & Per lepton   & MisID                   & 20--50\%*    & No  \\
DY-$\ttbar$ process dependence              & 5--25\%       & Per lepton   & MisID                   & 5--25\%      & Yes \\
Diboson jet multiplicity modeling           & $<$30\%       & Per event    & \WZ/\ZZ                 & $<$30\%      & No  \\
Diboson \pt modeling                        & $<$30\%       & Per event    & \WZ/\ZZ                 & 1--10\%      & No  \\
\end{scotch}
}
\end{table*}

\section{Results}\label{sec:Results}

\subsection{Model-independent results}\label{subsec:ModelIndependentResults}

\begin{figure*}[hbt!]
\centering
\includegraphics[width=0.49\textwidth]{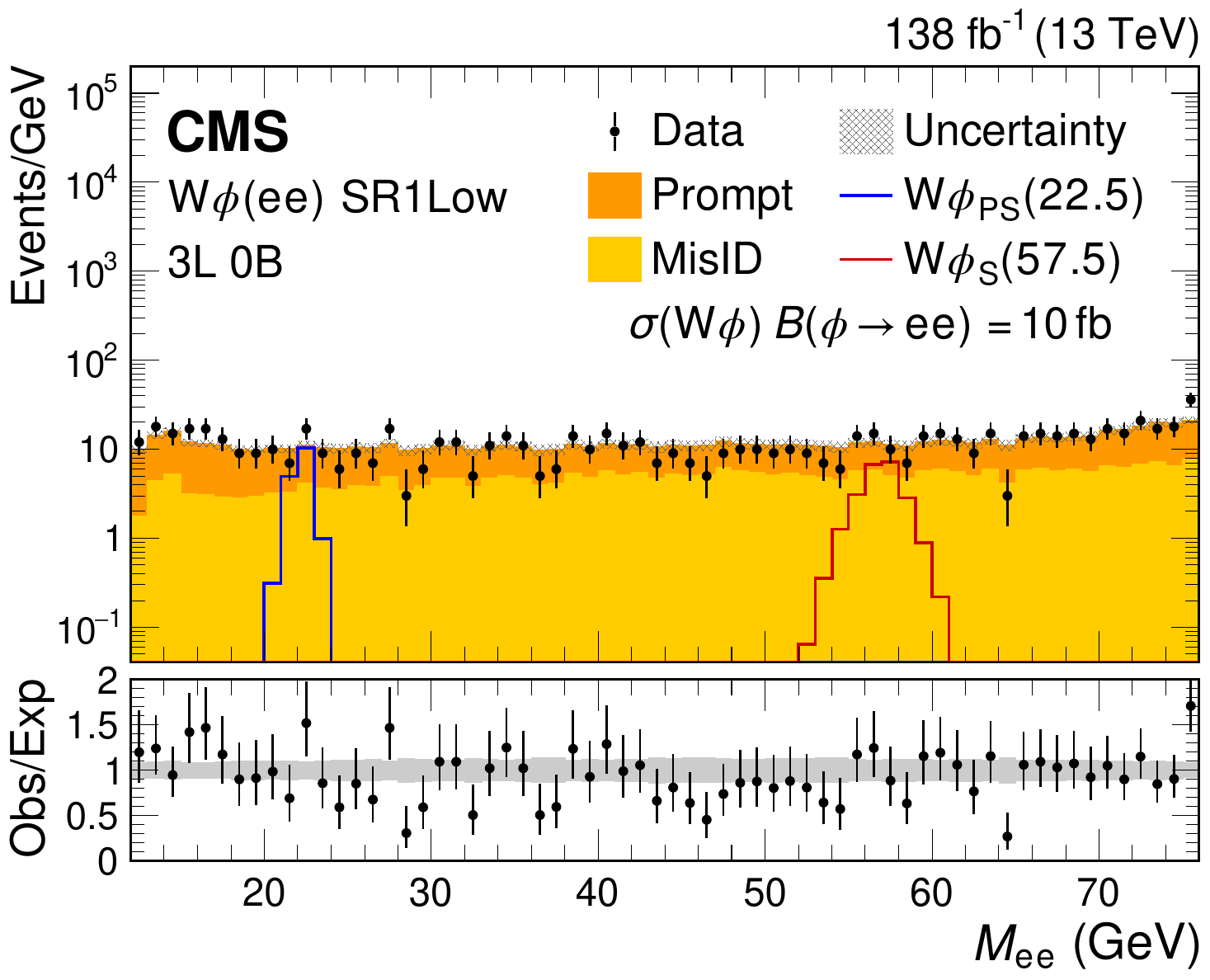}
\includegraphics[width=0.49\textwidth]{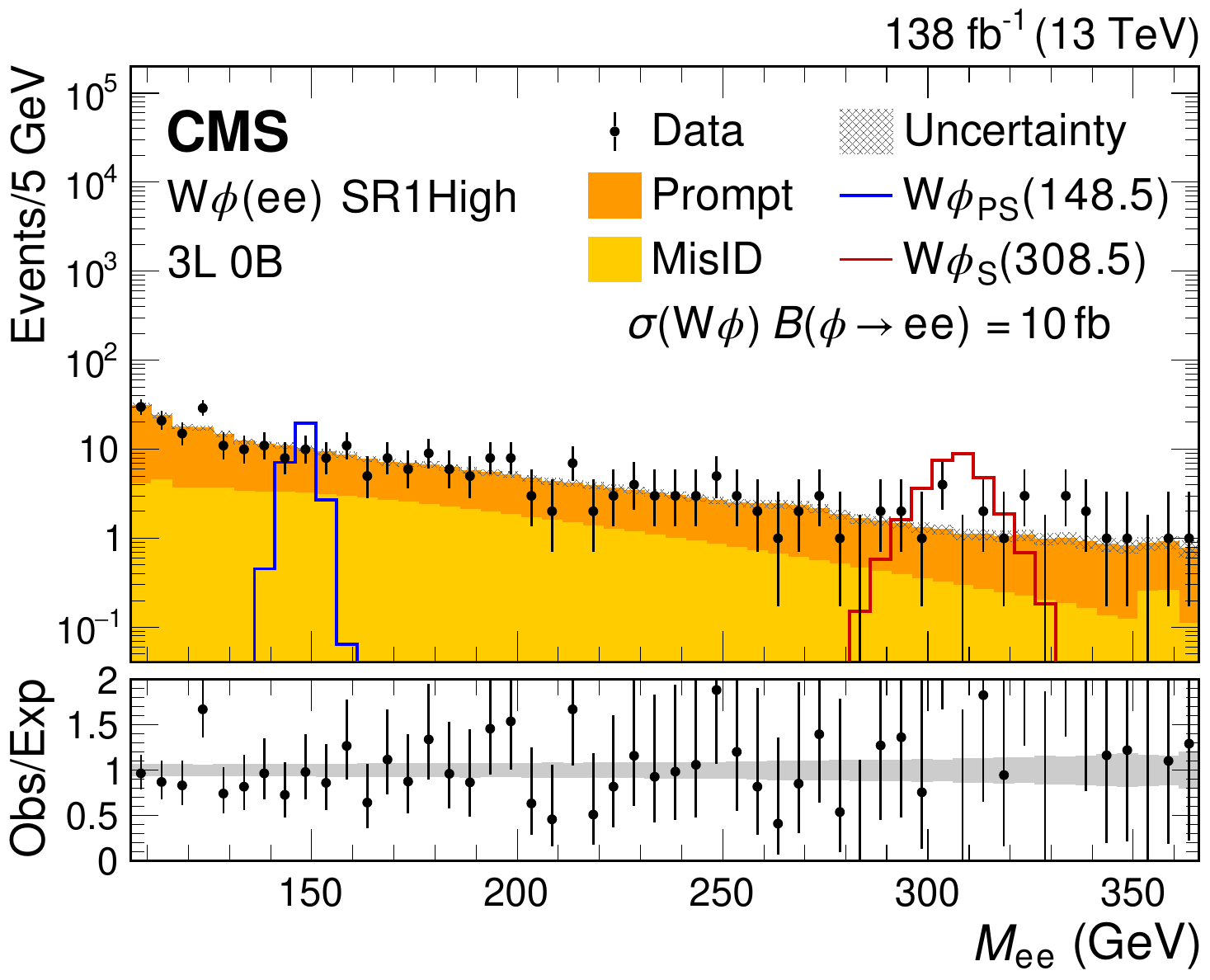}
\includegraphics[width=0.49\textwidth]{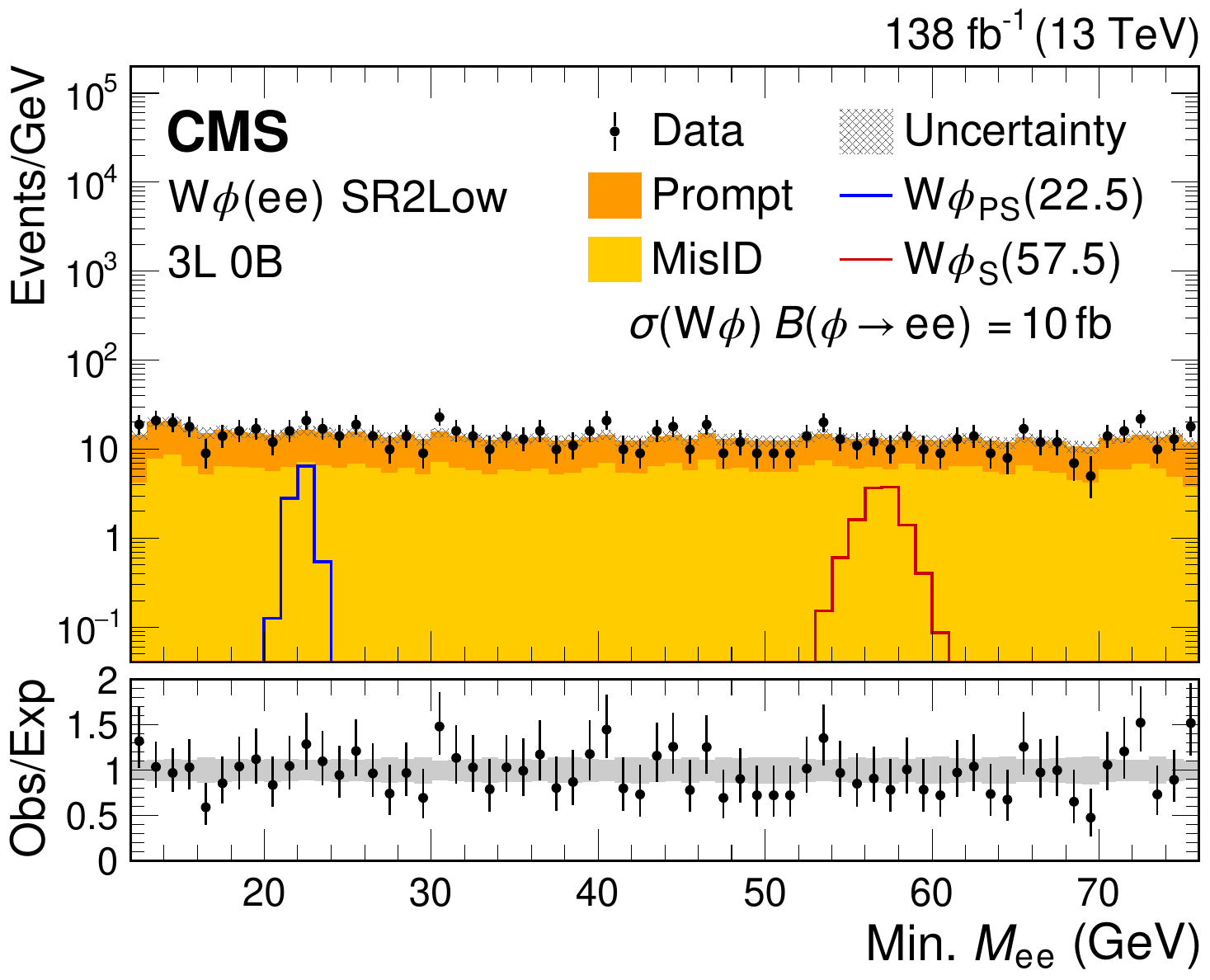}
\includegraphics[width=0.49\textwidth]{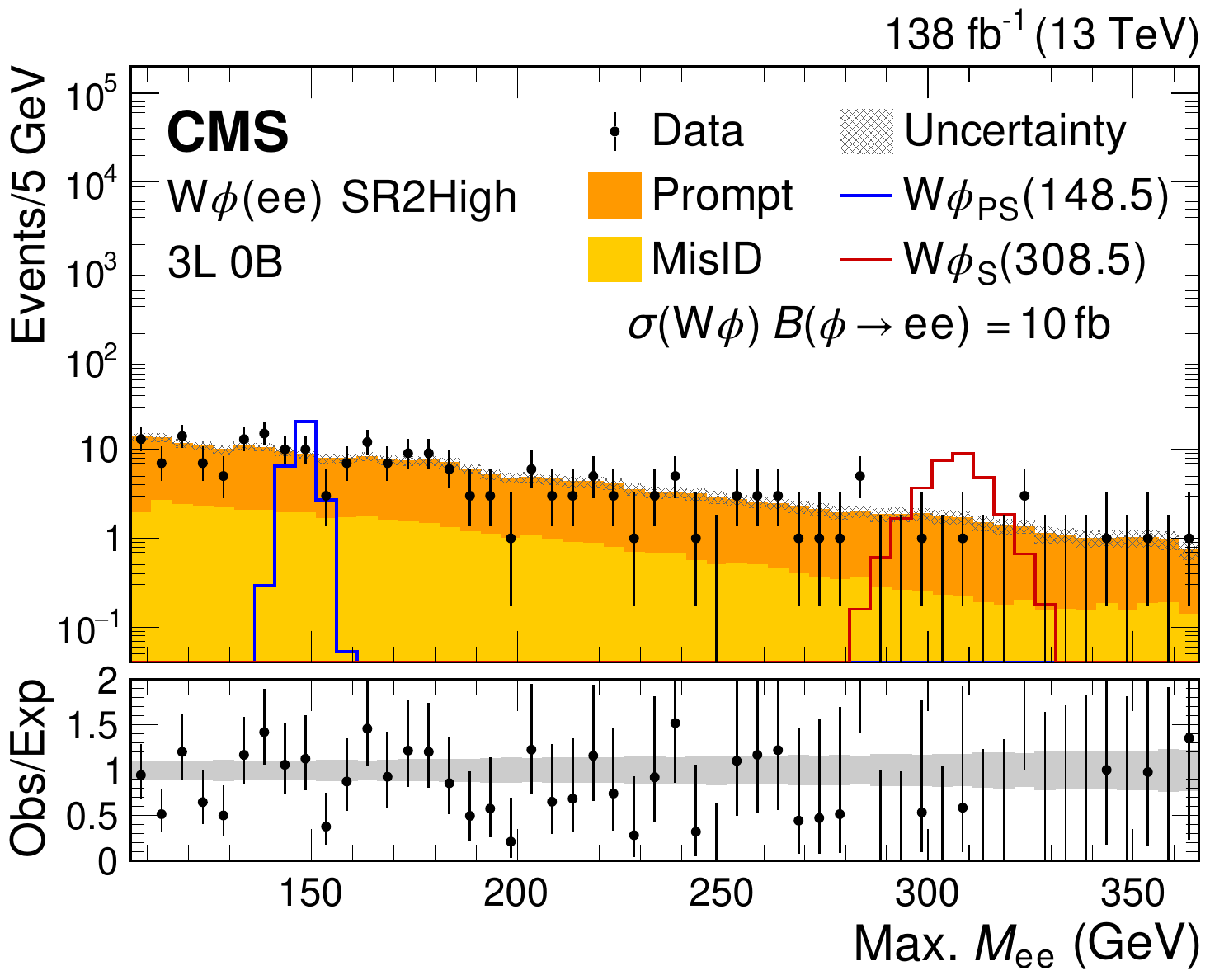}
\includegraphics[width=0.49\textwidth]{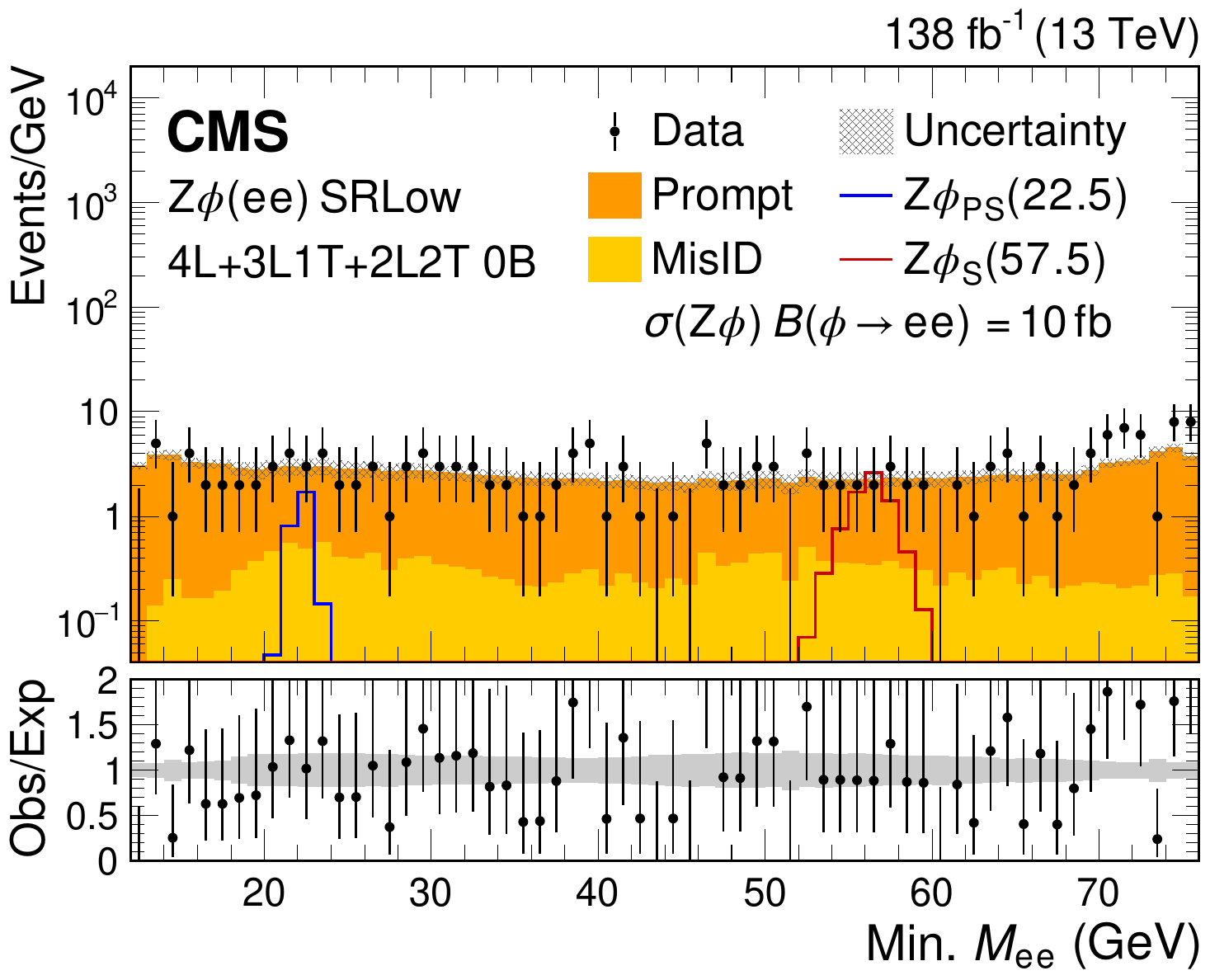}
\includegraphics[width=0.49\textwidth]{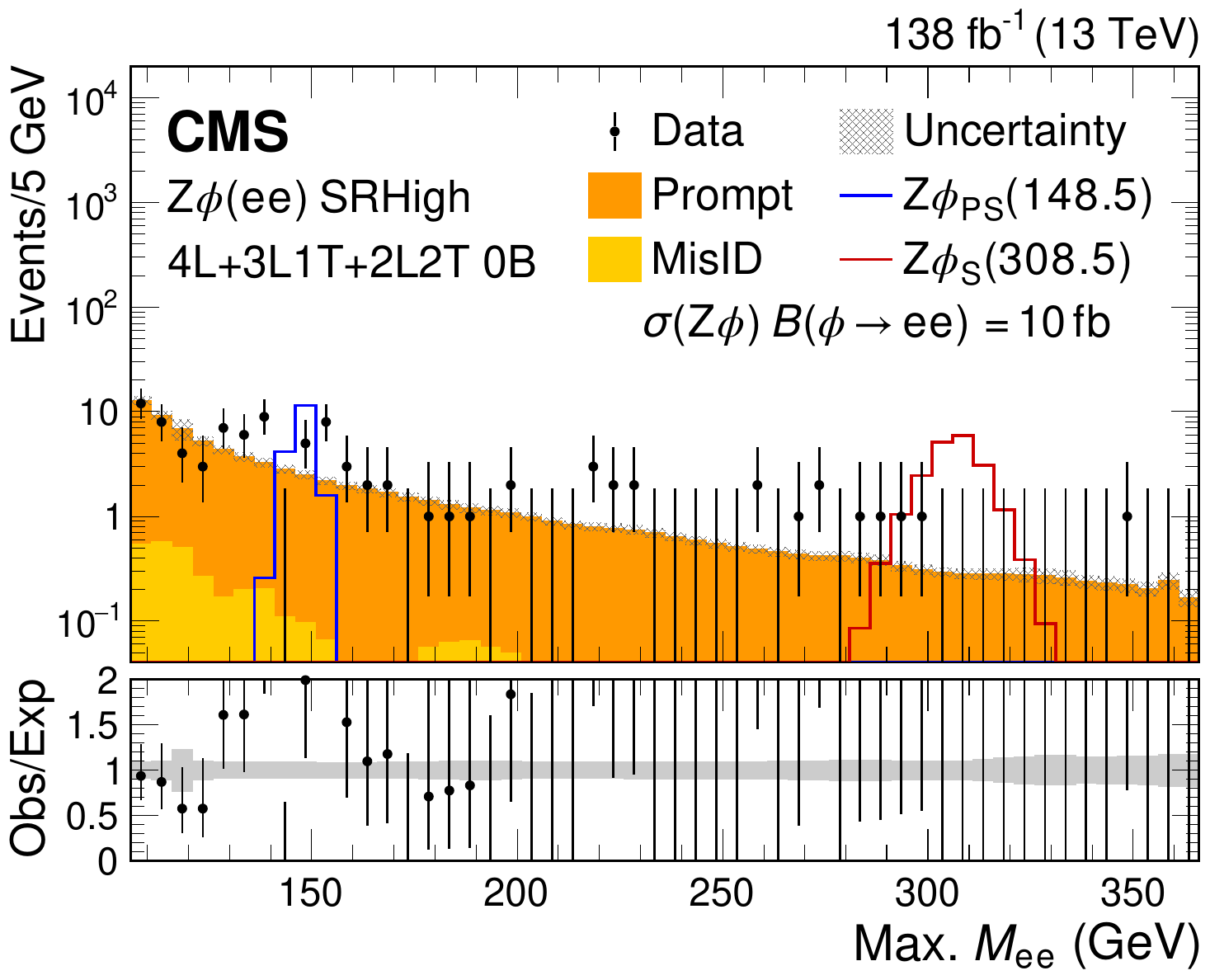}
\caption{\label{fig:WZPhiee} 
Dilepton mass spectra for the $\Wphi(\Pe\Pe)$ SR1 (upper), SR2 (middle), and for the $\Zphi(\Pe\Pe)$ SR (lower) event selections for the combined 2016--2018 data set.
The low (high) mass spectra are shown on the left (right).
The lower panel shows the ratio of observed events to the total expected SM background prediction (Obs/Exp), and the gray band represents the sum of statistical and systematic uncertainties in the background prediction. 
The expected background distributions and the uncertainties are shown after the data is fit under the background-only hypothesis.
For illustration, two example signal hypotheses for the production and decay of a scalar and a pseudoscalar $\phi$ boson are shown, and their masses (in units of \GeV) are indicated in the legend. 
The signals are normalized to the product of the cross section and branching fraction of 10\unit{fb}. 
}
\end{figure*}

\begin{figure*}[hbt!]
\centering
\includegraphics[width=0.49\textwidth]{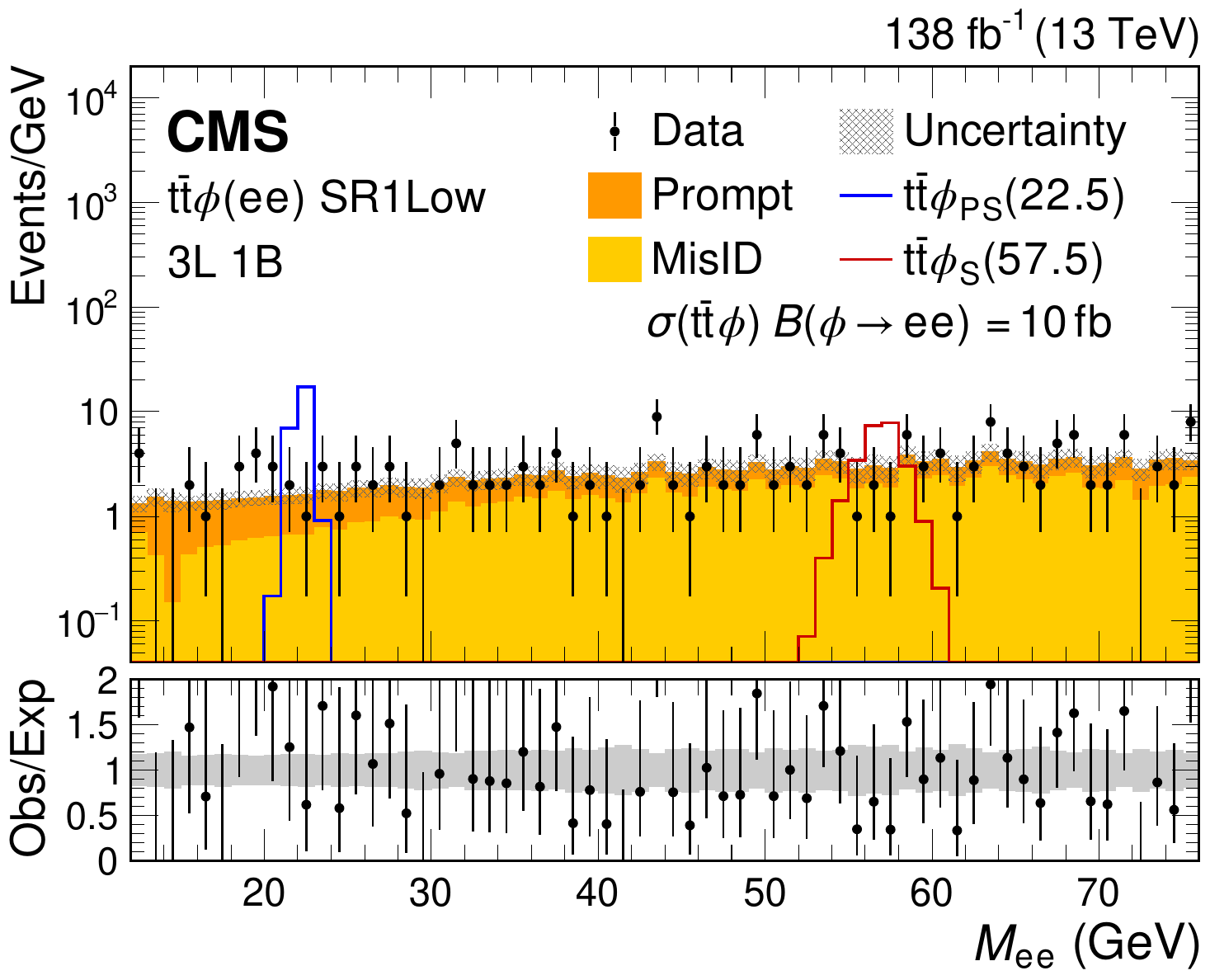}
\includegraphics[width=0.49\textwidth]{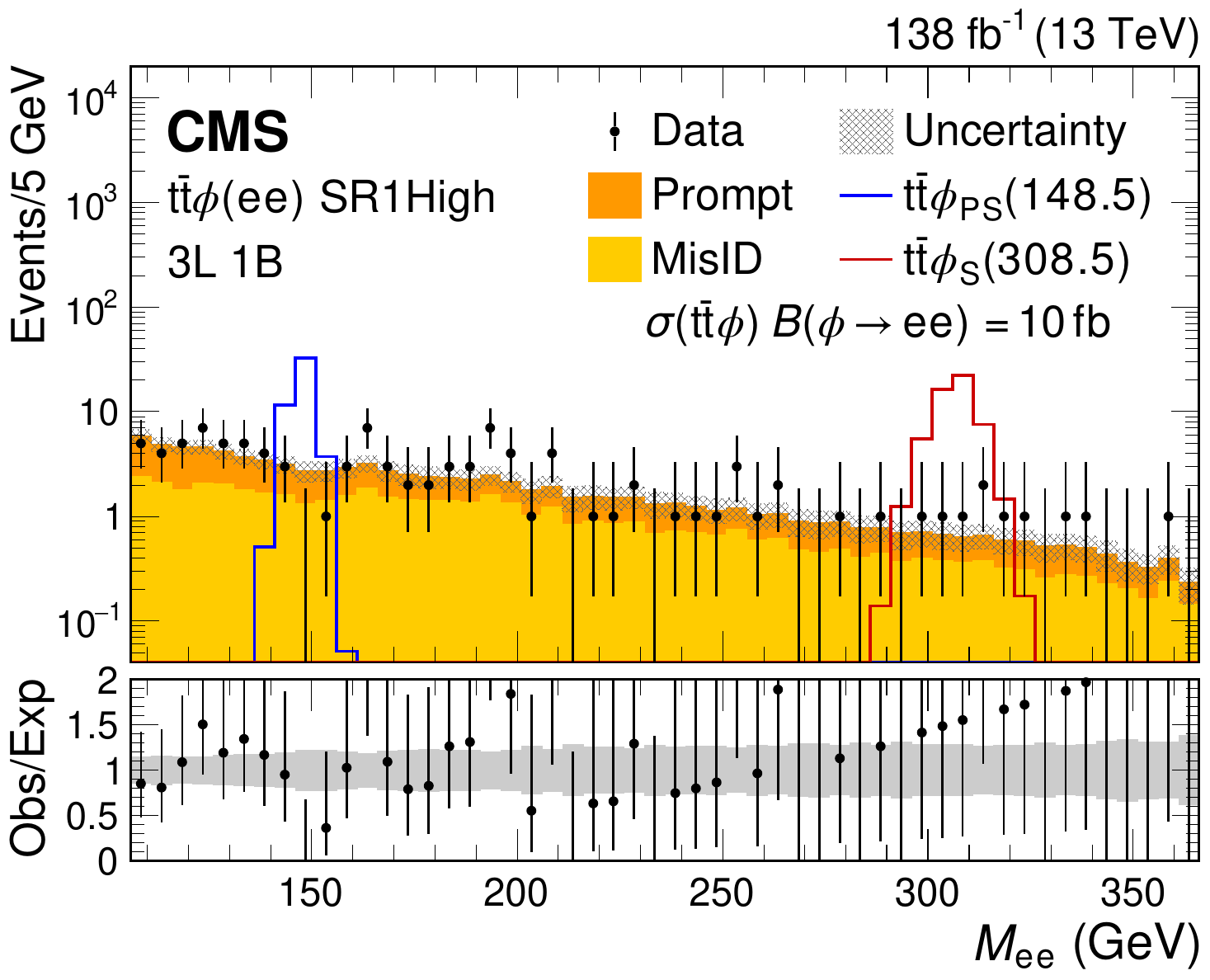}
\includegraphics[width=0.49\textwidth]{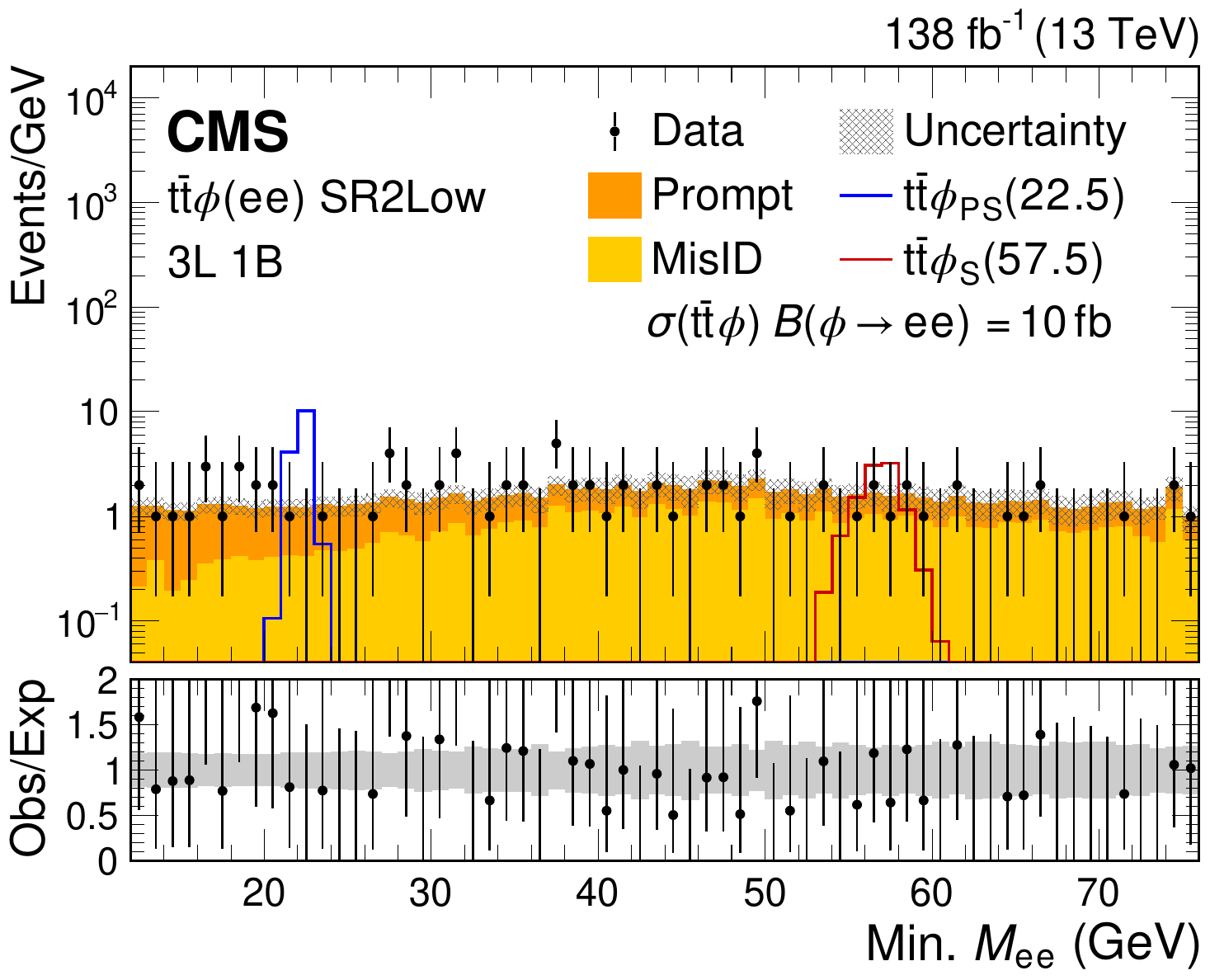}
\includegraphics[width=0.49\textwidth]{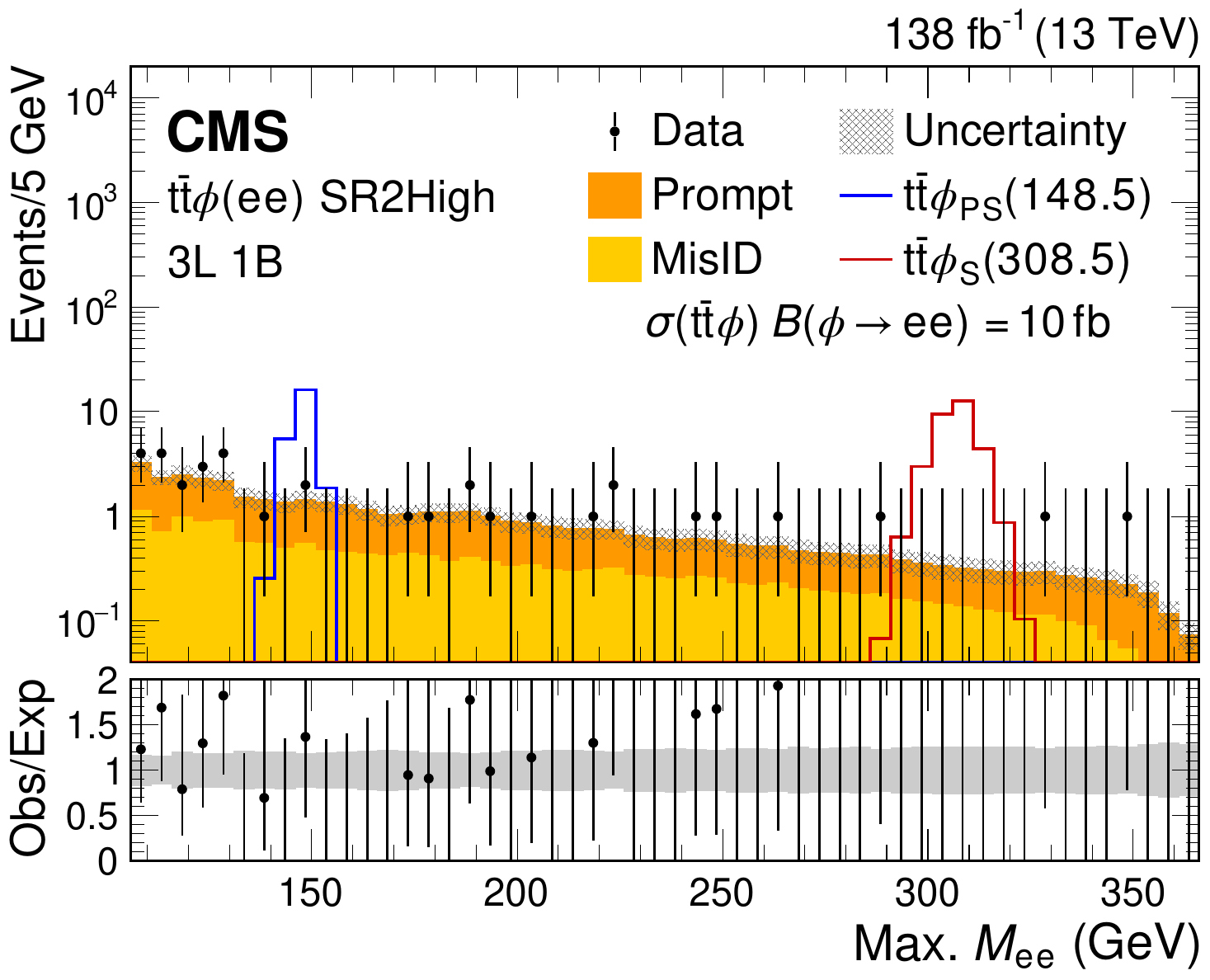}
\includegraphics[width=0.49\textwidth]{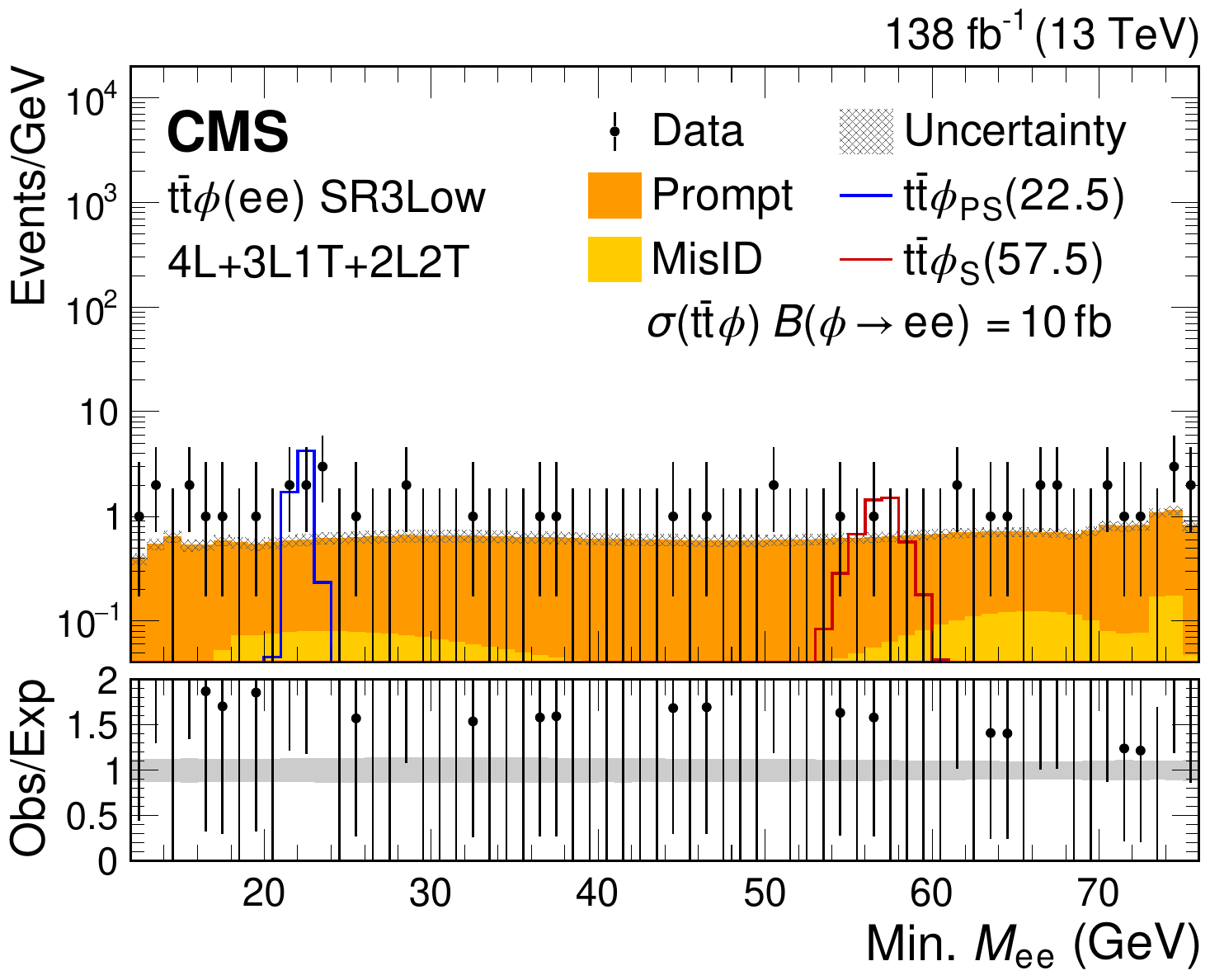}
\includegraphics[width=0.49\textwidth]{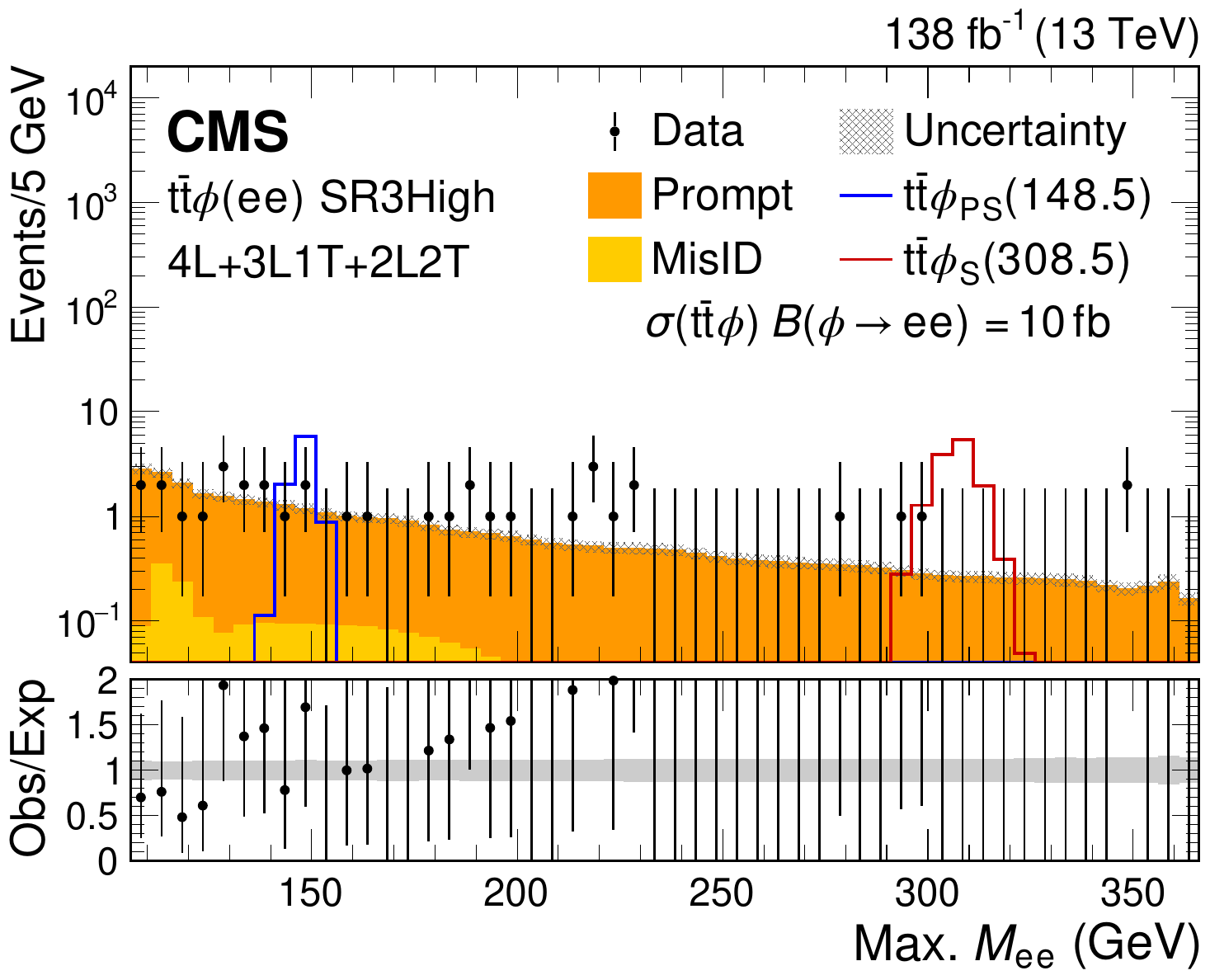}
\caption{\label{fig:ttPhiee} 
Dilepton mass spectra for the $\ttphi(\Pe\Pe)$ SR1 (upper), SR2 (middle), and SR3 (lower) event selections for the combined 2016--2018 data set.
The low (high) mass spectra are shown on the left (right).
The lower panel shows the ratio of observed events to the total expected SM background prediction (Obs/Exp), and the gray band represents the sum of statistical and systematic uncertainties in the background prediction. 
The expected background distributions and the uncertainties are shown after the data is fit under the background-only hypothesis.
For illustration, two example signal hypotheses for the production and decay of a scalar and a pseudoscalar $\phi$ boson are shown, and their masses (in units of \GeV) are indicated in the legend. The signals are normalized to the product of the cross section and branching fraction of 10\unit{fb}. 
}
\end{figure*}

\begin{figure*}[hbt!]
\centering
\includegraphics[width=0.49\textwidth]{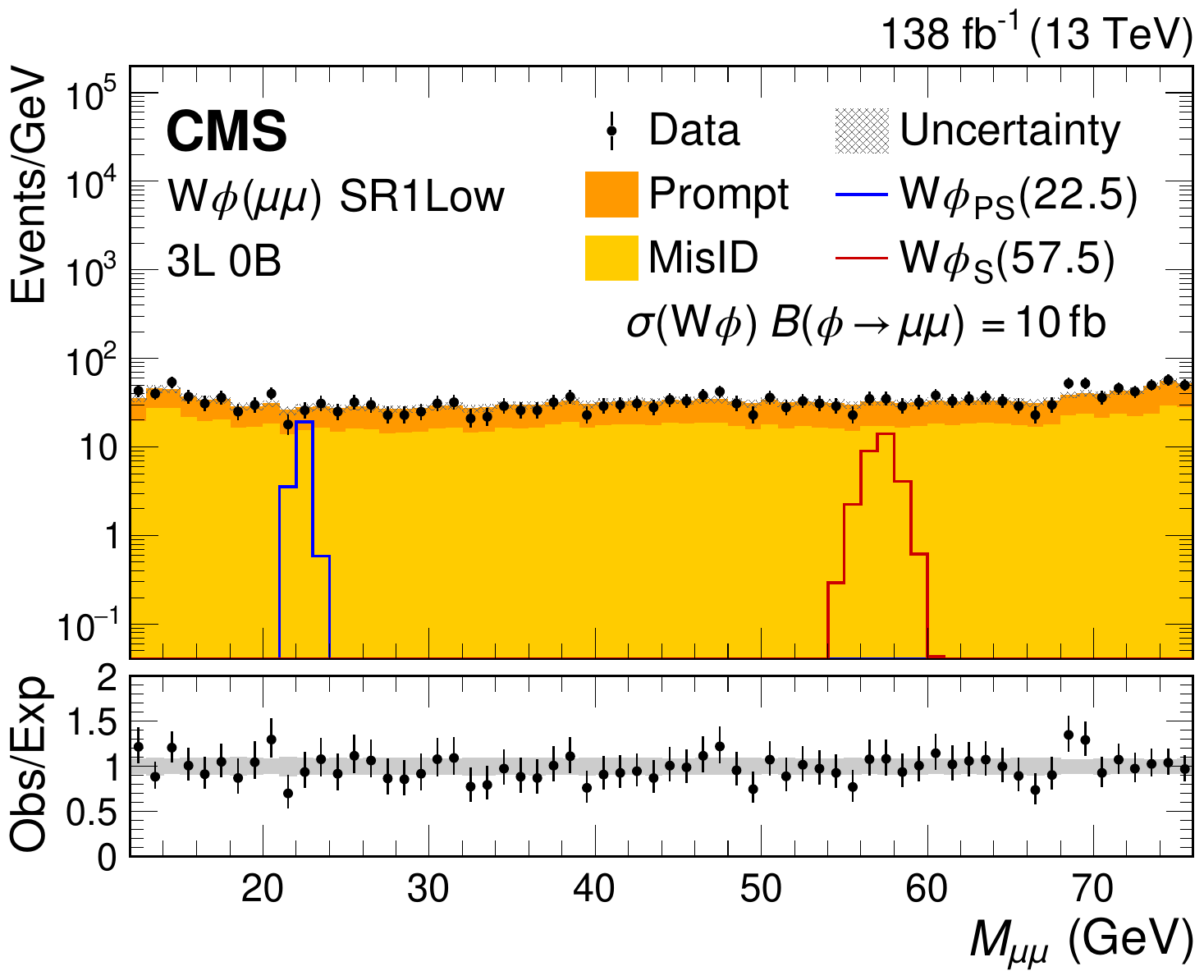}
\includegraphics[width=0.49\textwidth]{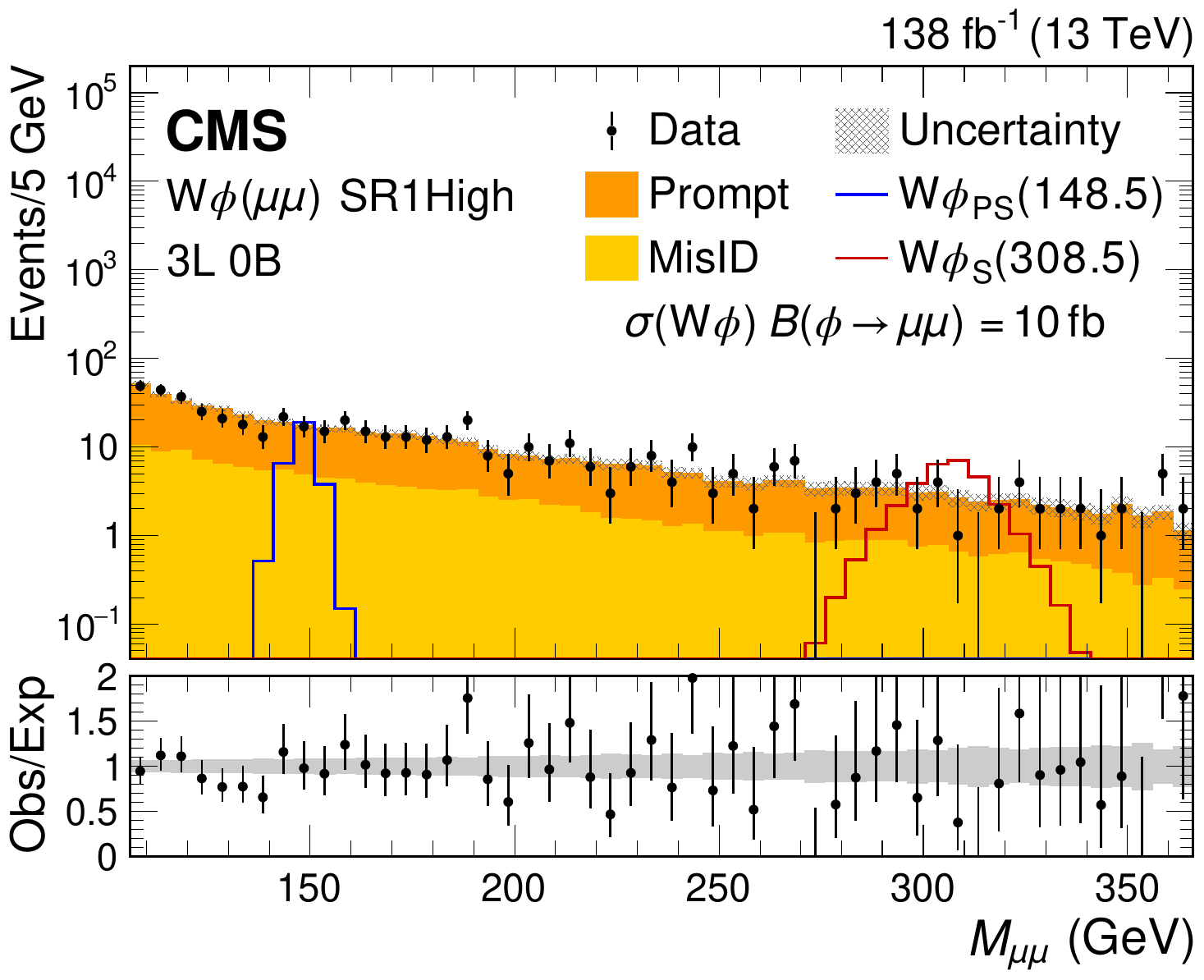}
\includegraphics[width=0.49\textwidth]{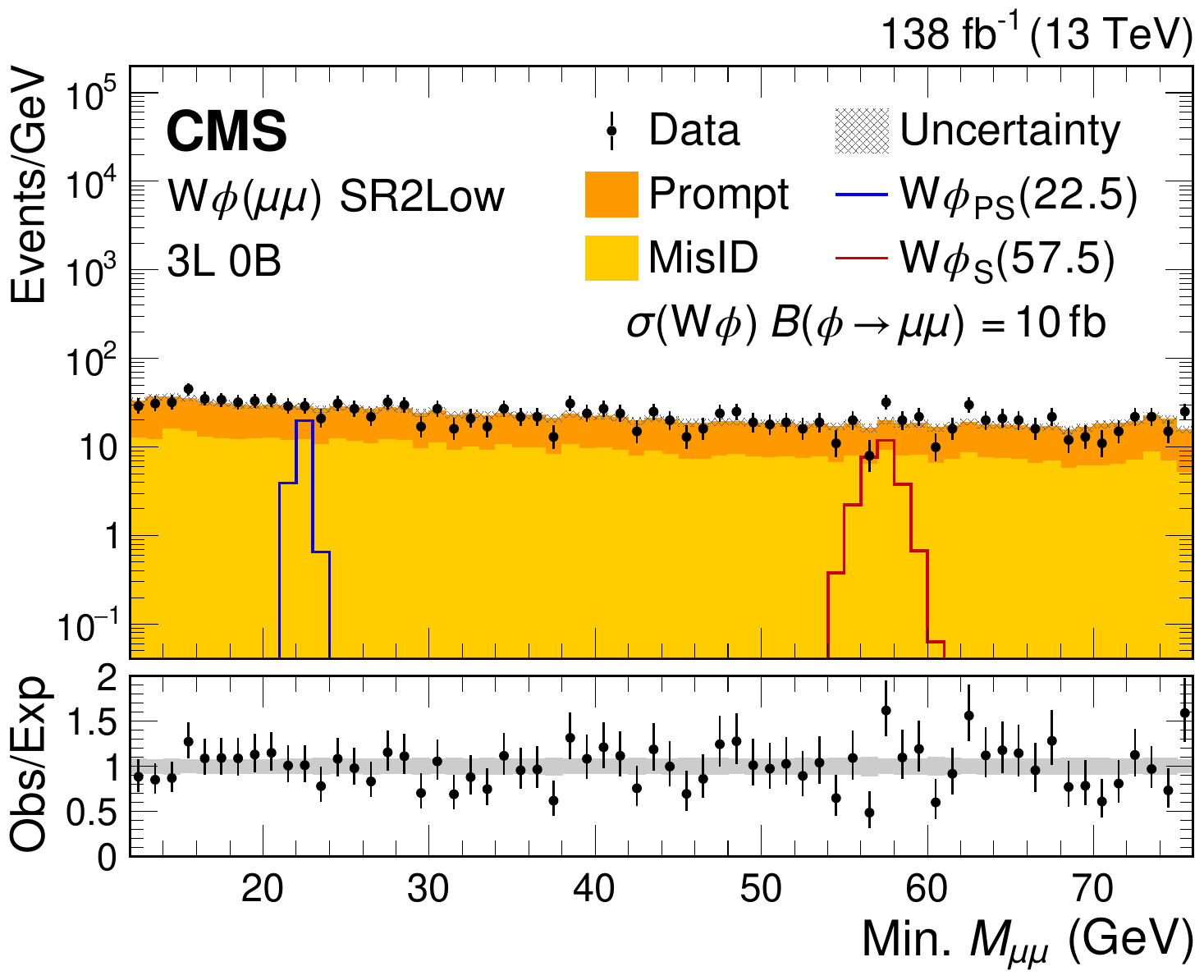}
\includegraphics[width=0.49\textwidth]{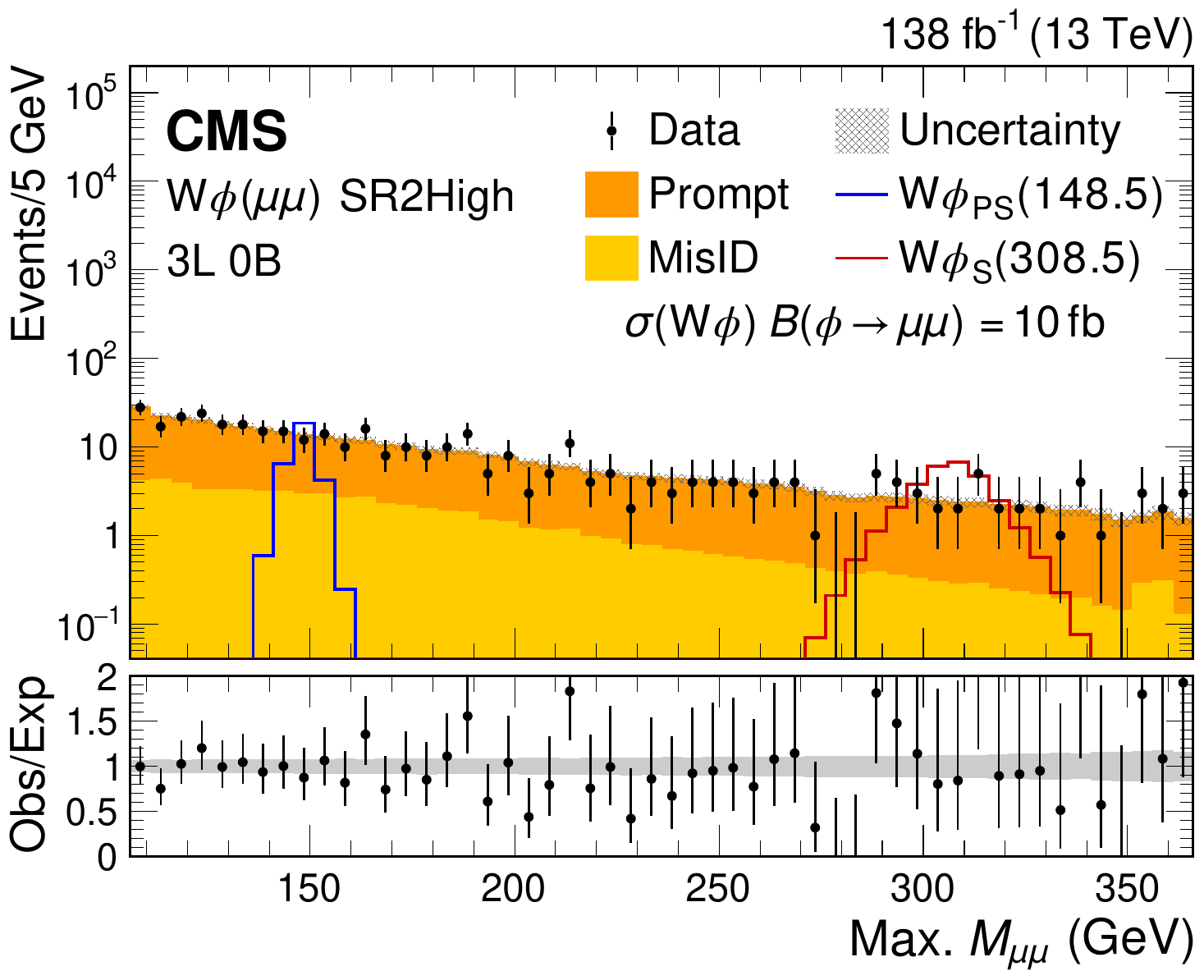}
\includegraphics[width=0.49\textwidth]{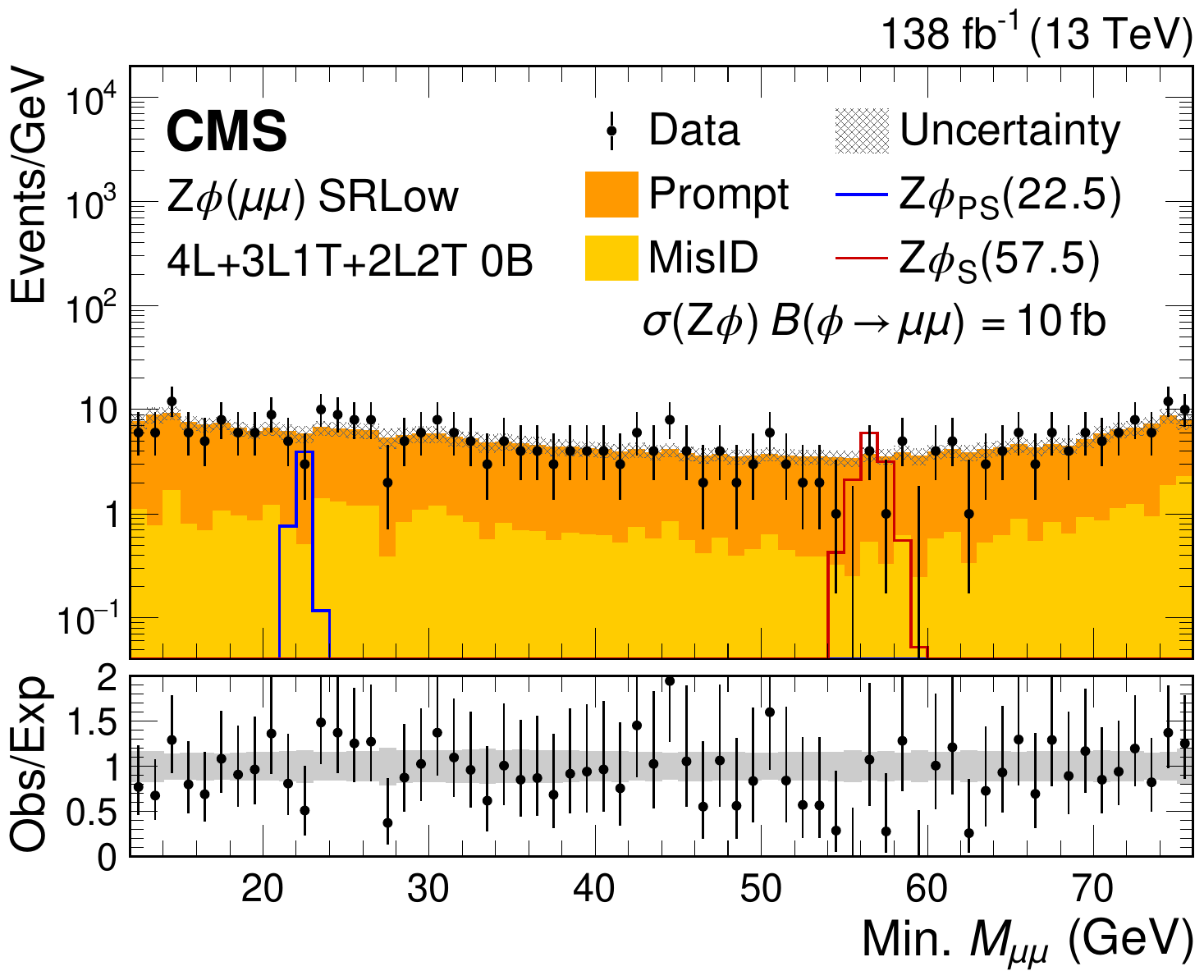}
\includegraphics[width=0.49\textwidth]{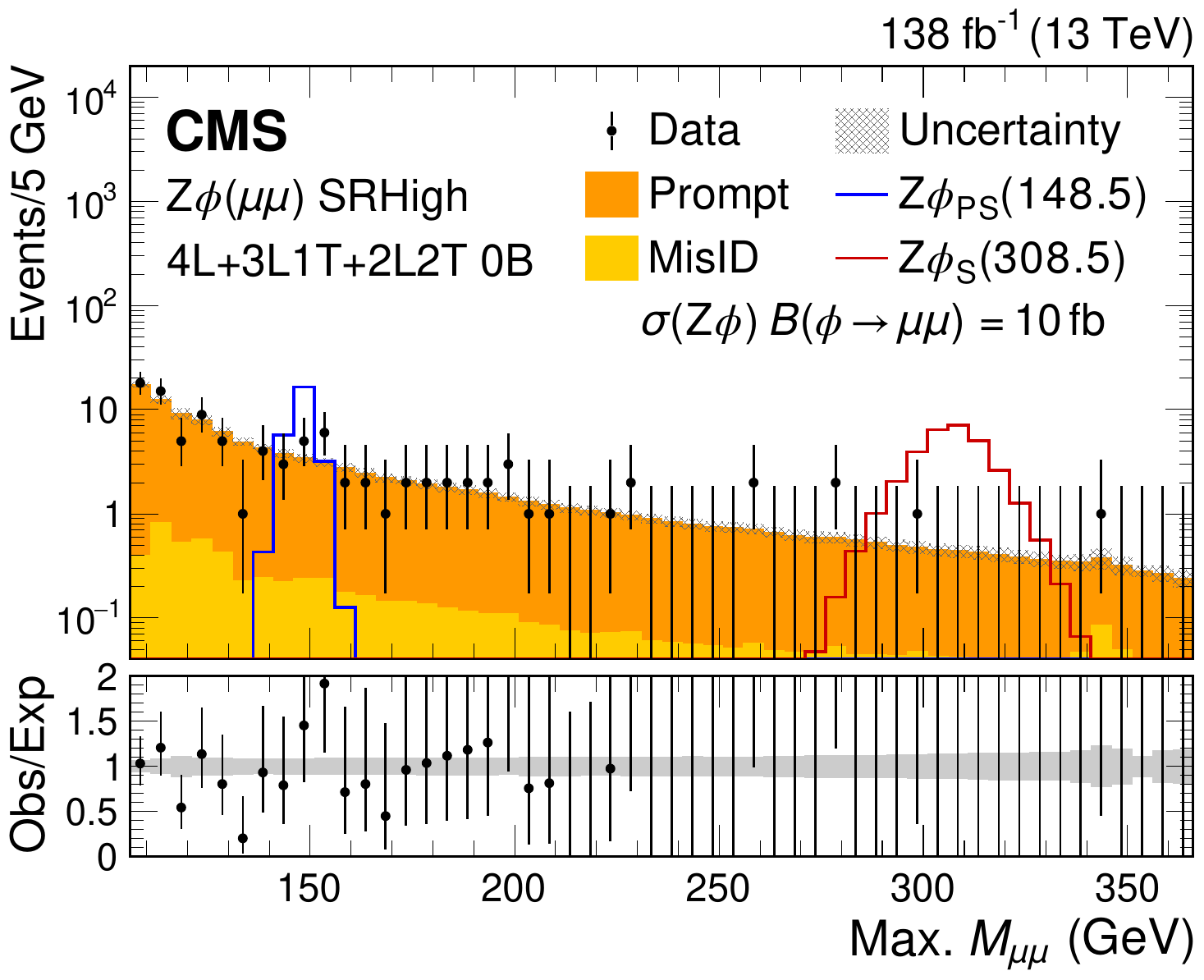}
\caption{\label{fig:WZPhimm} 
Dilepton mass spectra for the $\Wphi(\PGm\PGm)$ SR1 (upper), SR2 (middle), and $\Zphi(\PGm\PGm)$ SR (lower) event selections for the combined 2016--2018 data set.
The low (high) mass spectra are shown on the left (right).
The lower panel shows the ratio of observed events to the total expected SM background prediction (Obs/Exp), and the gray band represents the sum of statistical and systematic uncertainties in the background prediction. 
The expected background distributions and the uncertainties are shown after the data is fit under the background-only hypothesis.
For illustration, two example signal hypotheses for the production and decay of a scalar and a pseudoscalar $\phi$ boson are shown, and their masses (in units of \GeV) are indicated in the legend. The signals are normalized to the product of the cross section and branching fraction of 10\unit{fb}. 
}
\end{figure*}

\begin{figure*}[hbt!]
\centering
\includegraphics[width=0.49\textwidth]{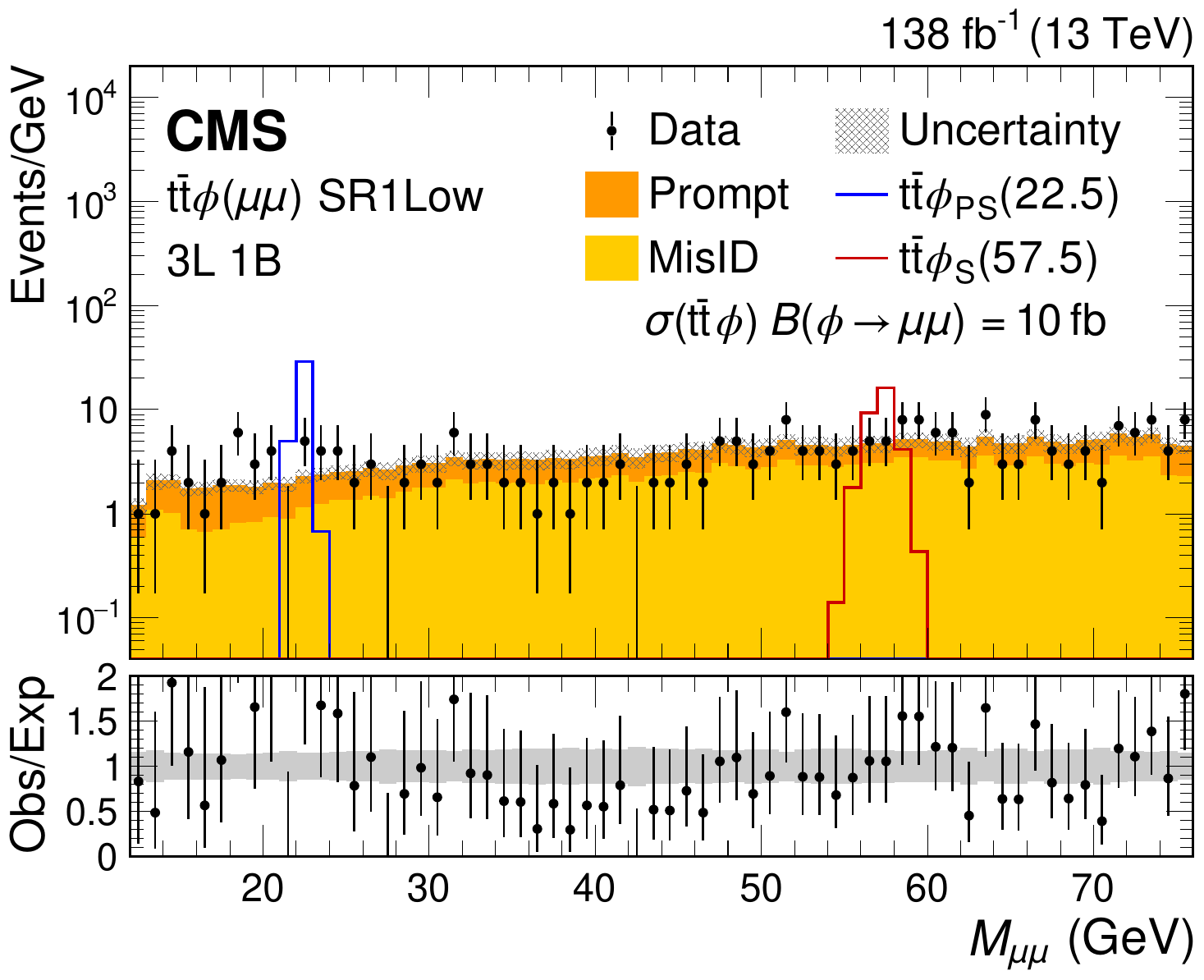}
\includegraphics[width=0.49\textwidth]{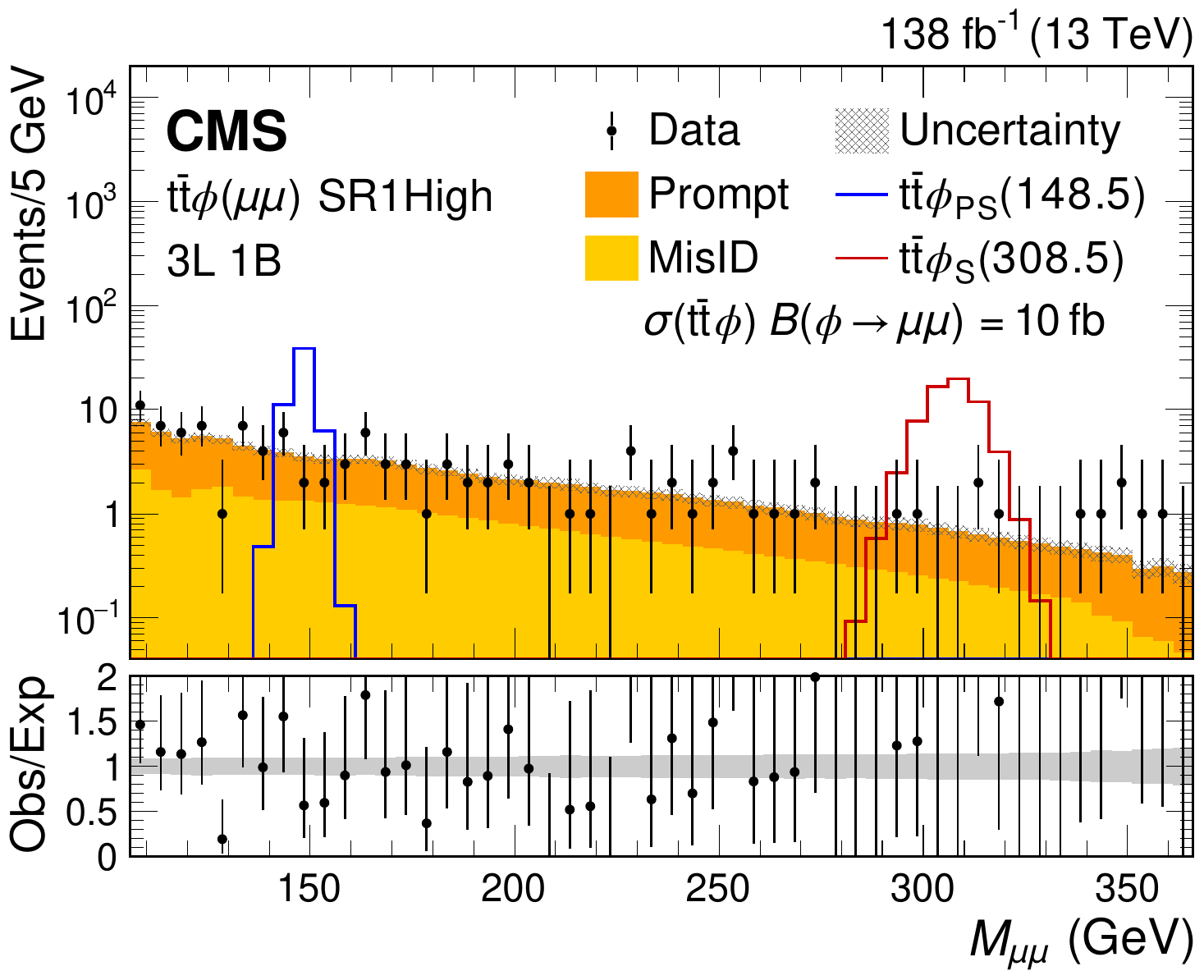}
\includegraphics[width=0.49\textwidth]{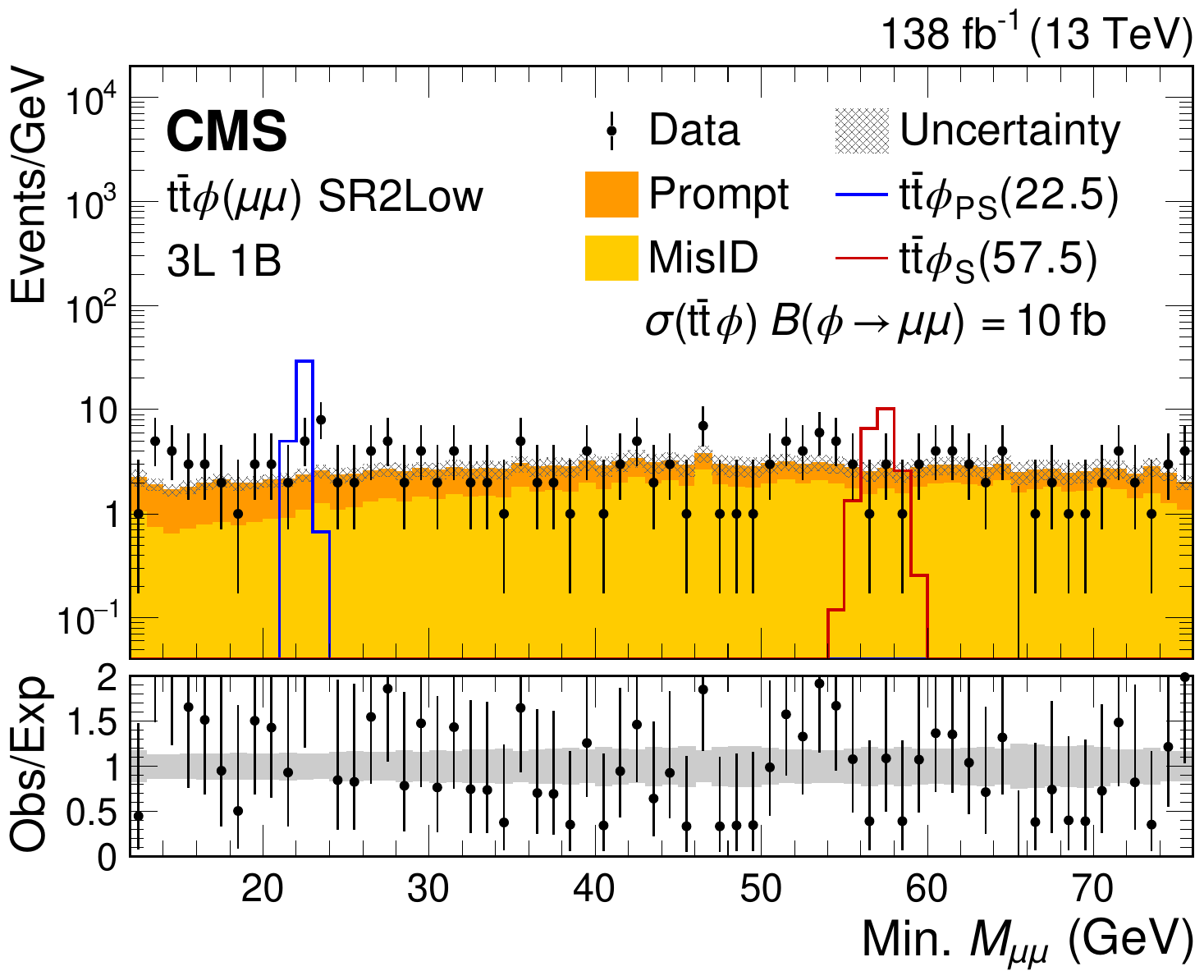}
\includegraphics[width=0.49\textwidth]{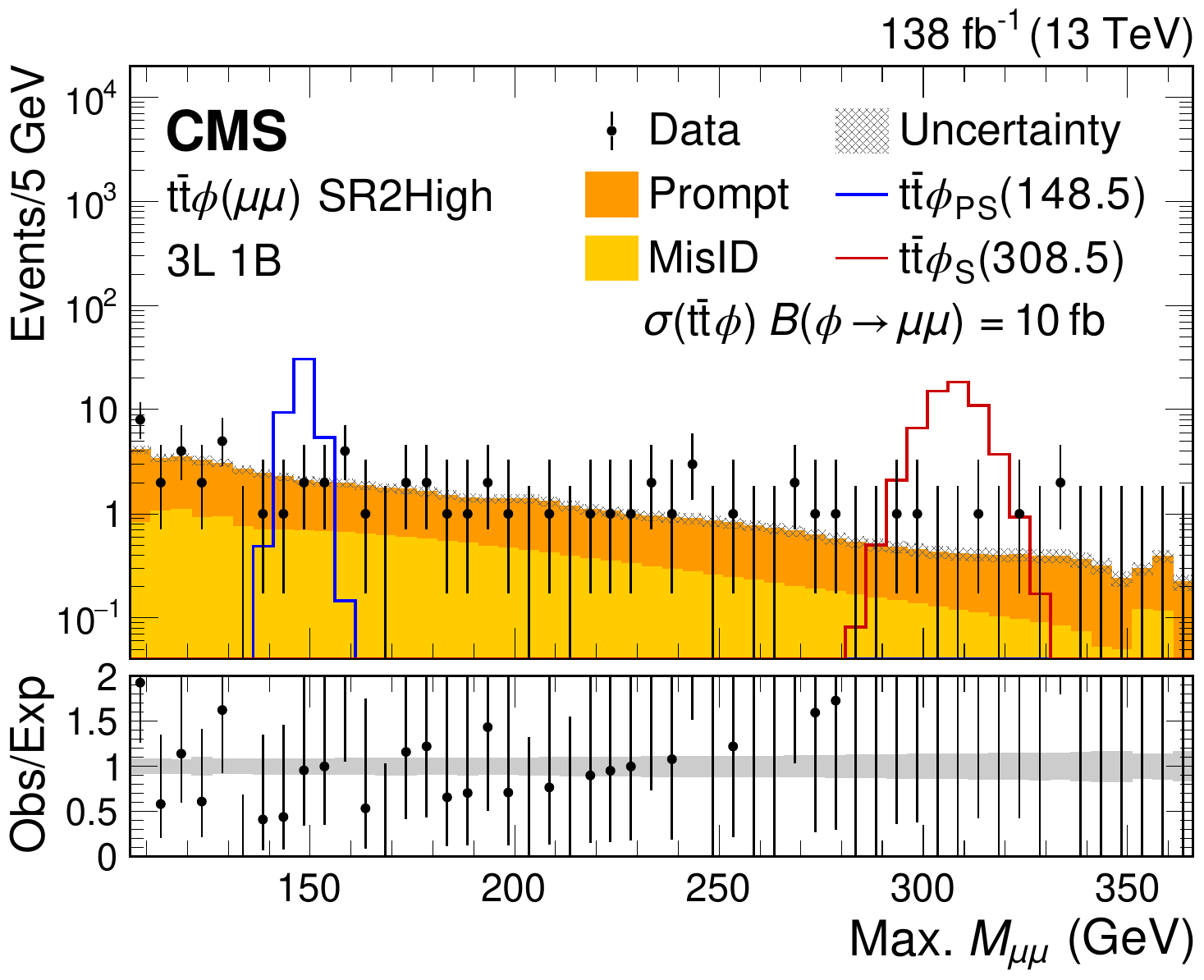}
\includegraphics[width=0.49\textwidth]{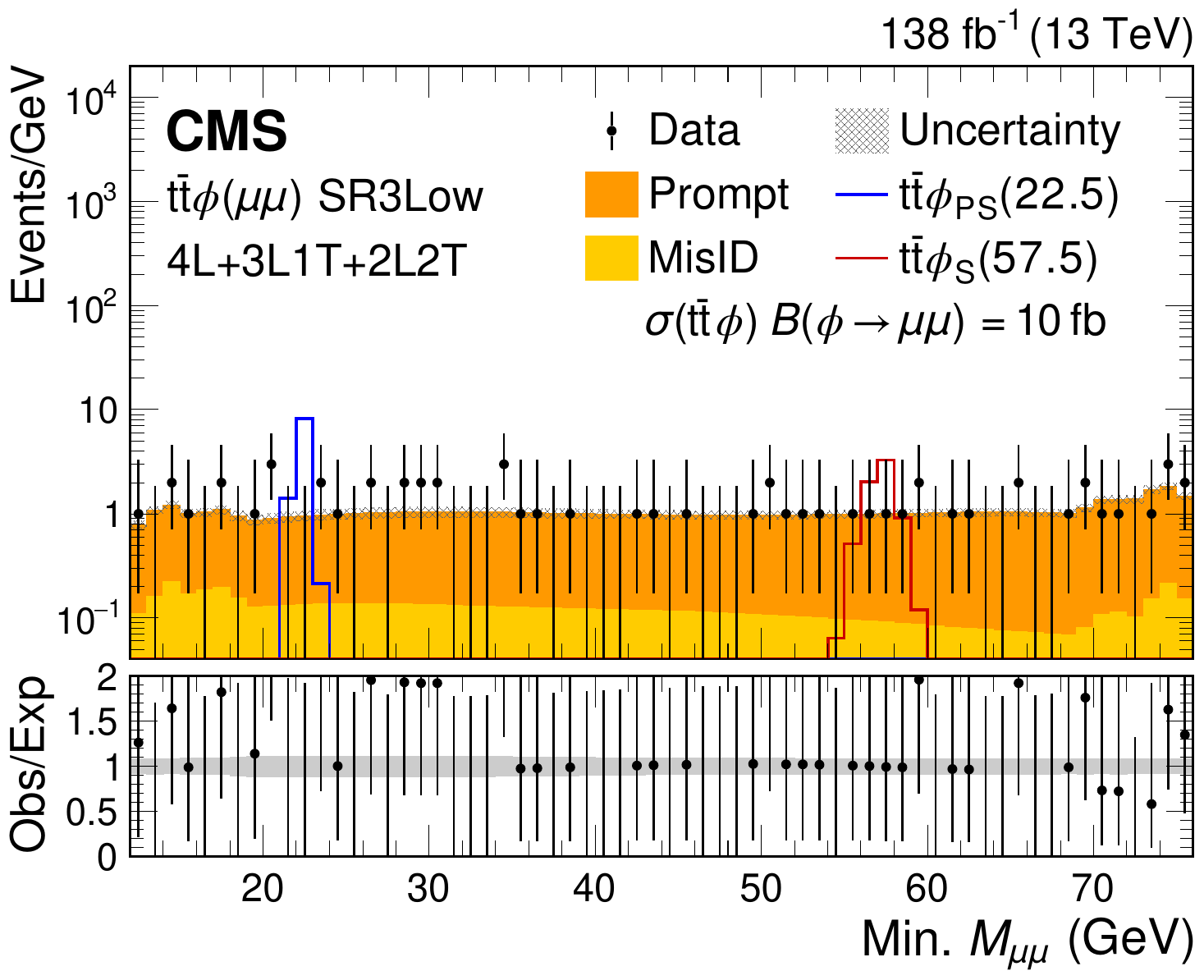}
\includegraphics[width=0.49\textwidth]{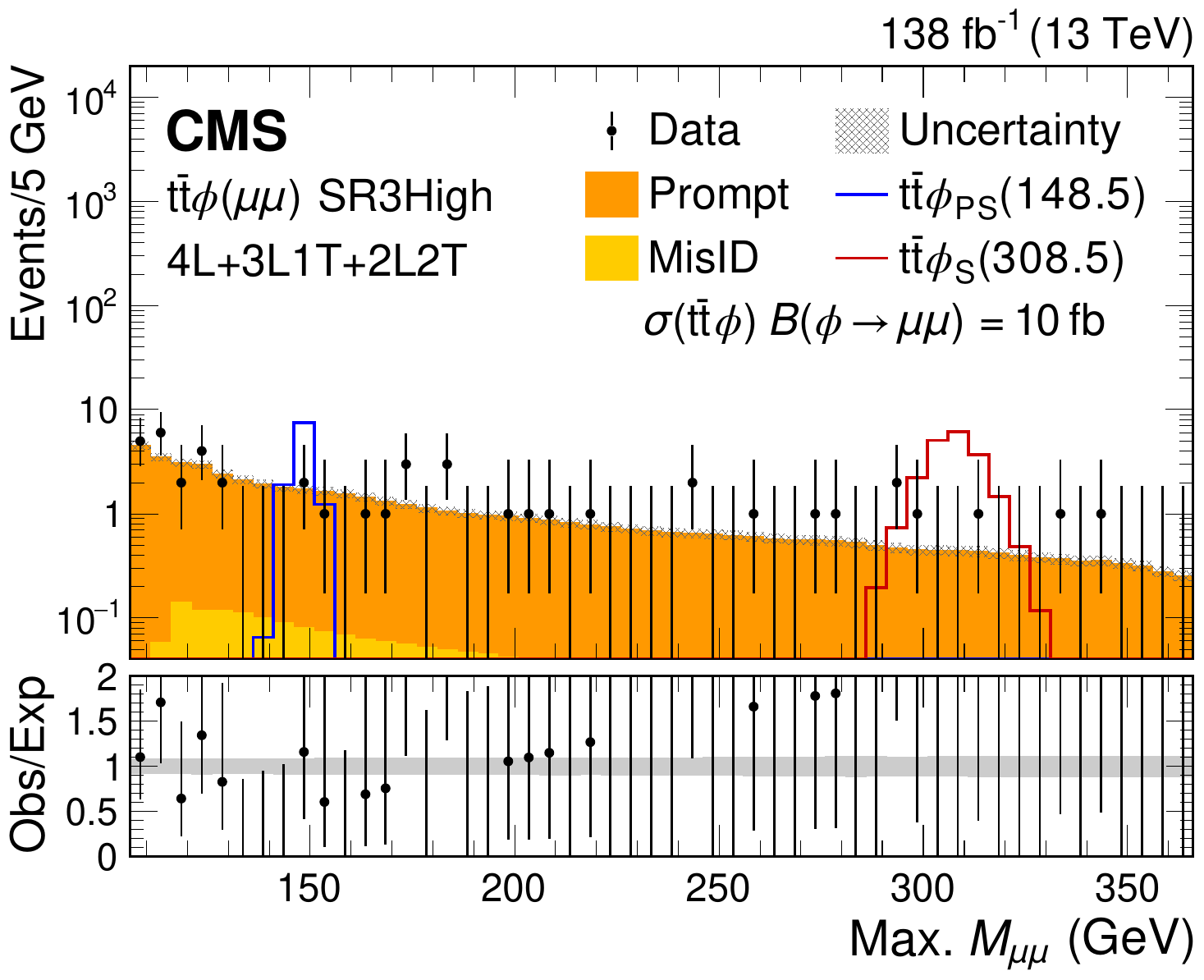}
\caption{\label{fig:ttPhimm} 
Dilepton mass spectra for the $\ttphi(\PGm\PGm)$ SR1 (upper), SR2 (middle), and SR3 (lower) event selections for the combined 2016--2018 data set.
The low (high) mass spectra are shown on the left (right).
The lower panel shows the ratio of observed events to the total expected SM background prediction (Obs/Exp), and the gray band represents the sum of statistical and systematic uncertainties in the background prediction. 
The expected background distributions and the uncertainties are shown after the data is fit under the background-only hypothesis.
For illustration, two example signal hypotheses for the production and decay of a scalar and a pseudoscalar $\phi$ boson are shown, and their masses (in units of \GeV) are indicated in the legend. The signals are normalized to the product of the cross section and branching fraction of 10\unit{fb}. 
}
\end{figure*}

\begin{figure*}[hbt!]
\centering
\includegraphics[width=0.49\textwidth]{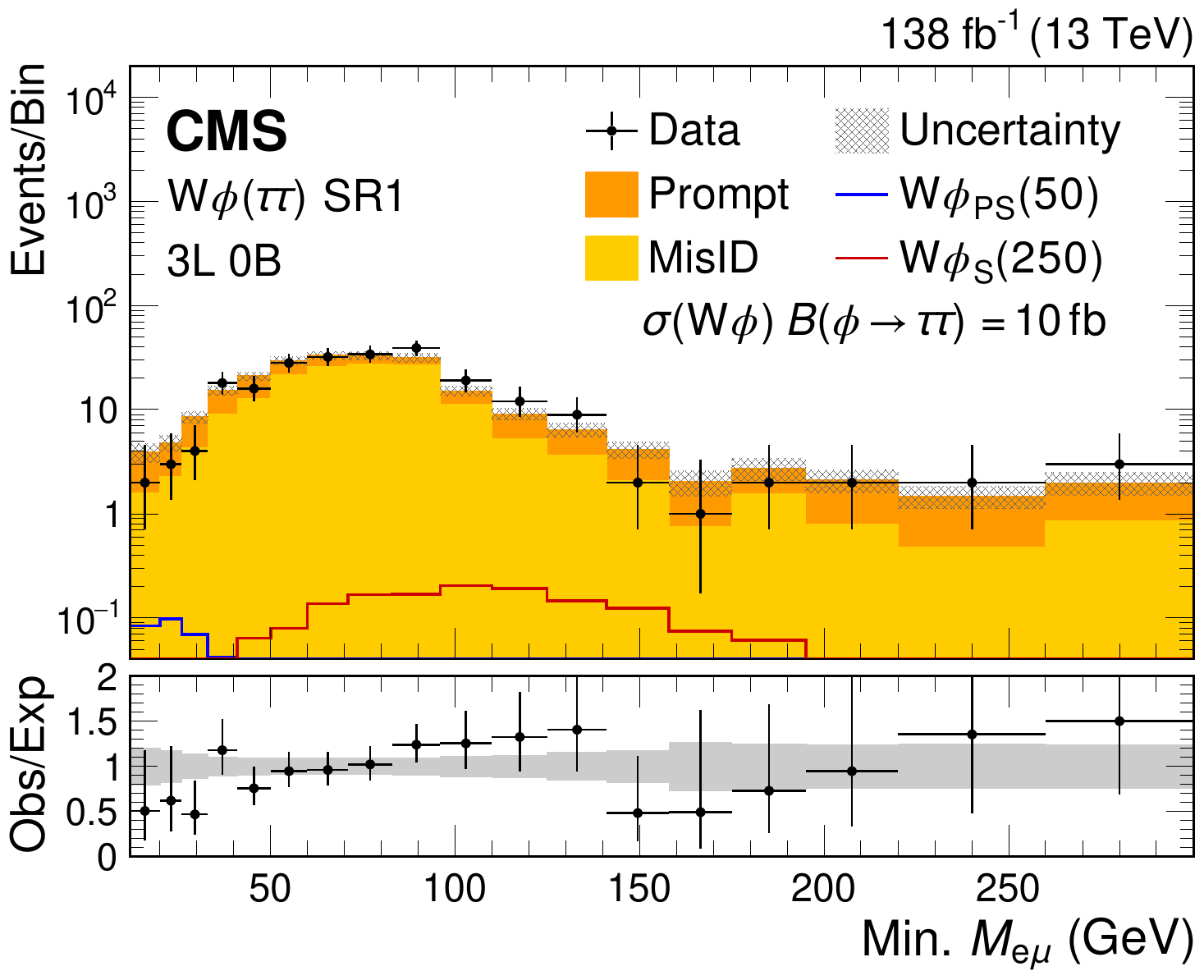}
\includegraphics[width=0.49\textwidth]{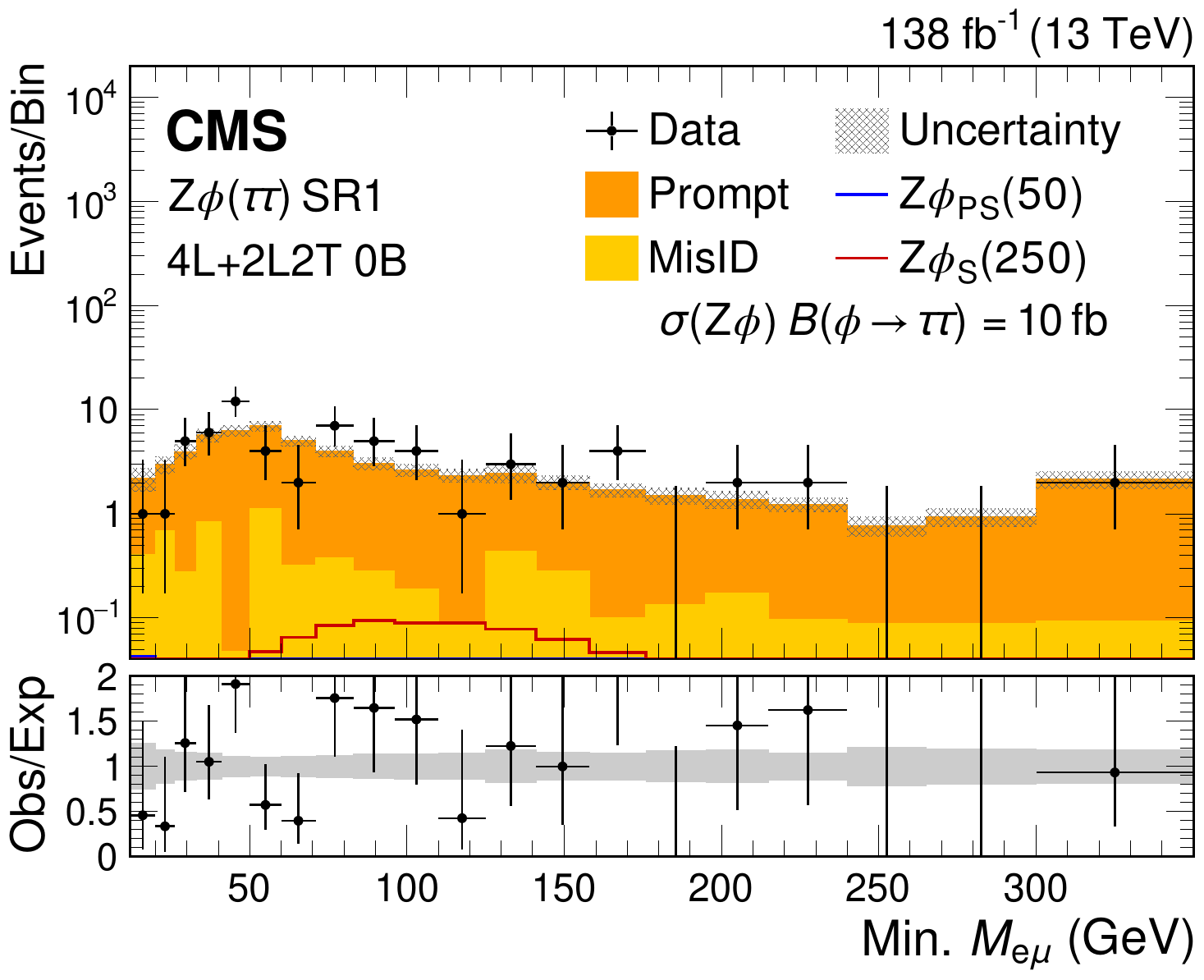}
\includegraphics[width=0.49\textwidth]{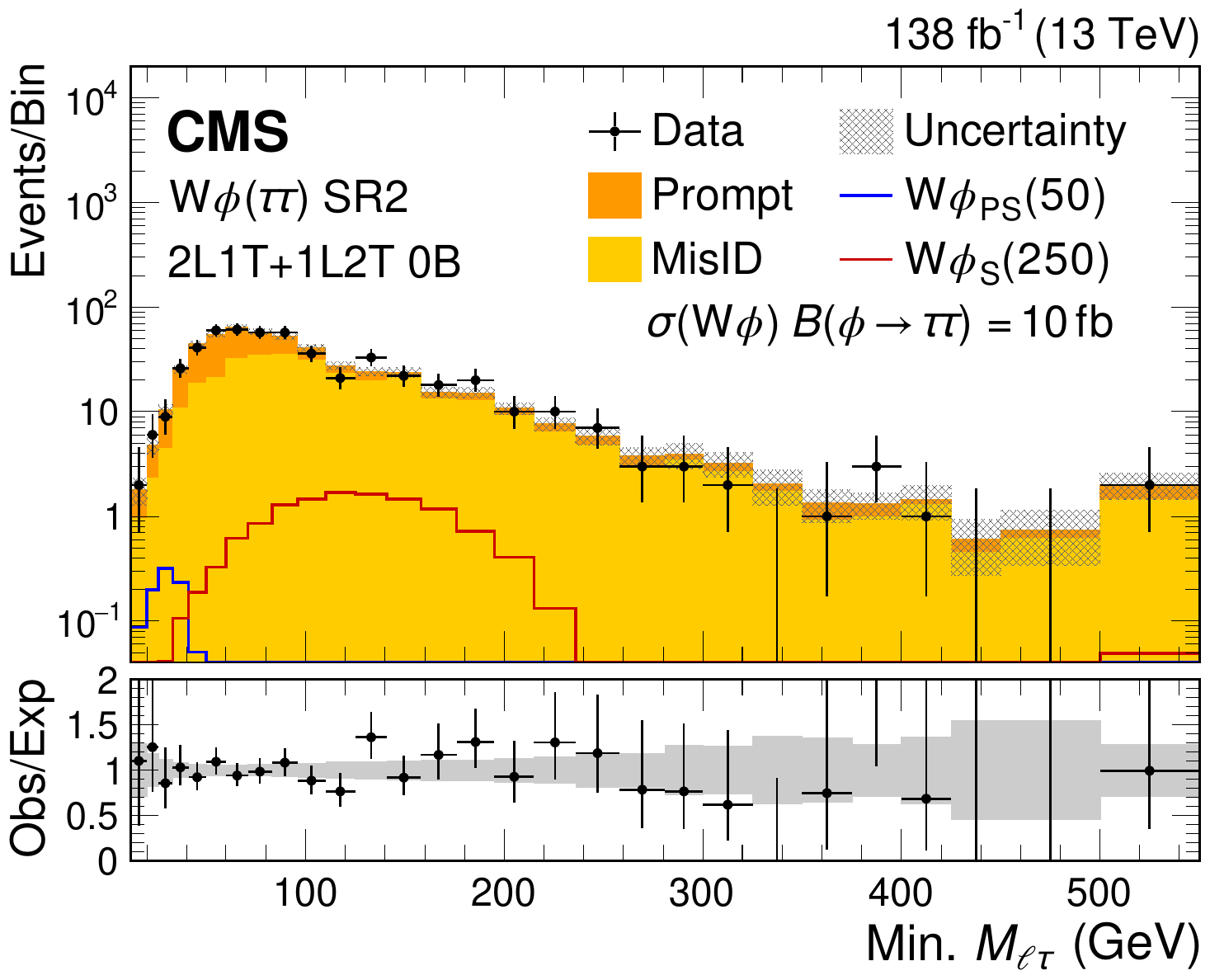}
\includegraphics[width=0.49\textwidth]{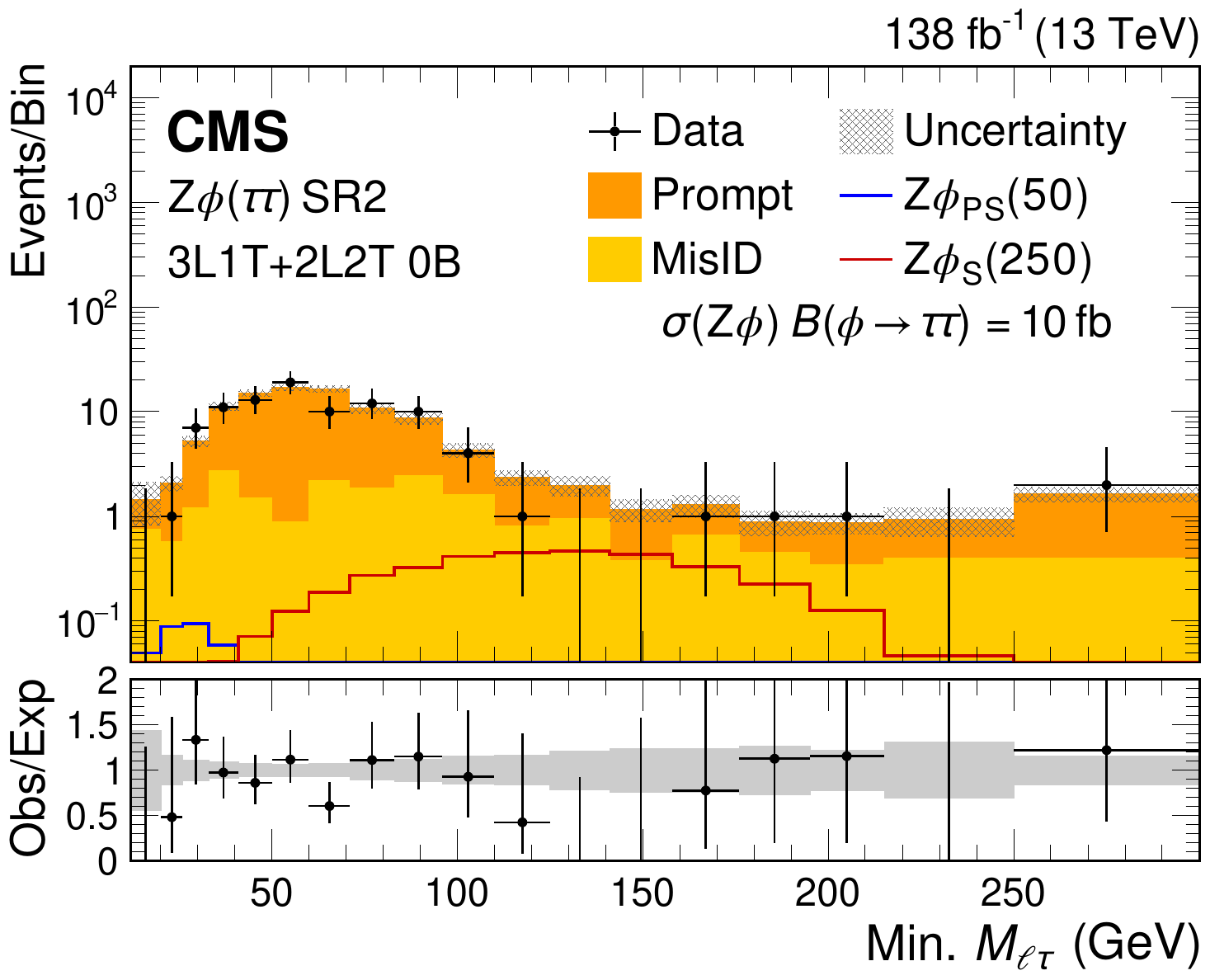}
\includegraphics[width=0.49\textwidth]{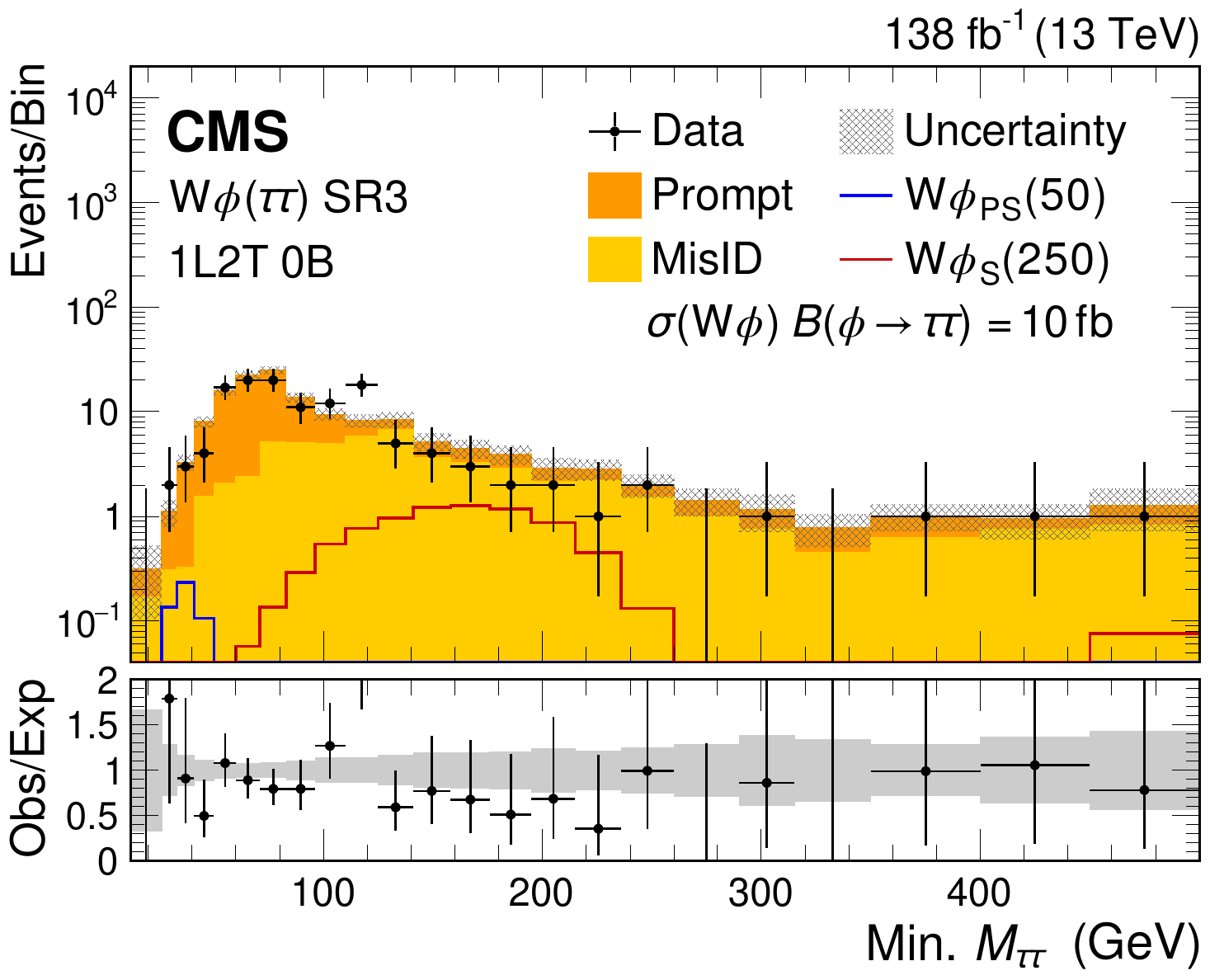}
\includegraphics[width=0.49\textwidth]{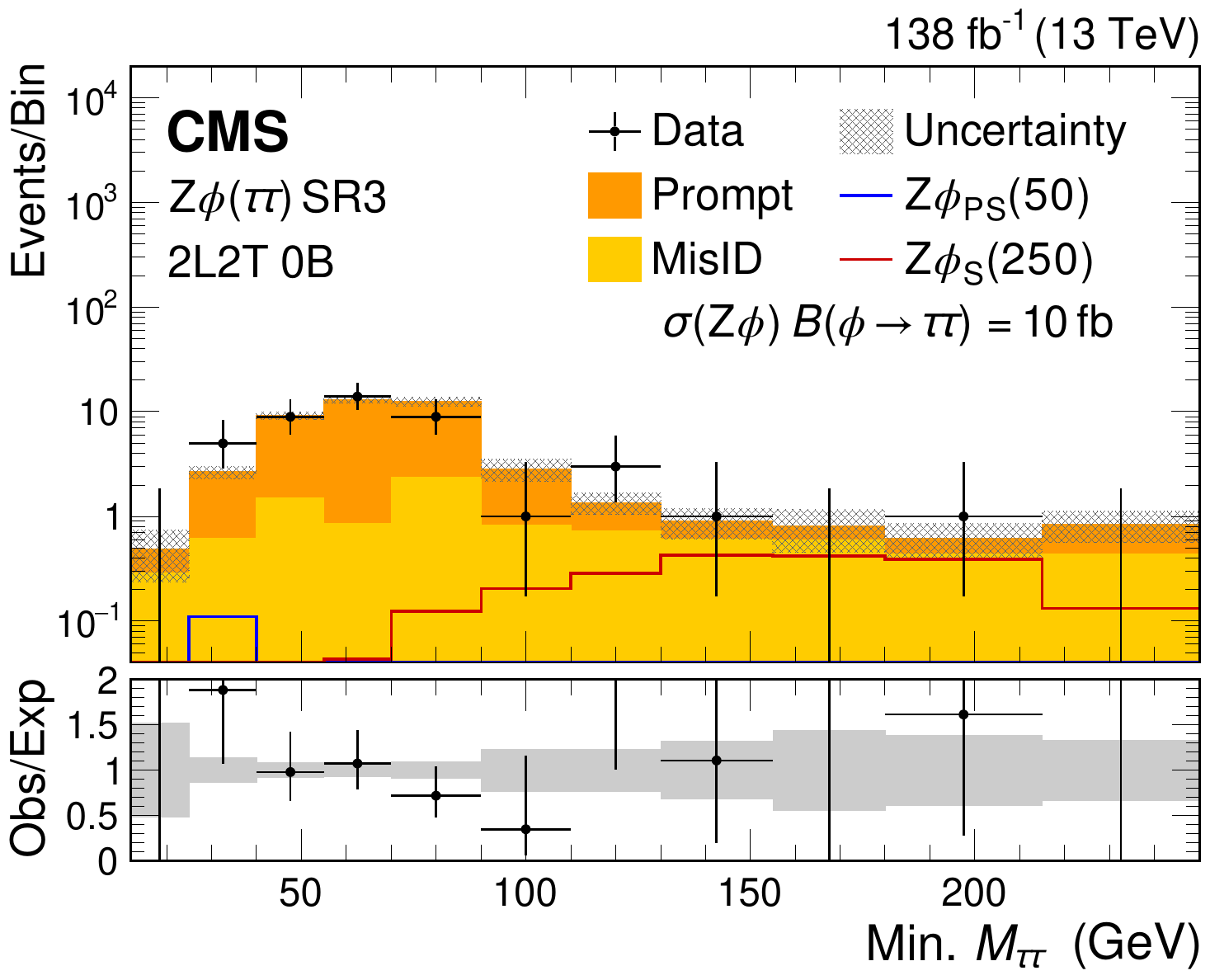}
\caption{\label{fig:WZPhitautau} 
Dilepton mass spectra for the $\Wphi(\PGt\PGt)$ SR (left) and $\Zphi(\PGt\PGt)$ SR (right) event selections for the combined 2016--2018 data set.
The lower panel shows the ratio of observed events to the total expected SM background prediction (Obs/Exp), and the gray band represents the sum of statistical and systematic uncertainties in the background prediction. 
The rightmost bins contain the overflow events in each distribution. 
The expected background distributions and the uncertainties are shown after the data is fit under the background-only hypothesis. 
For illustration, two example signal hypotheses for the production and decay of a scalar and a pseudoscalar $\phi$ boson are shown, and their masses (in units of \GeV) are indicated in the legend. 
The signals are normalized to the product of the cross section and branching fraction of 10\unit{fb}. 
}
\end{figure*}

\begin{figure*}[hbt!]
\centering
\includegraphics[width=0.49\textwidth]{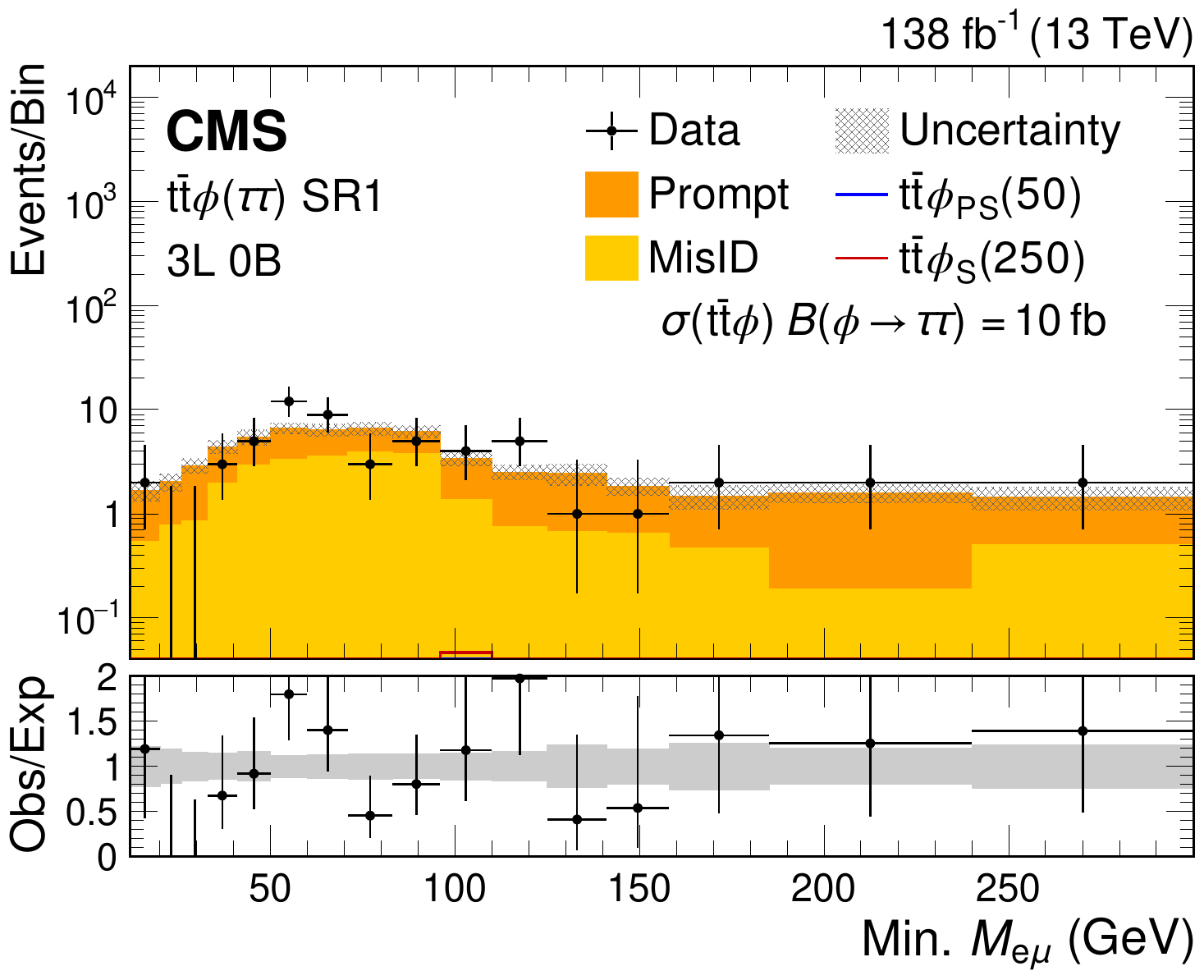}
\includegraphics[width=0.49\textwidth]{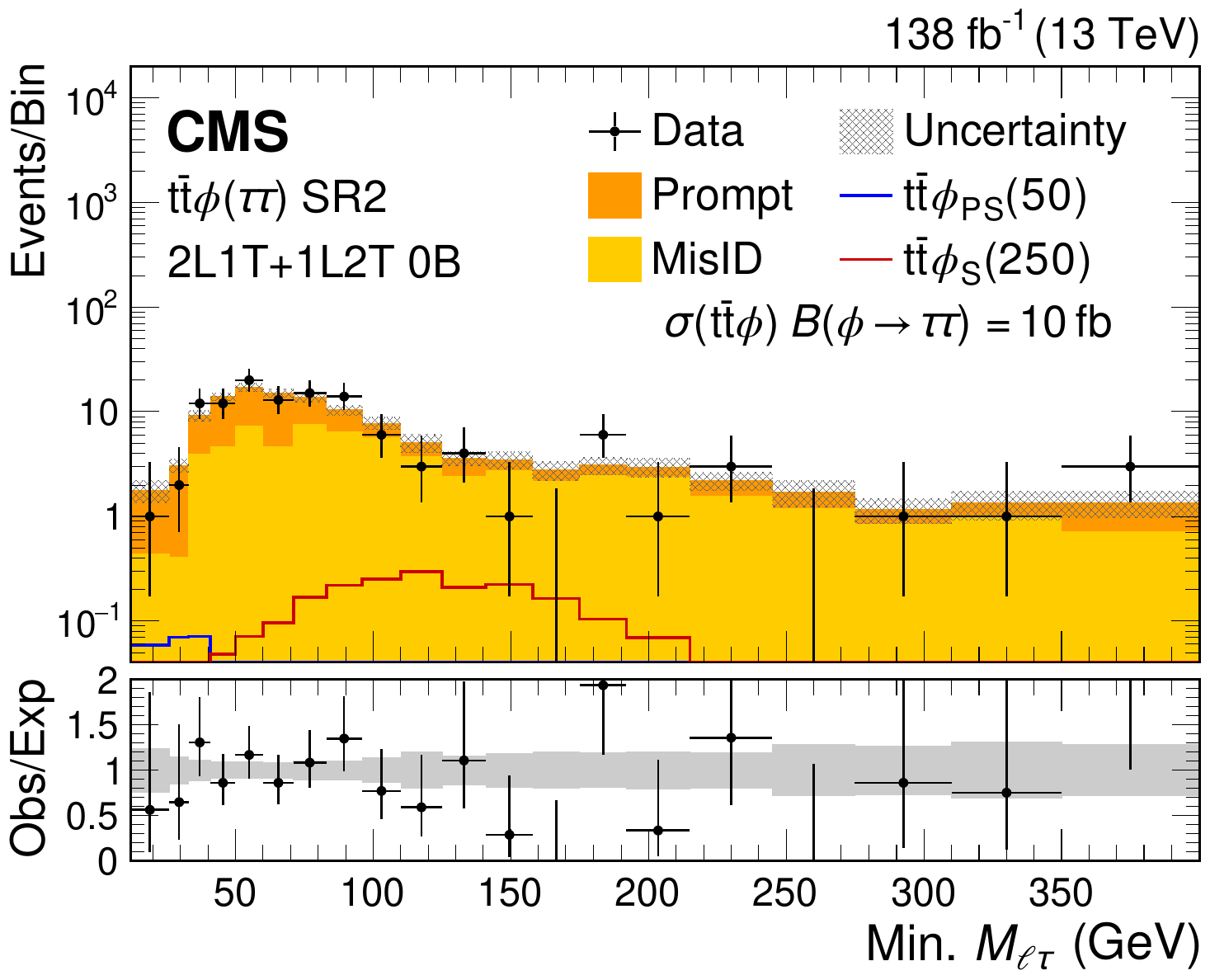}
\includegraphics[width=0.49\textwidth]{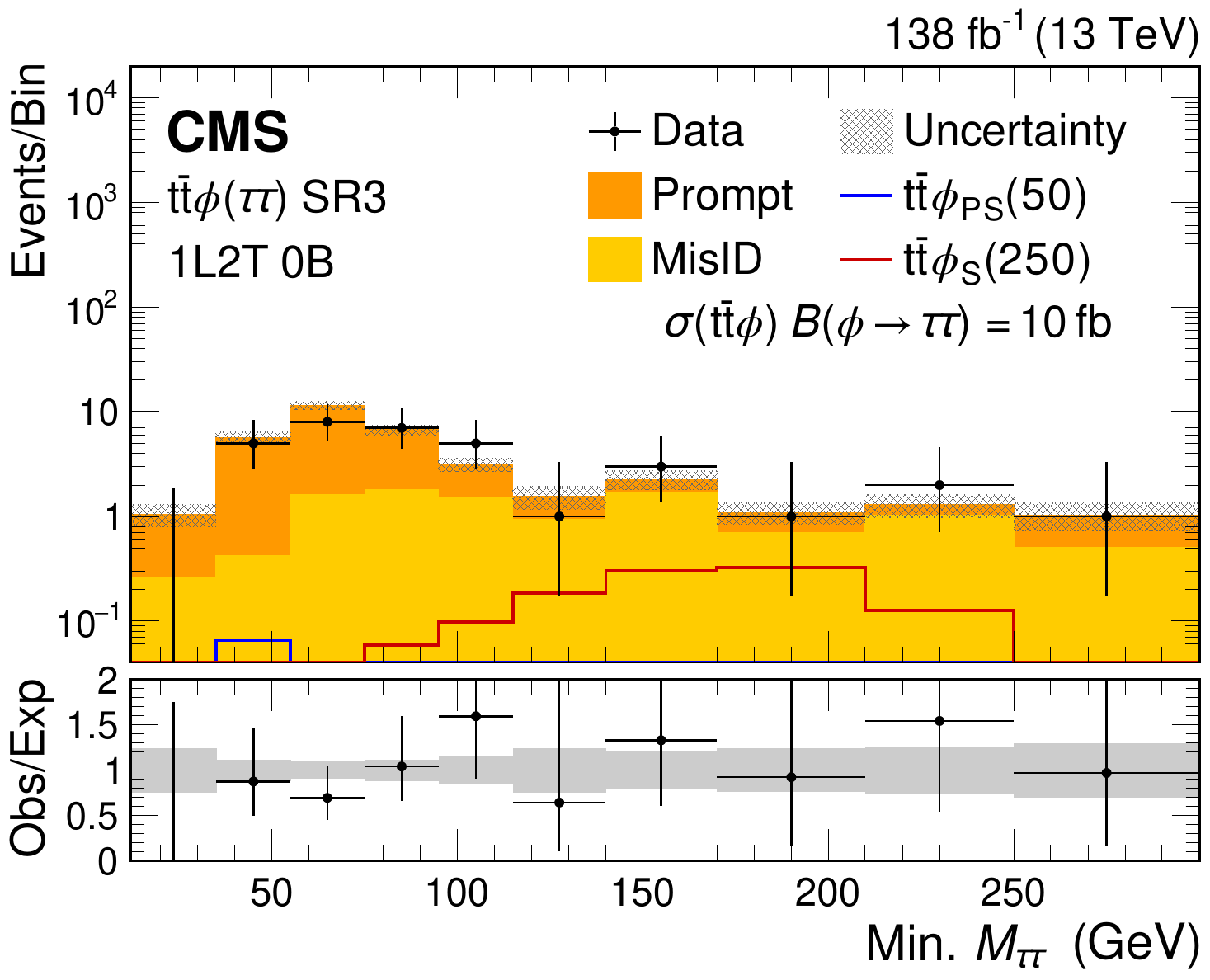}
\includegraphics[width=0.49\textwidth]{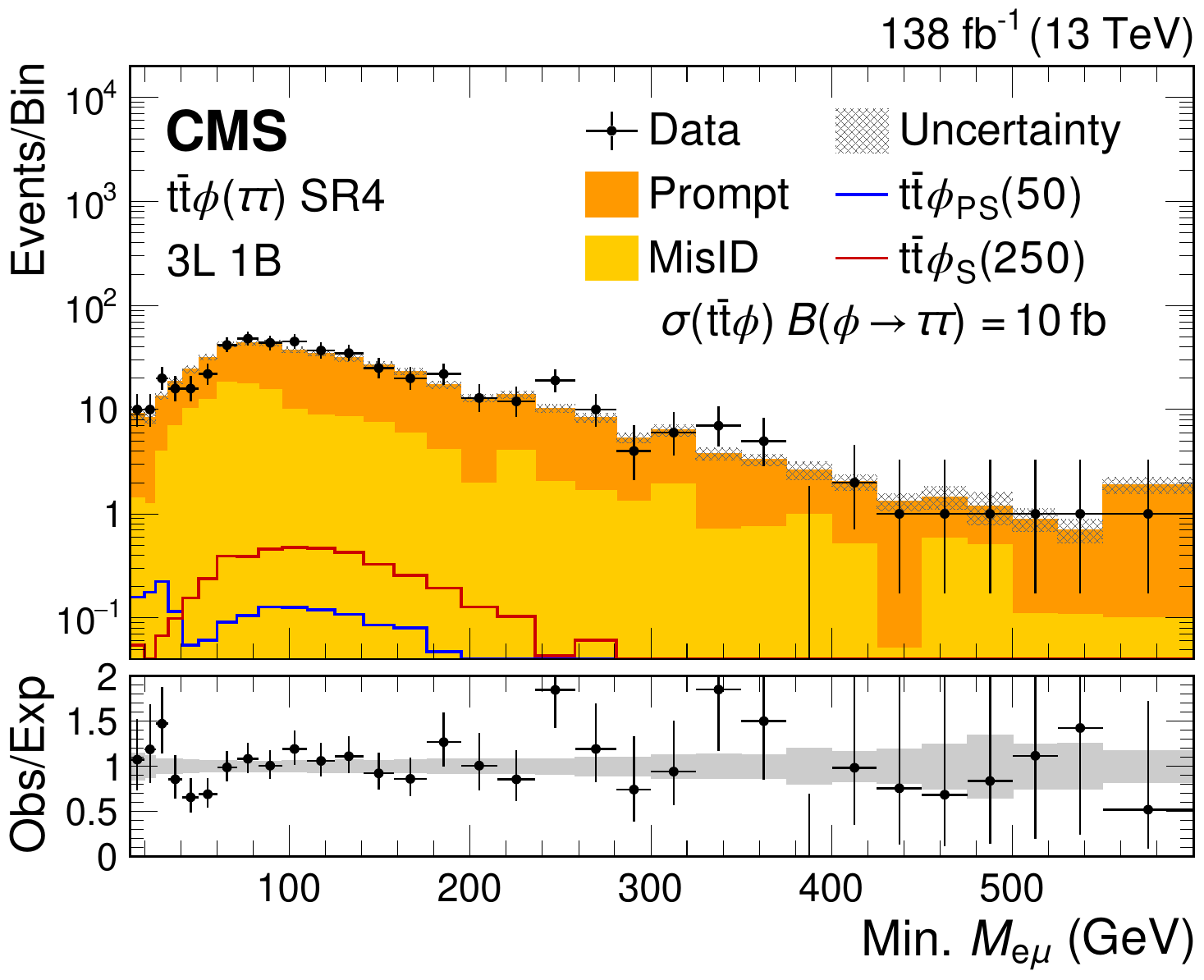}
\includegraphics[width=0.49\textwidth]{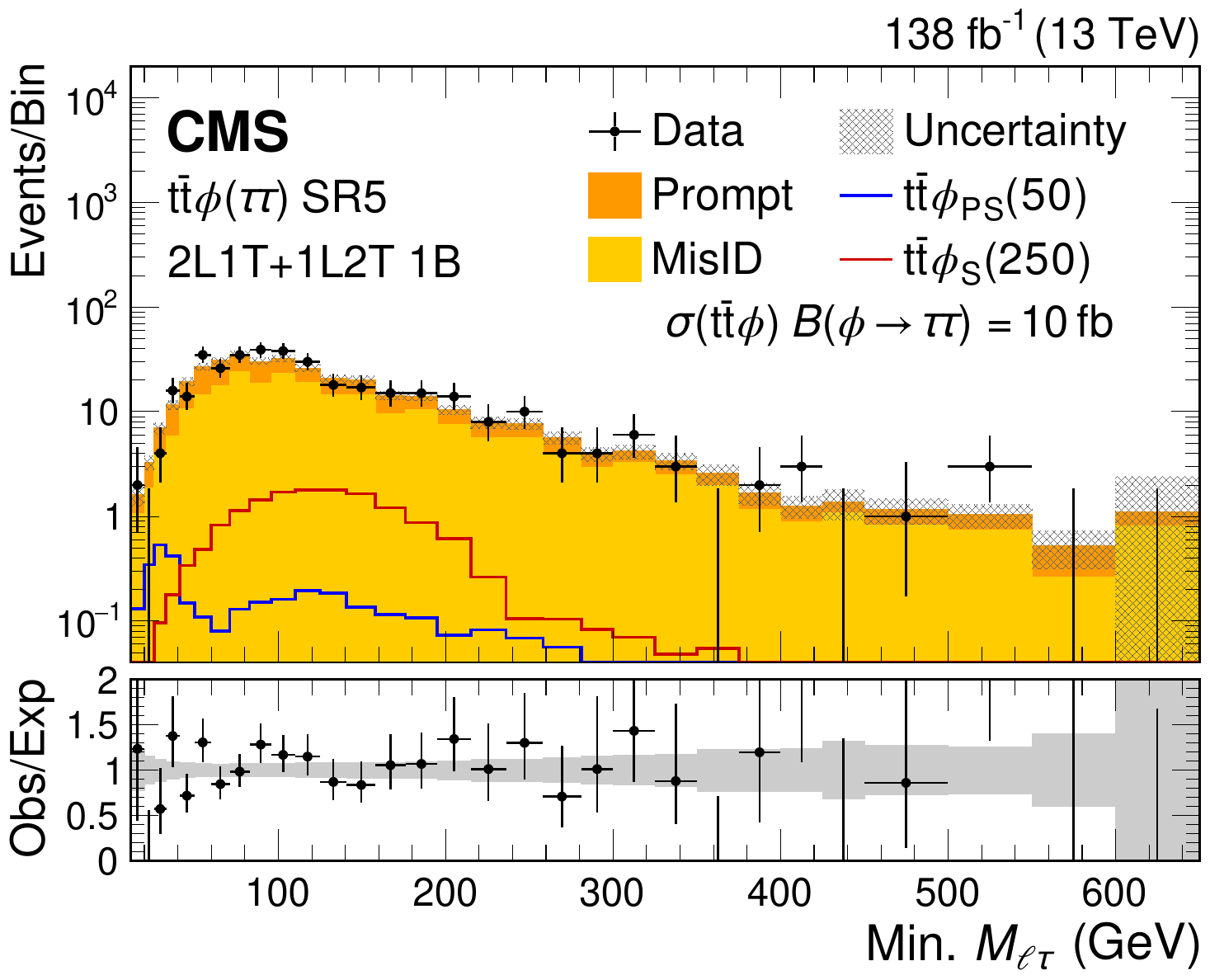}
\includegraphics[width=0.49\textwidth]{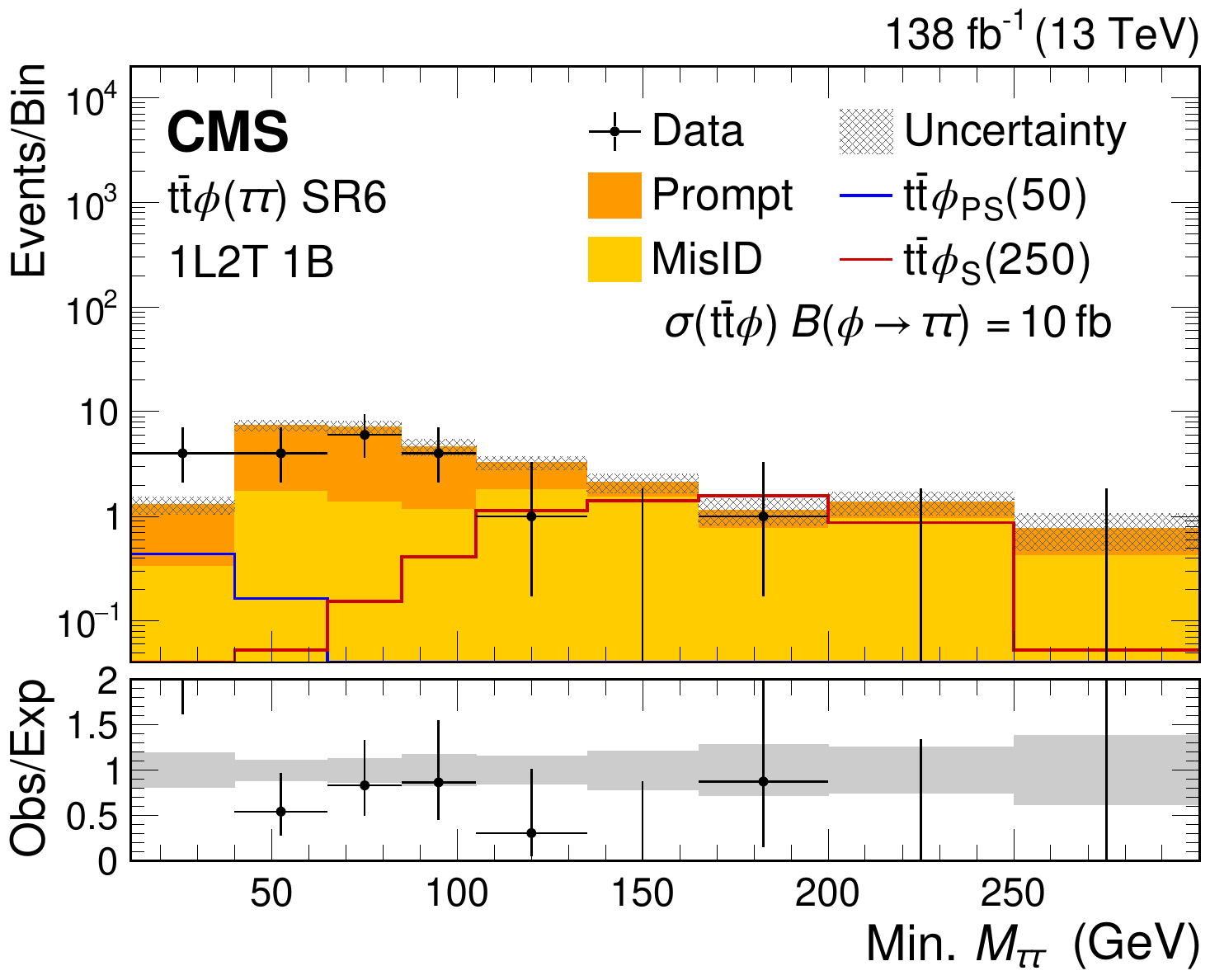}
\caption{\label{fig:ttPhitautau1} Dilepton mass spectra for the $\ttphi(\PGt\PGt)$ SR1-6 event selections for the combined 2016--2018 data set. 
The lower panel shows the ratio of observed events to the total expected SM background prediction (Obs/Exp), and the gray band represents the sum of statistical and systematic uncertainties in the background prediction. 
The rightmost bins contain the overflow events in each distribution. 
The expected background distributions and the uncertainties are shown after the data is fit under the background-only hypothesis. 
For illustration, two example signal hypotheses for the production and decay of a scalar and a pseudoscalar $\phi$ boson are shown, and their masses (in units of \GeV) are indicated in the legend. 
The signals are normalized to the product of the cross section and branching fraction of 10\unit{fb}. 
}
\end{figure*}

\begin{figure*}[hbt!]
\centering
\includegraphics[width=0.49\textwidth]{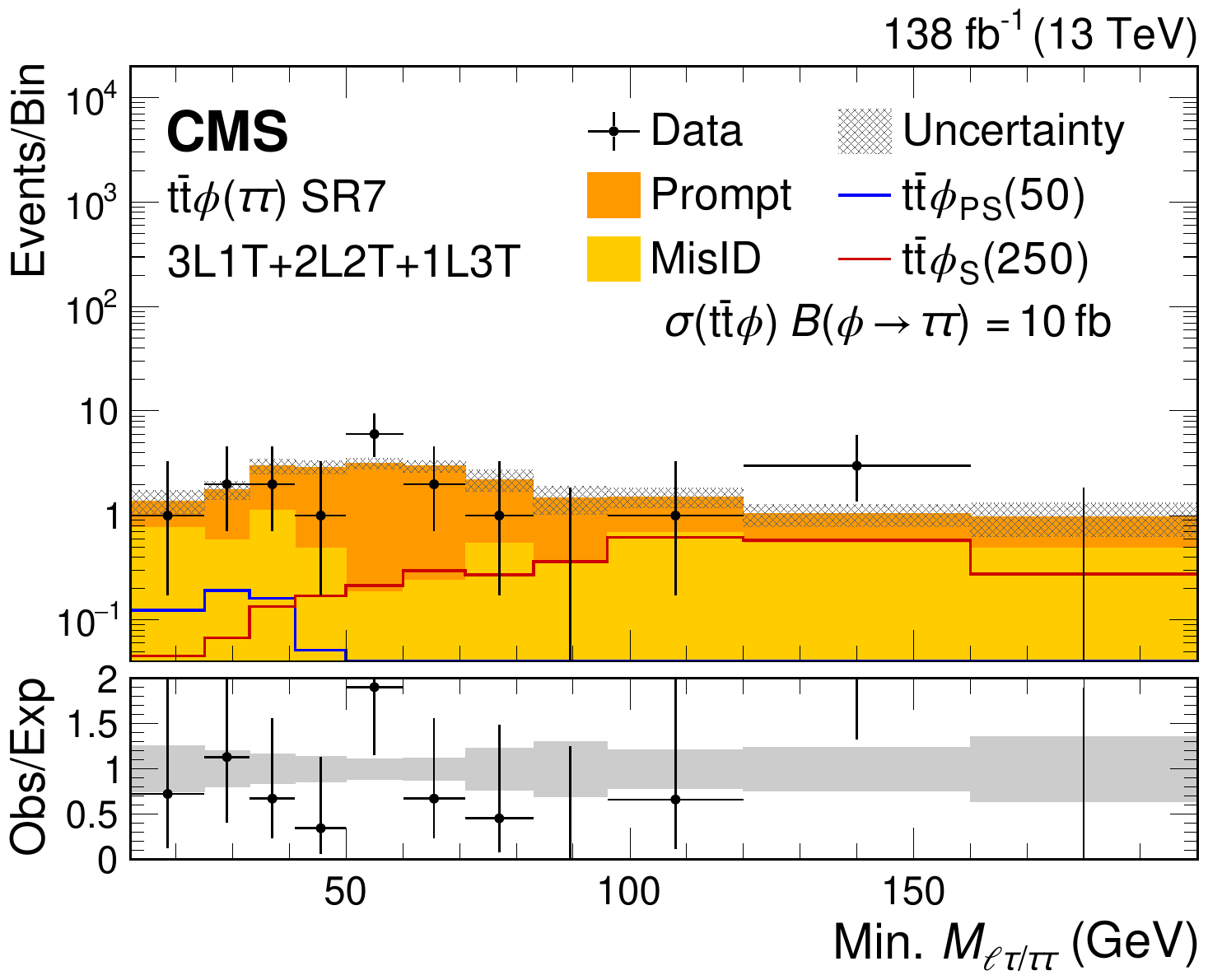}
\caption{\label{fig:ttPhitautau2} 
Dilepton mass spectra for the $\ttphi(\PGt\PGt)$ SR7 event selection for the combined 2016--2018 data set.
The lower panel shows the ratio of observed events to the total expected SM background prediction (Obs/Exp), and the gray band represents the sum of statistical and systematic uncertainties in the background prediction. 
The rightmost bins contain the overflow events in each distribution. 
The expected background distributions and the uncertainties are shown after the data is fit under the background-only hypothesis. 
For illustration, two example signal hypotheses for the production and decay of a scalar and a pseudoscalar $\phi$ boson are shown, and their masses (in units of \GeV) are indicated in the legend. 
The signals are normalized to the product of the cross section and branching fraction of 10\unit{fb}. 
}
\end{figure*}

\begin{figure*}[hbt!]
\centering
\includegraphics[width=0.41\textwidth]{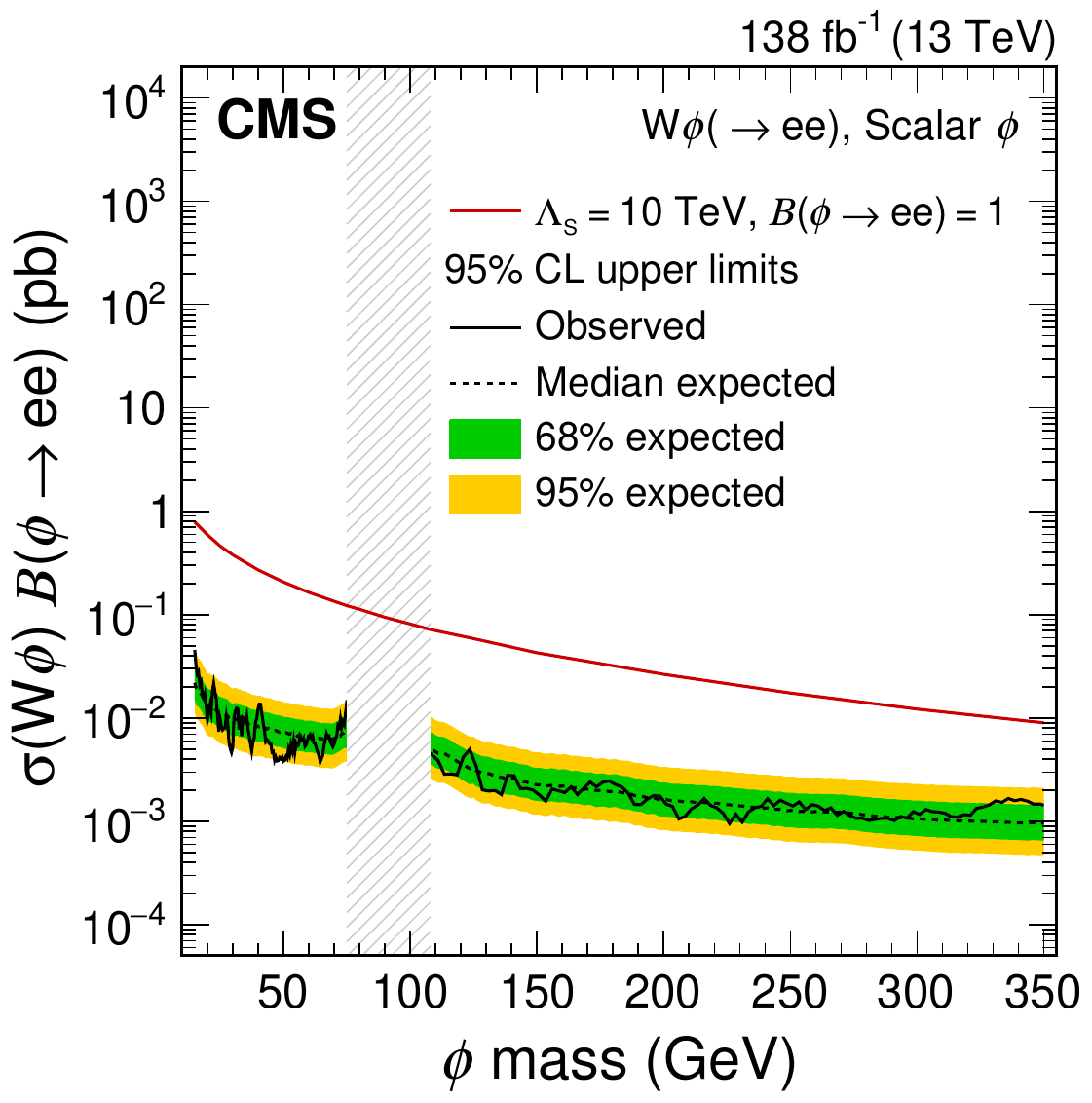}
\includegraphics[width=0.41\textwidth]{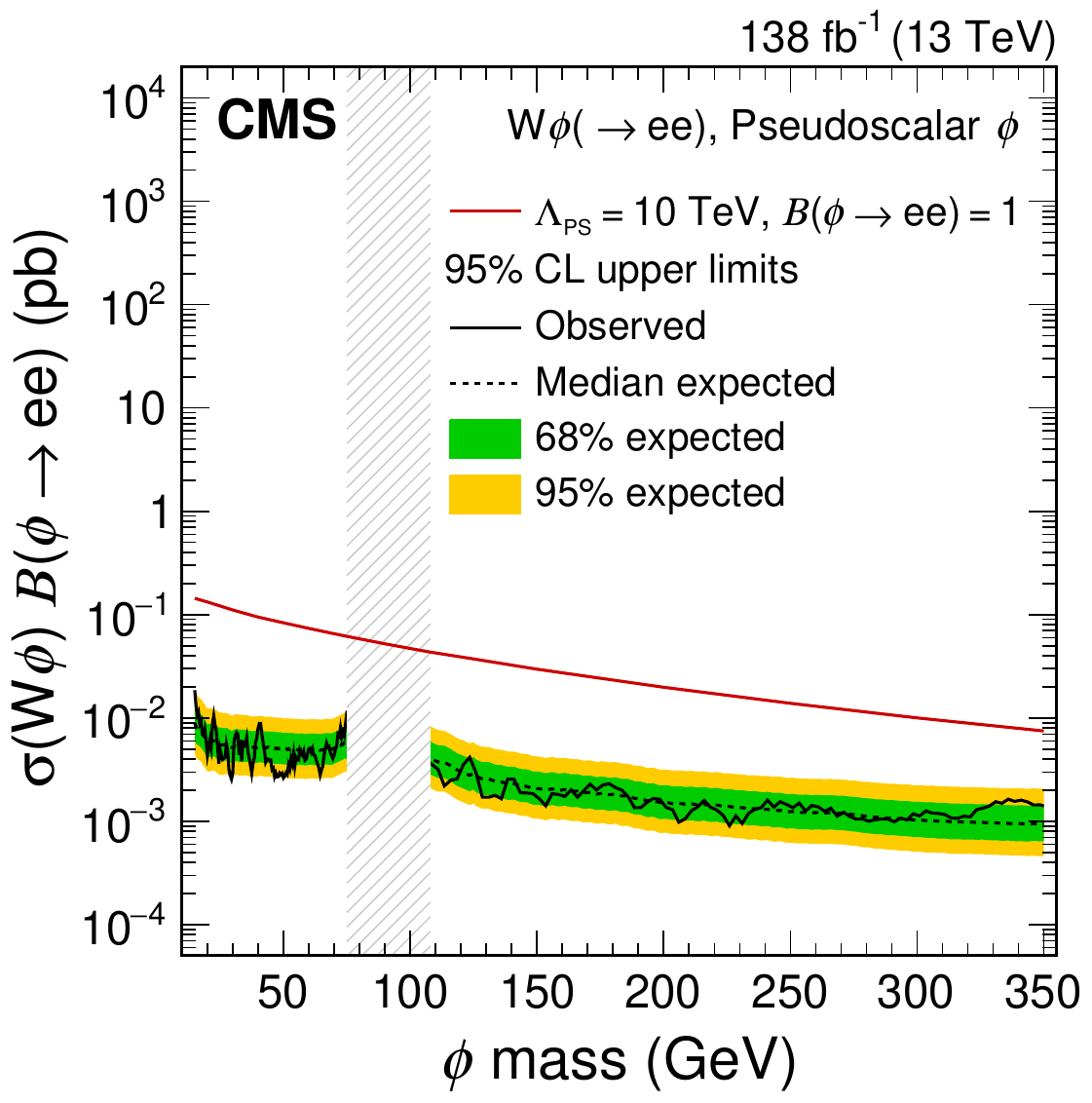}
\includegraphics[width=0.41\textwidth]{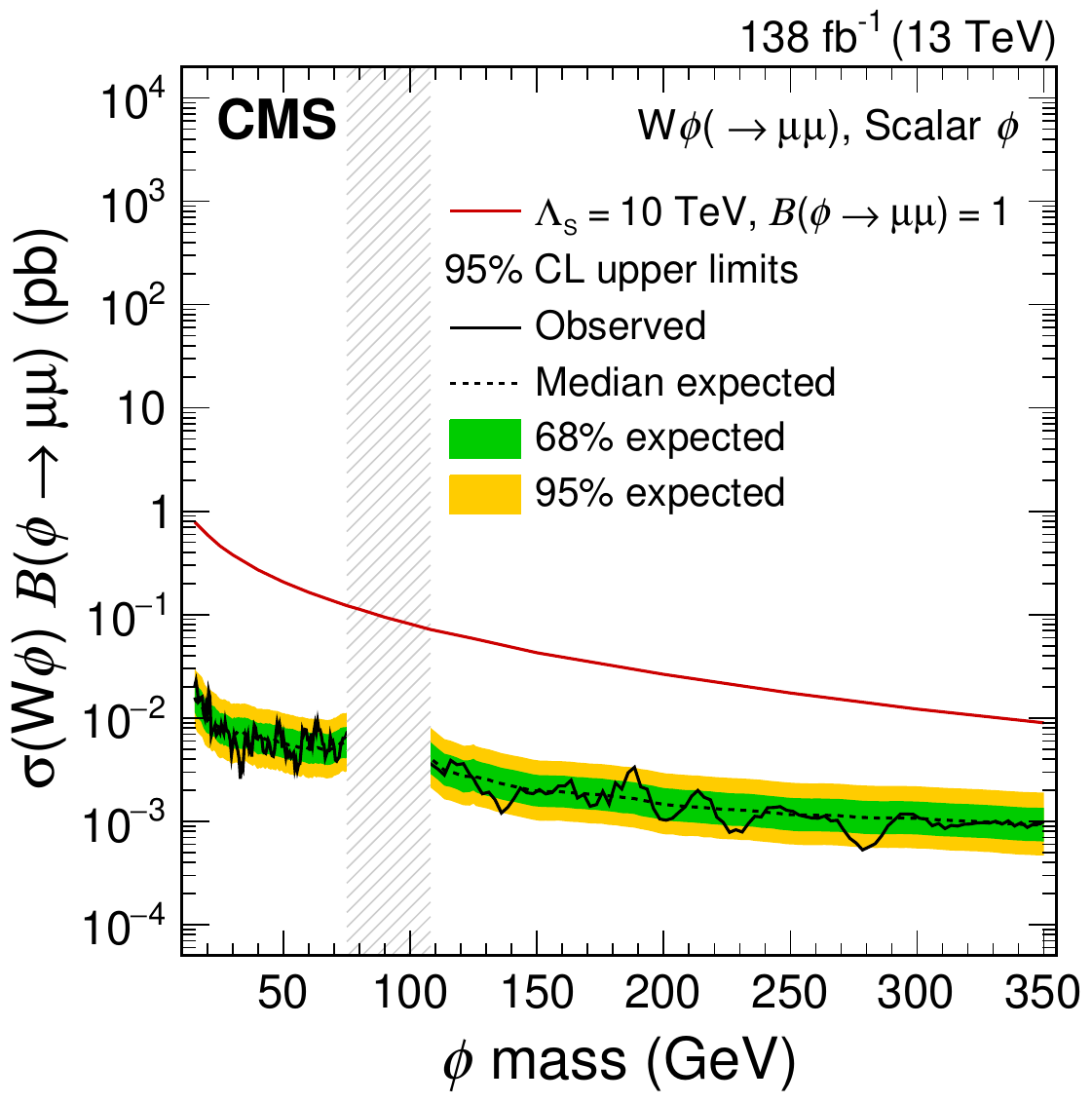}
\includegraphics[width=0.41\textwidth]{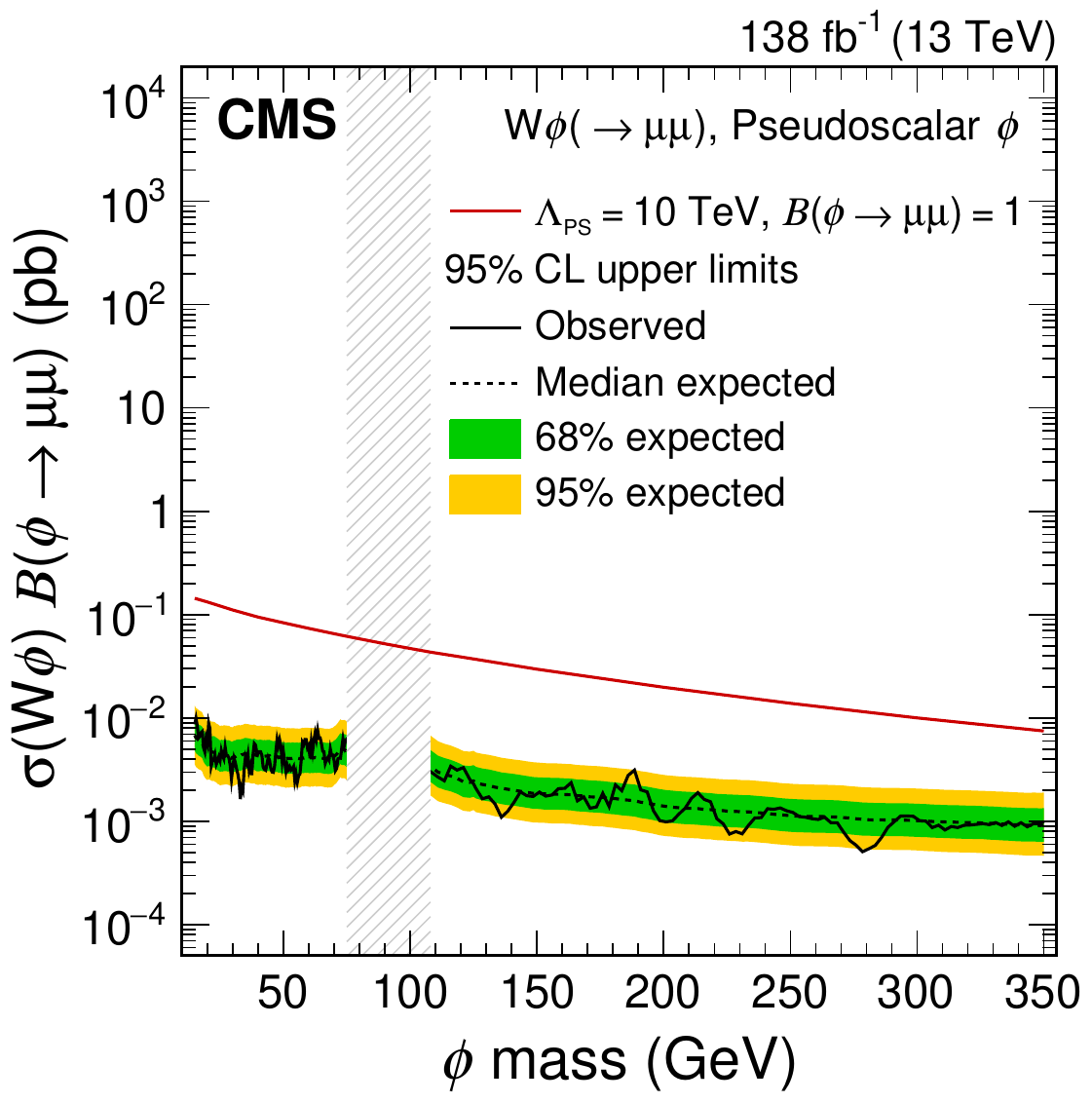}
\includegraphics[width=0.41\textwidth]{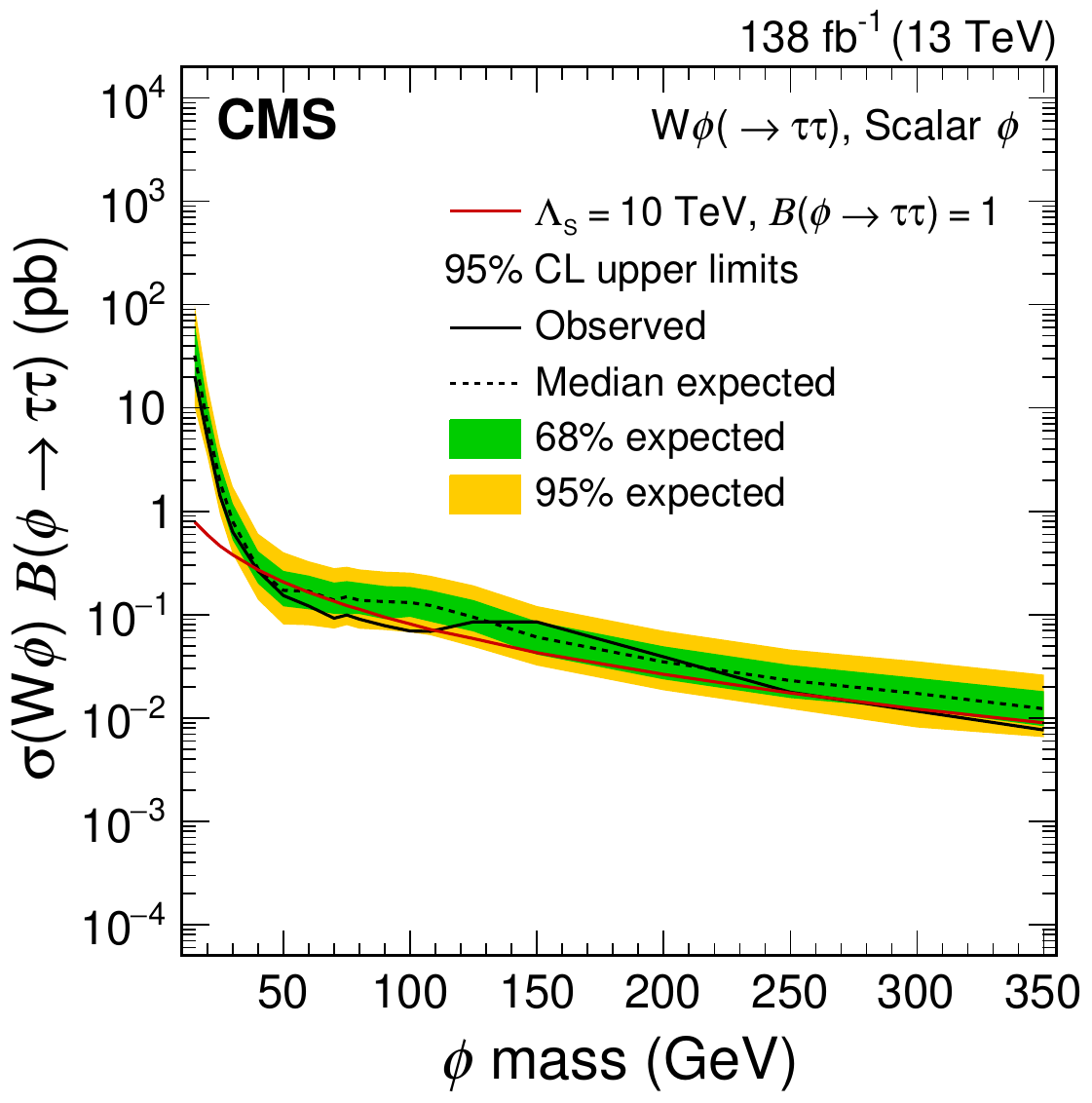}
\includegraphics[width=0.41\textwidth]{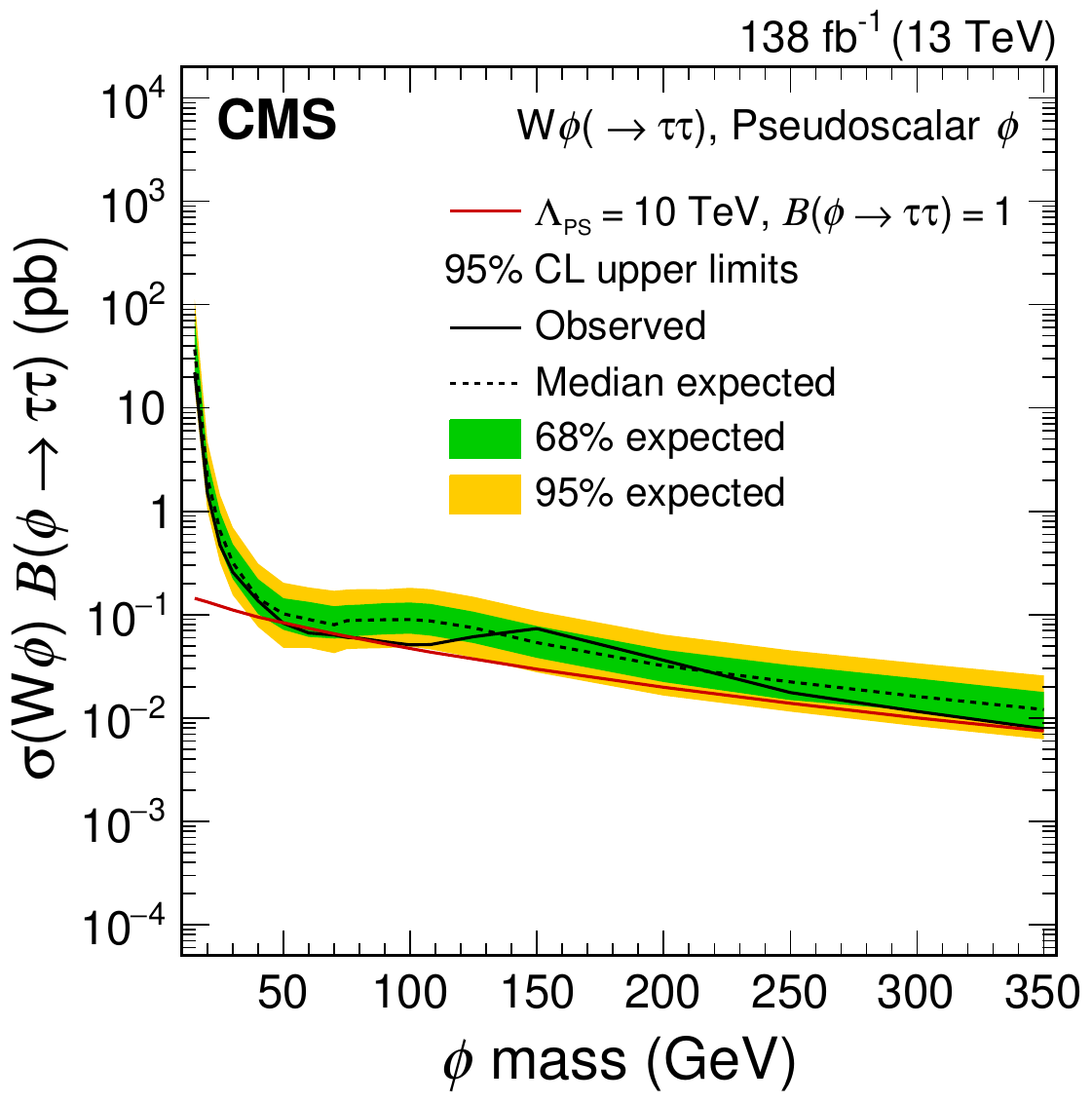}
\caption{\label{fig:WPhiSPSLimitPlots} 
The 95\% confidence level upper limits on the product of the production cross section and branching fraction of the $\Wphi$ signal in the $\Pe\Pe$ (upper), $\PGm\PGm$ (middle), and $\PGt\PGt$ (lower) decay scenarios. 
The results for the scalar coupling are shown on the left and pseudoscalar on the right. 
The vertical gray band indicates the mass region not considered in the analysis.
The red line is the theoretical prediction for the product of the production cross section and branching fraction of the $\Wphi$ signal.
}
\end{figure*}

\begin{figure*}[hbt!]
\centering
\includegraphics[width=0.41\textwidth]{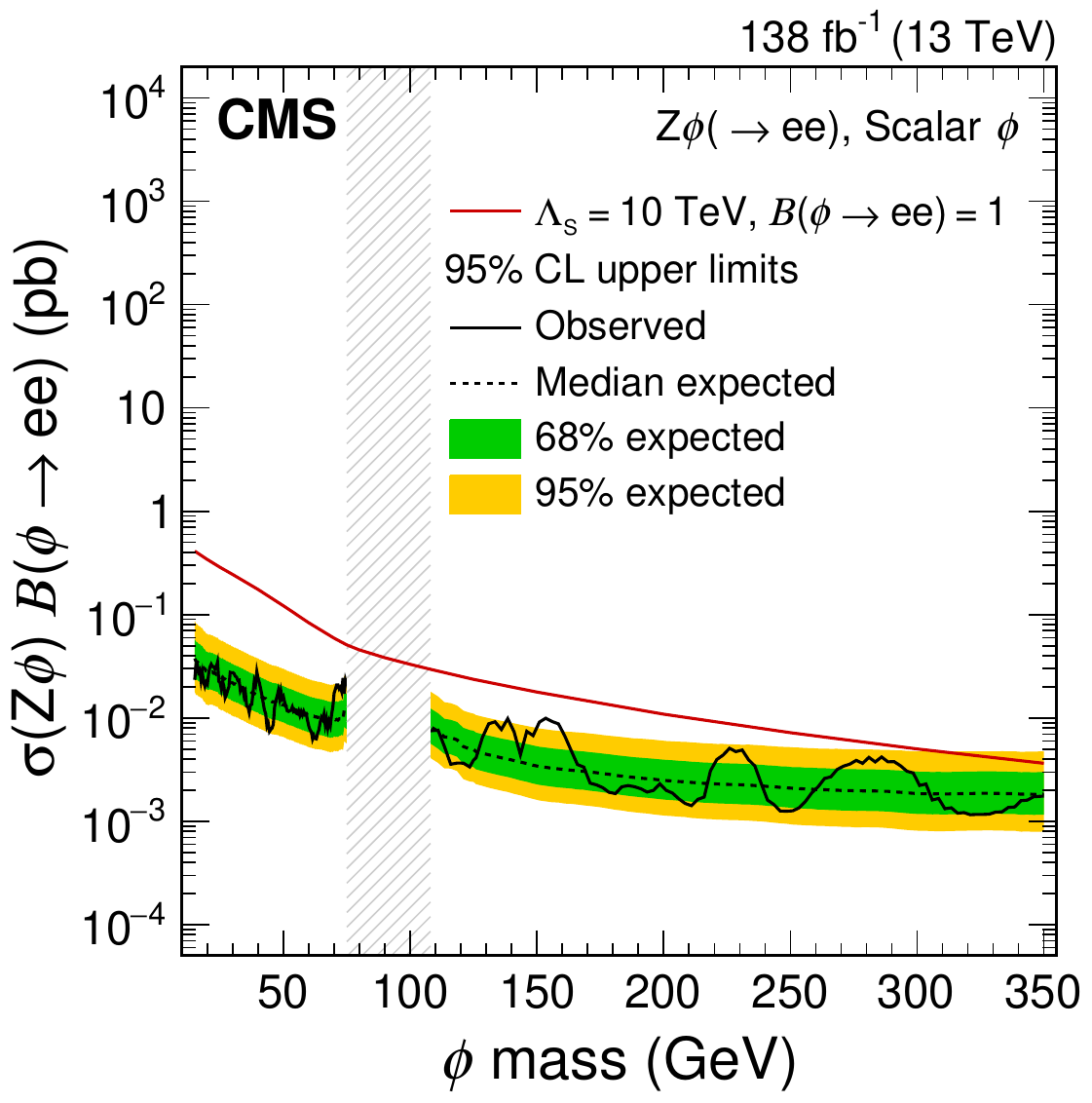}
\includegraphics[width=0.41\textwidth]{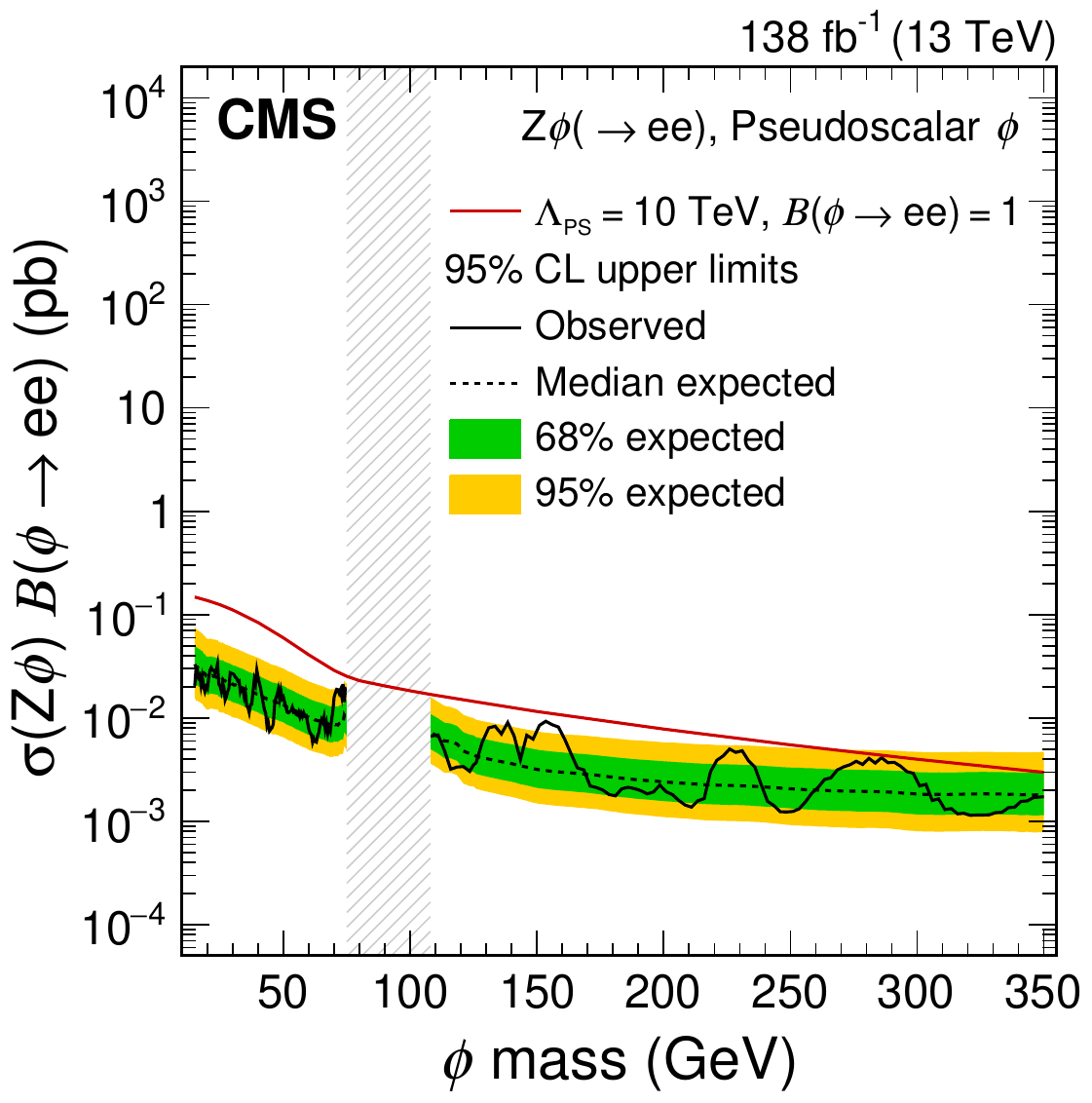}
\includegraphics[width=0.41\textwidth]{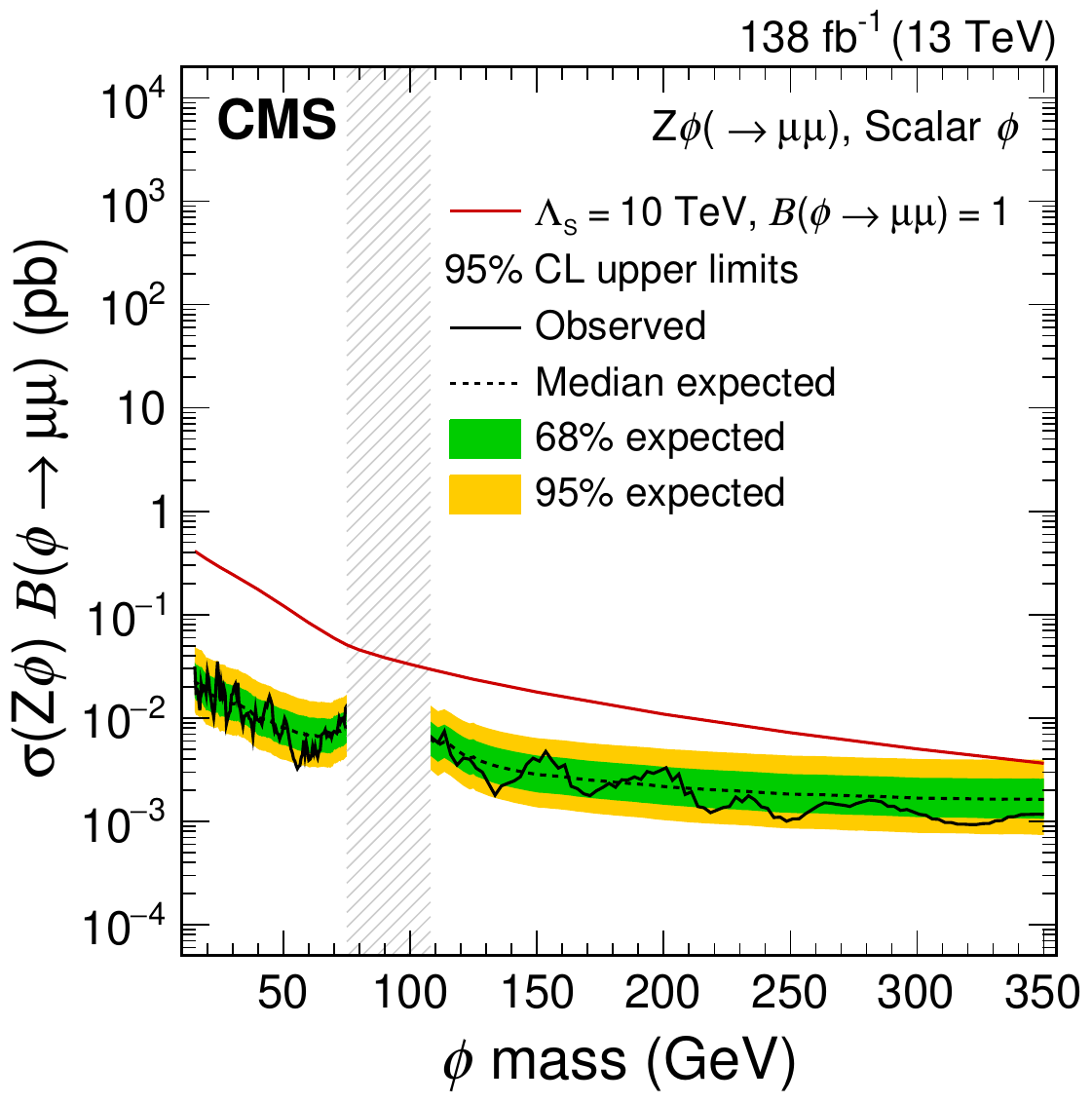}
\includegraphics[width=0.41\textwidth]{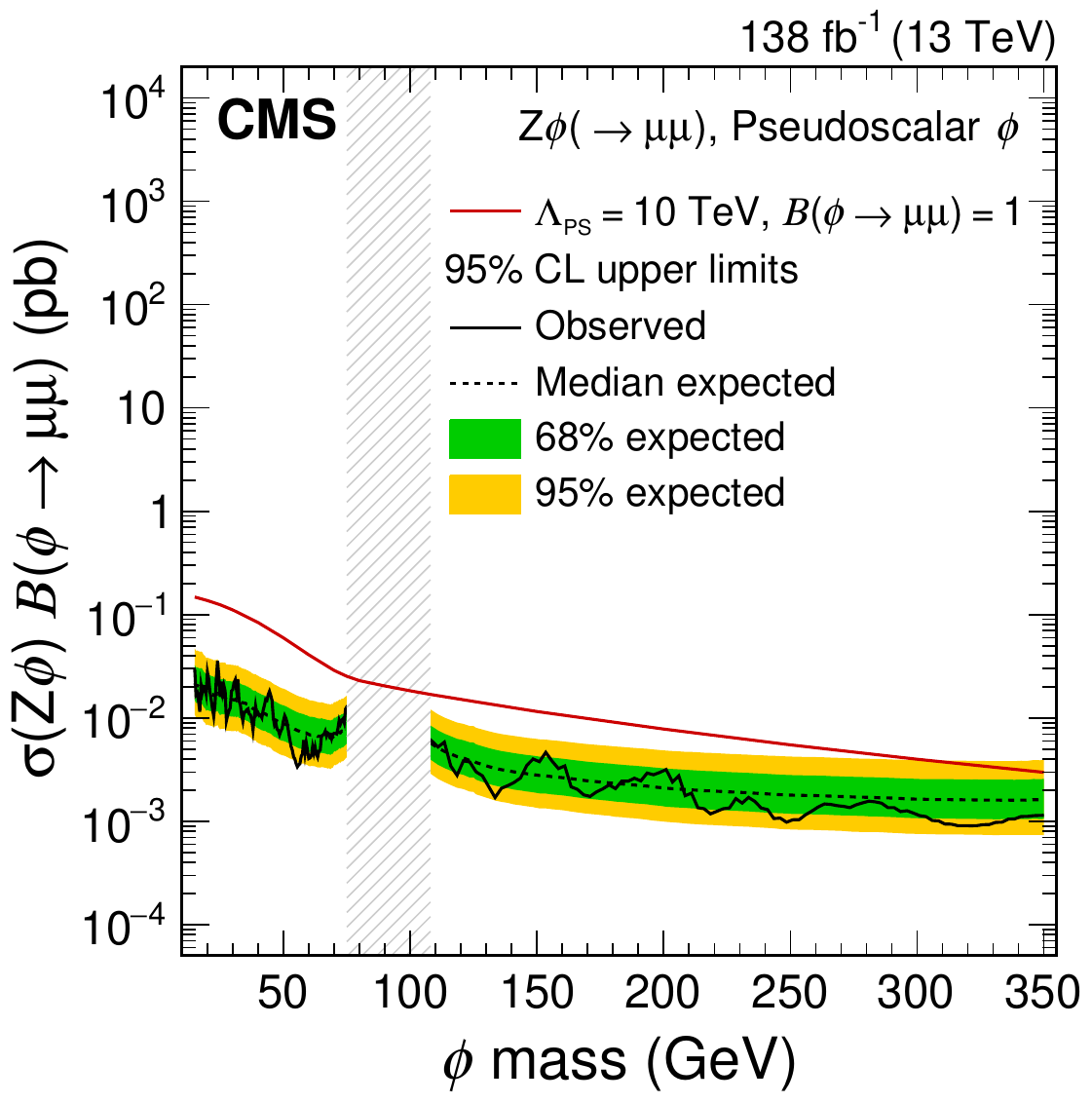}
\includegraphics[width=0.41\textwidth]{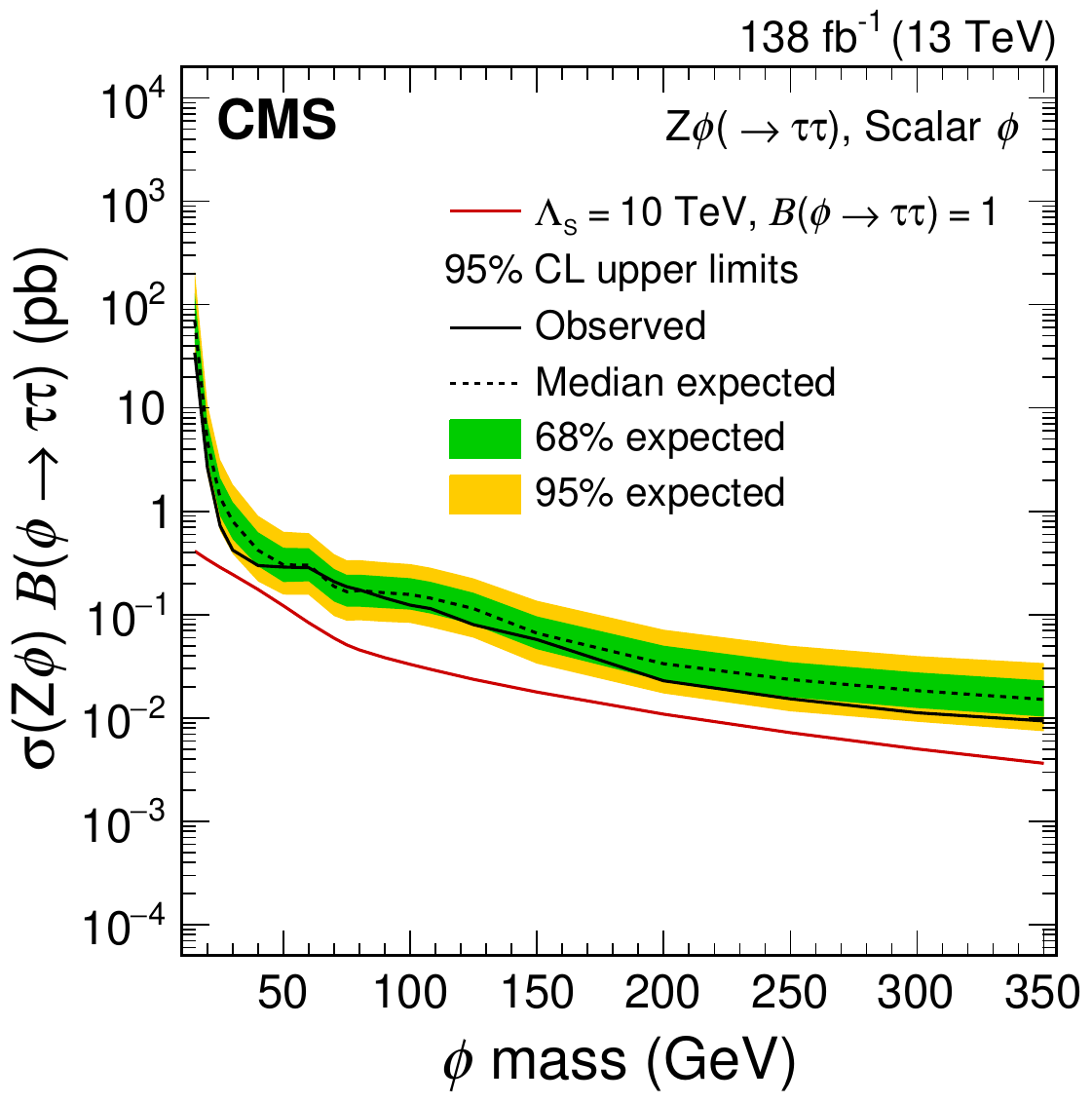}
\includegraphics[width=0.41\textwidth]{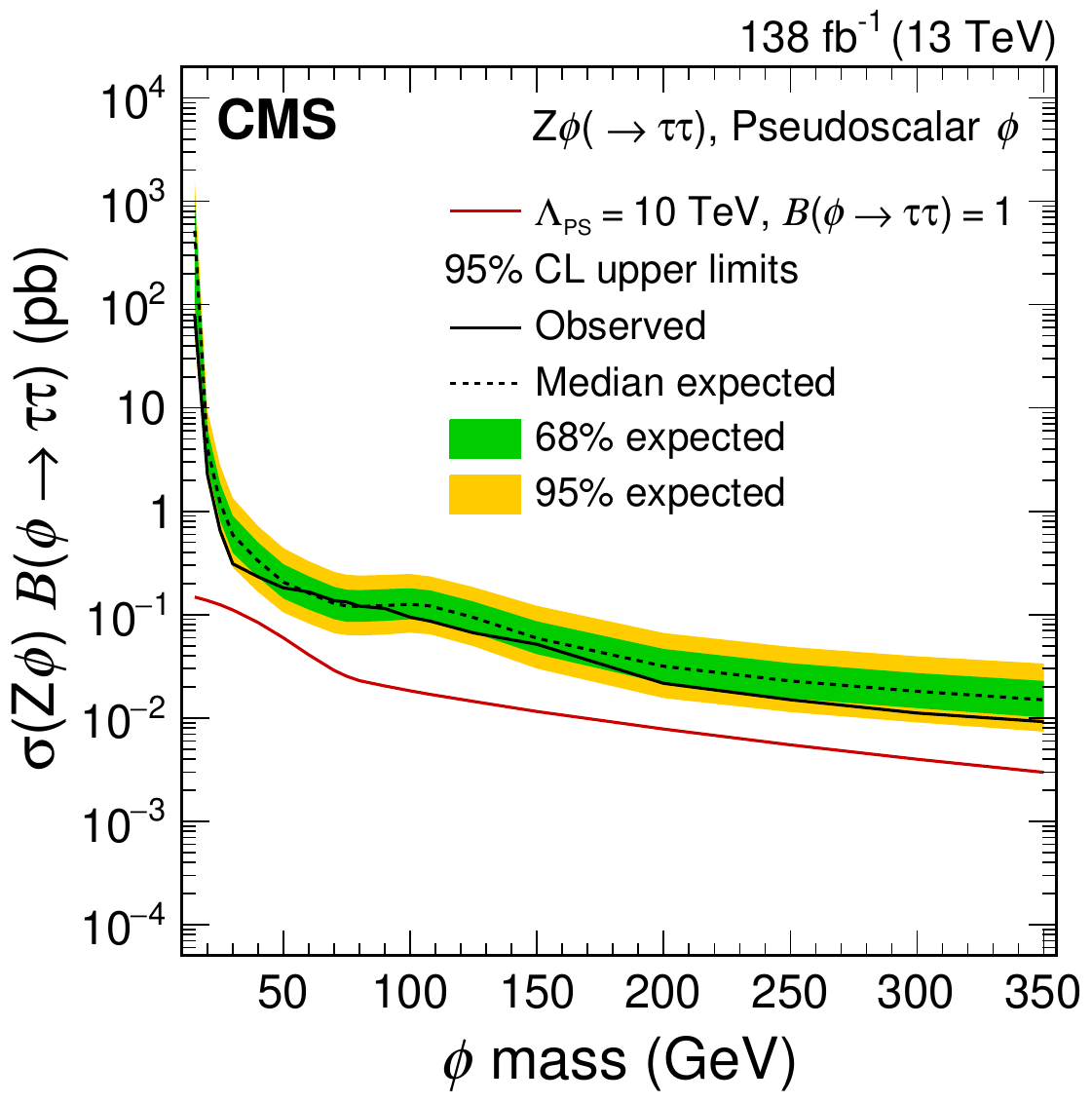}
\caption{\label{fig:ZPhiSPSLimitPlots} 
The 95\% confidence level upper limits on the product of the production cross section and branching fraction of the $\Zphi$ signal in the $\Pe\Pe$ (upper), $\PGm\PGm$ (middle) and $\PGt\PGt$ (lower) decay scenarios. 
The results for the scalar coupling are shown on the left and pseudoscalar on the right. 
The vertical gray band indicates the mass region not considered in the analysis.
The red line is the theoretical prediction for the product of the production cross section and branching fraction of the $\Zphi$ signal.
}
\end{figure*}

\begin{figure*}[hbt!]
\centering
\includegraphics[width=0.41\textwidth]{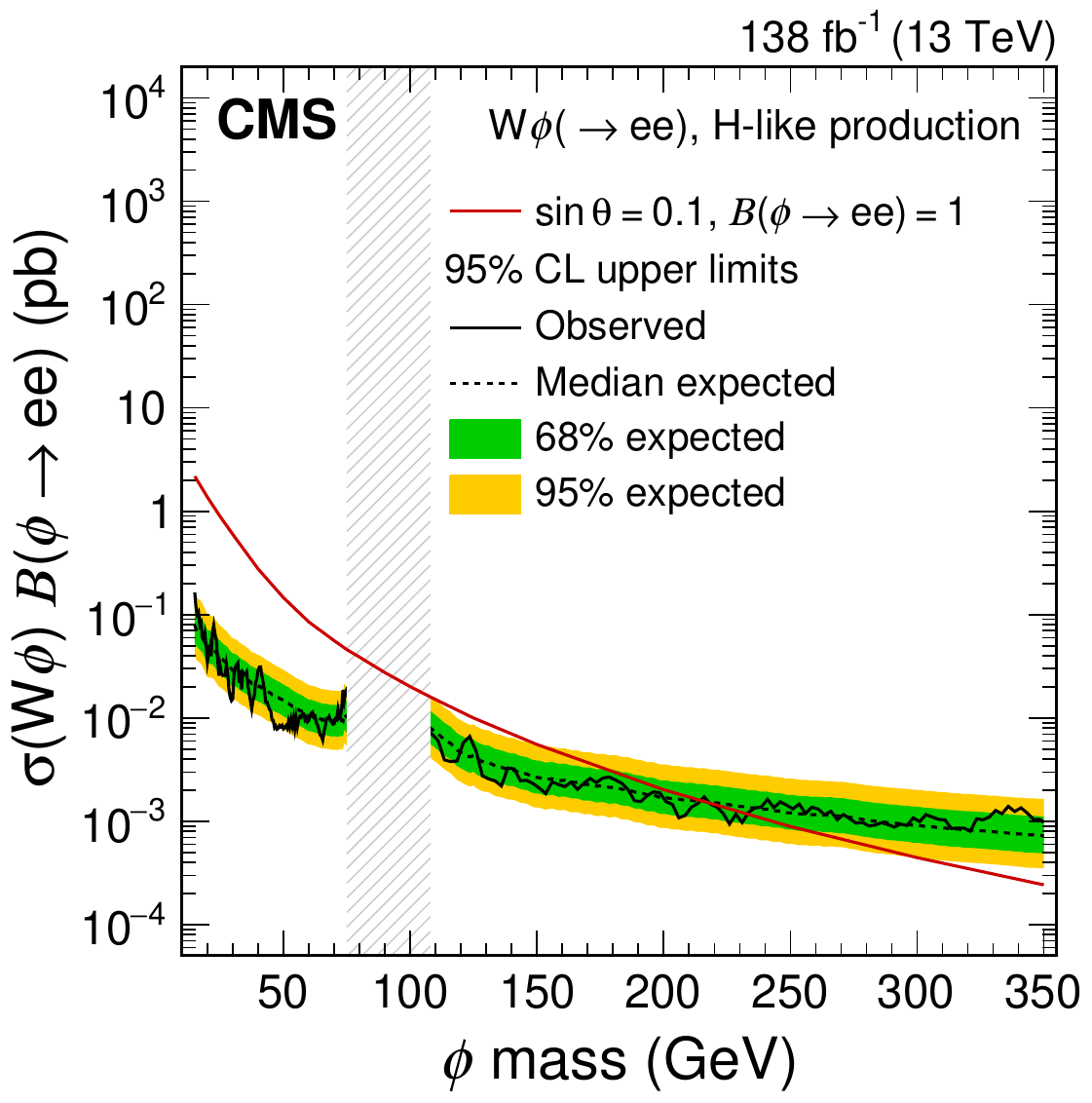}
\includegraphics[width=0.41\textwidth]{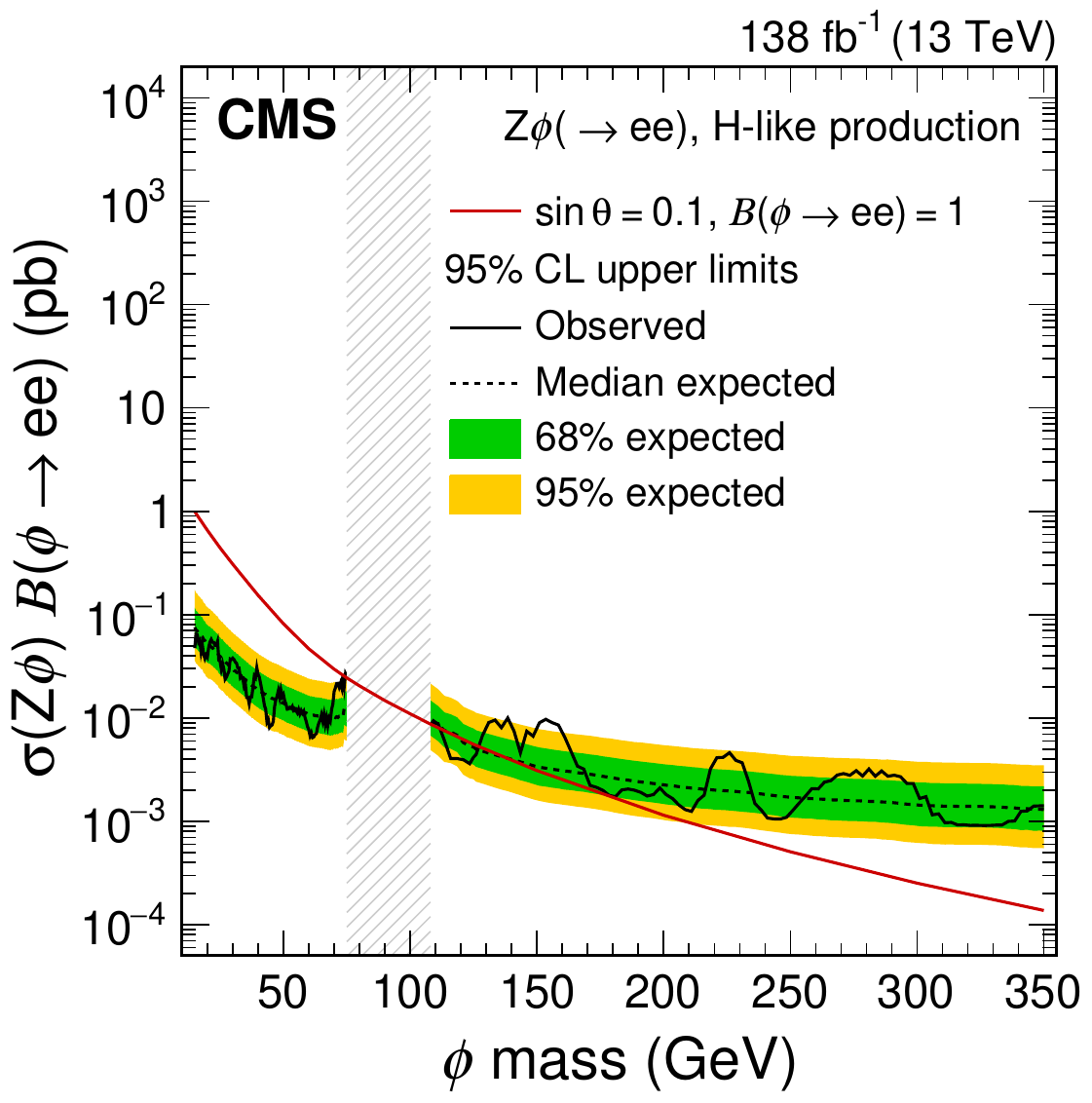}
\includegraphics[width=0.41\textwidth]{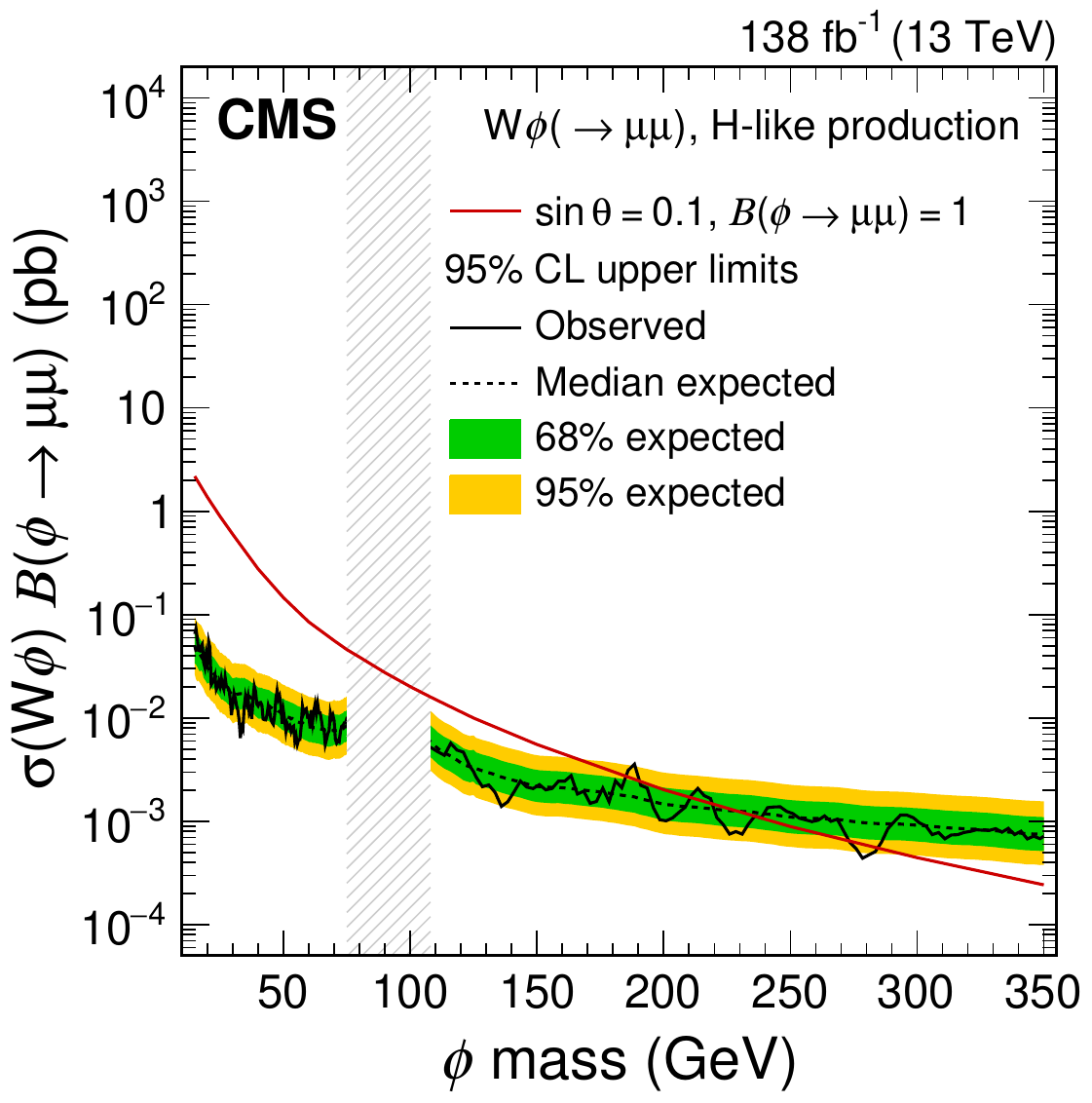}
\includegraphics[width=0.41\textwidth]{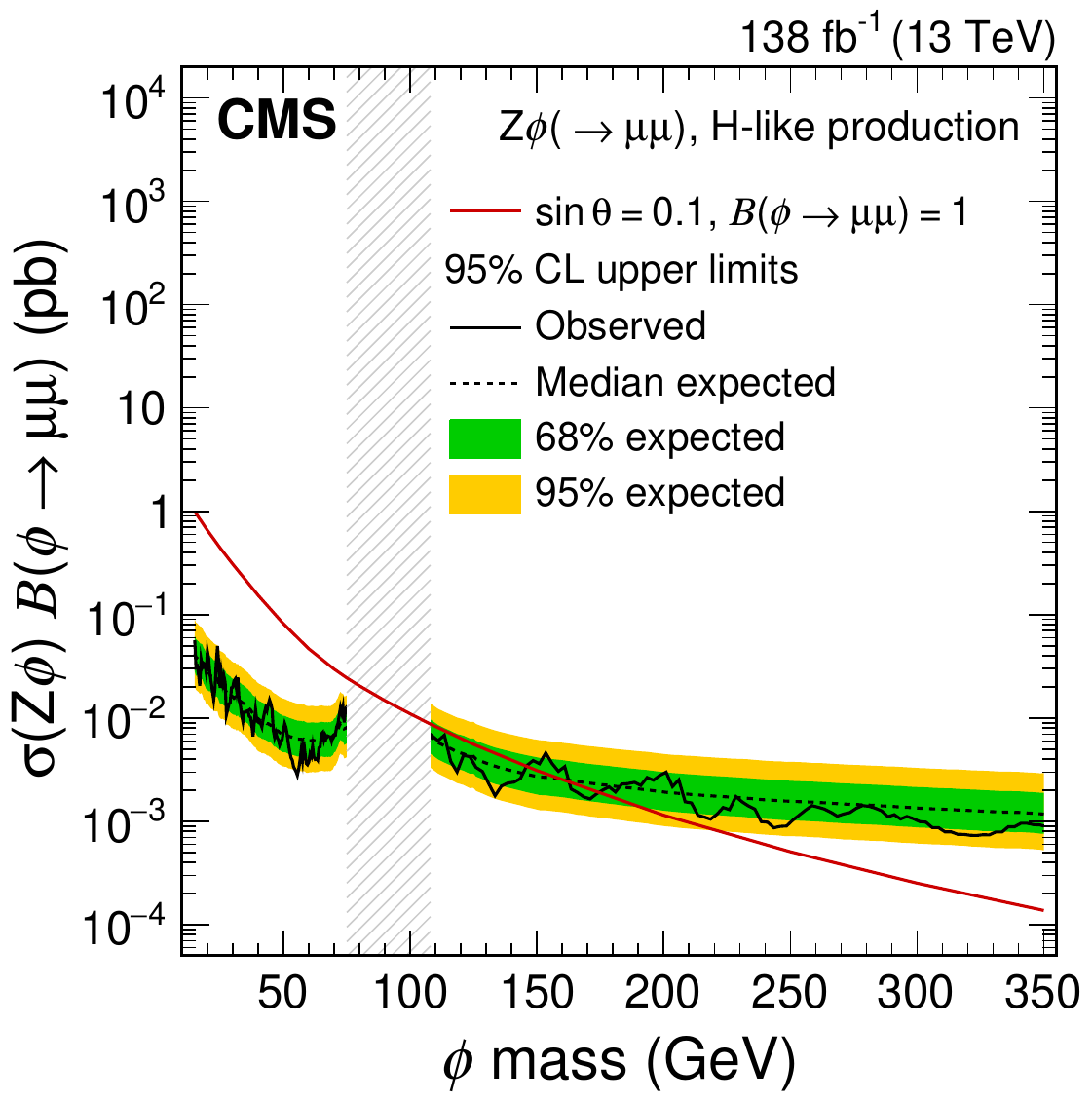}
\includegraphics[width=0.41\textwidth]{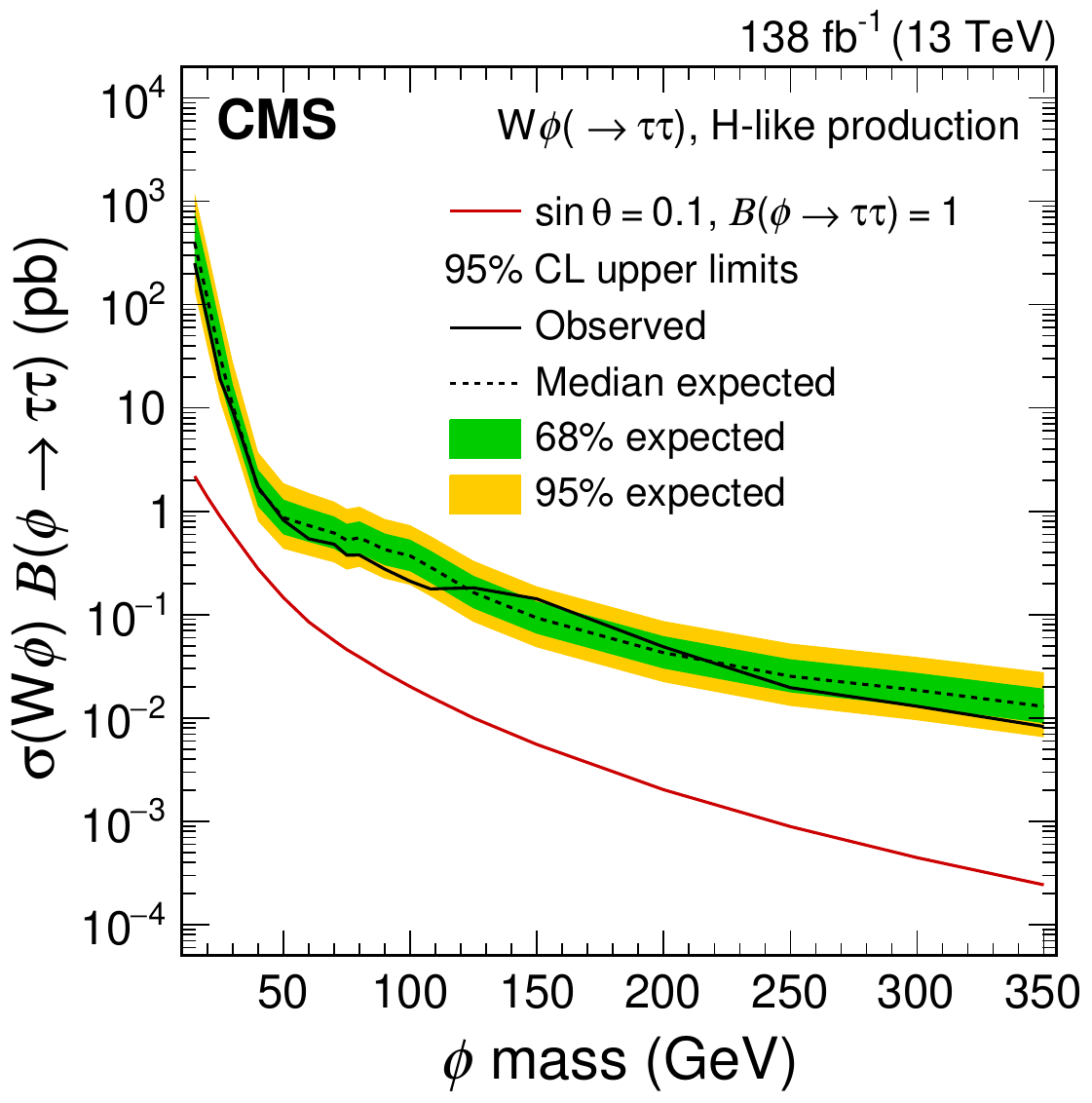}
\includegraphics[width=0.41\textwidth]{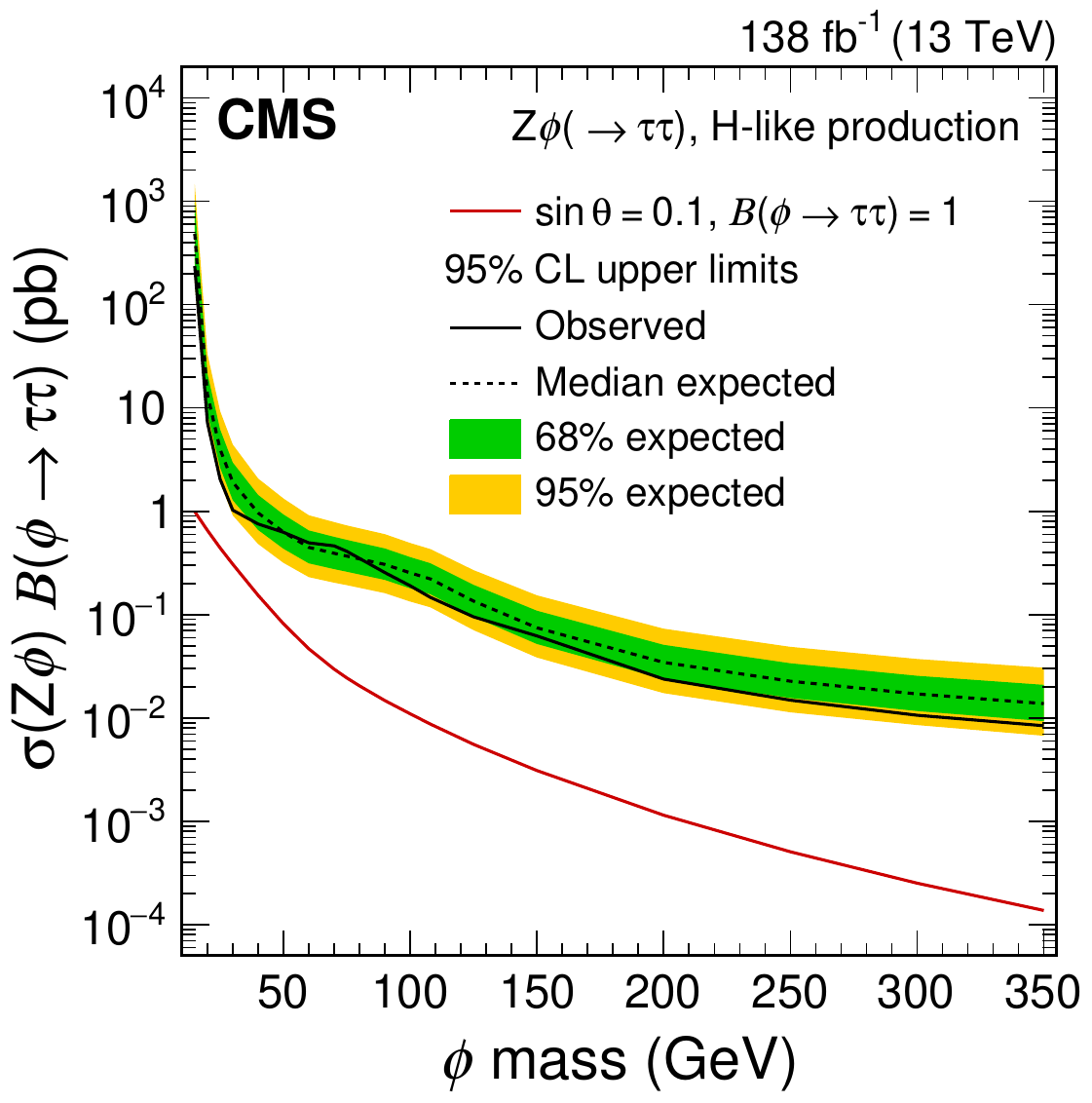}
\caption{\label{fig:WZPhiHLimitPlots} 
The 95\% confidence level upper limits on the product of the production cross section and branching fraction of the $\Wphi$ signal on the left and the $\Zphi$ signal on the right with H-like couplings in the $\Pe\Pe$ (upper), $\PGm\PGm$ (middle) and $\PGt\PGt$ (lower) decay scenarios. 
The vertical gray band indicates the mass region not considered in the analysis.
The red line is the theoretical prediction for the product of the production cross section and branching fraction of the $\Wphi$ and $\Zphi$ signals.
}
\end{figure*}

\begin{figure*}[hbt!]
\centering
\includegraphics[width=0.41\textwidth]{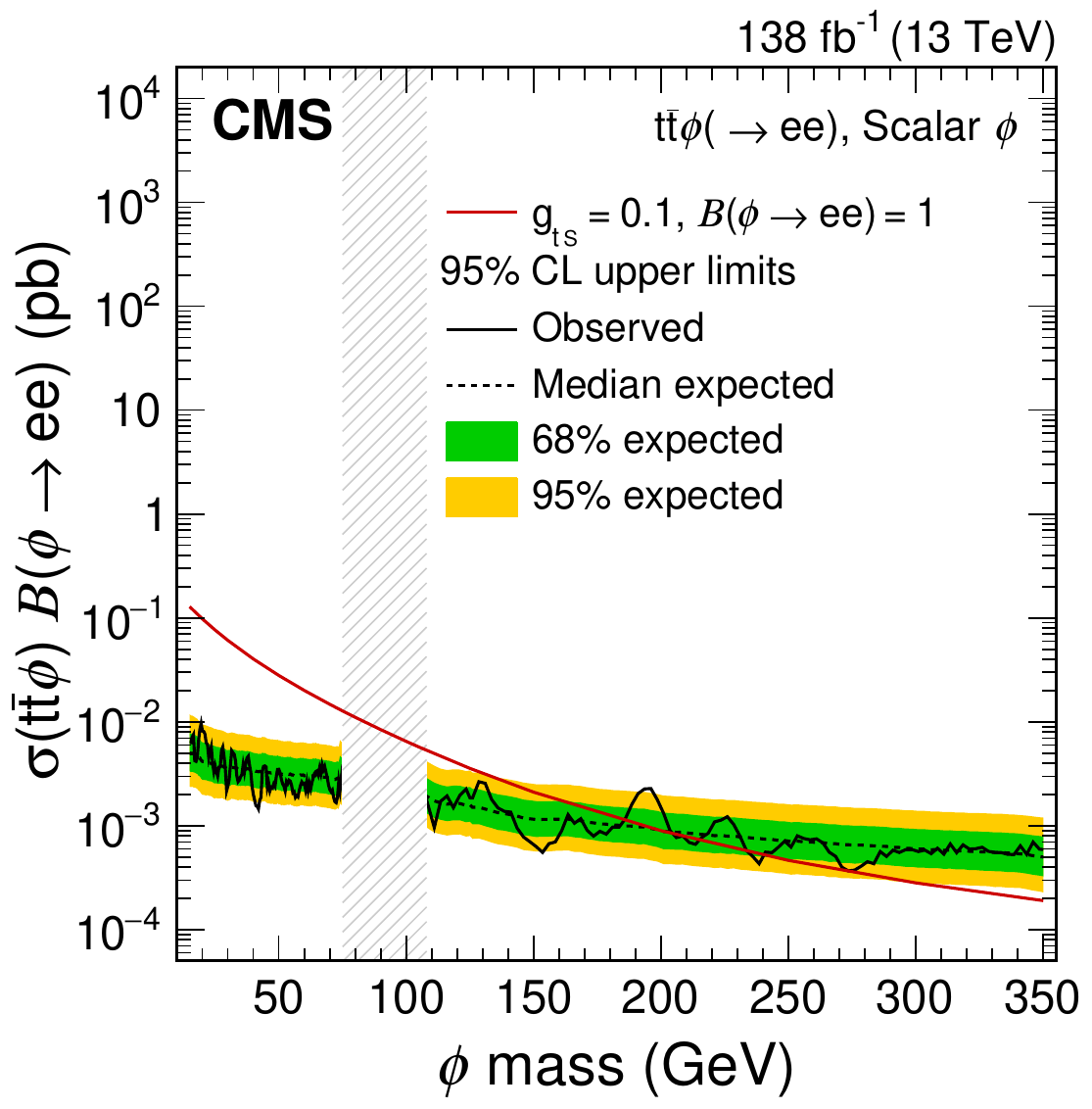}
\includegraphics[width=0.41\textwidth]{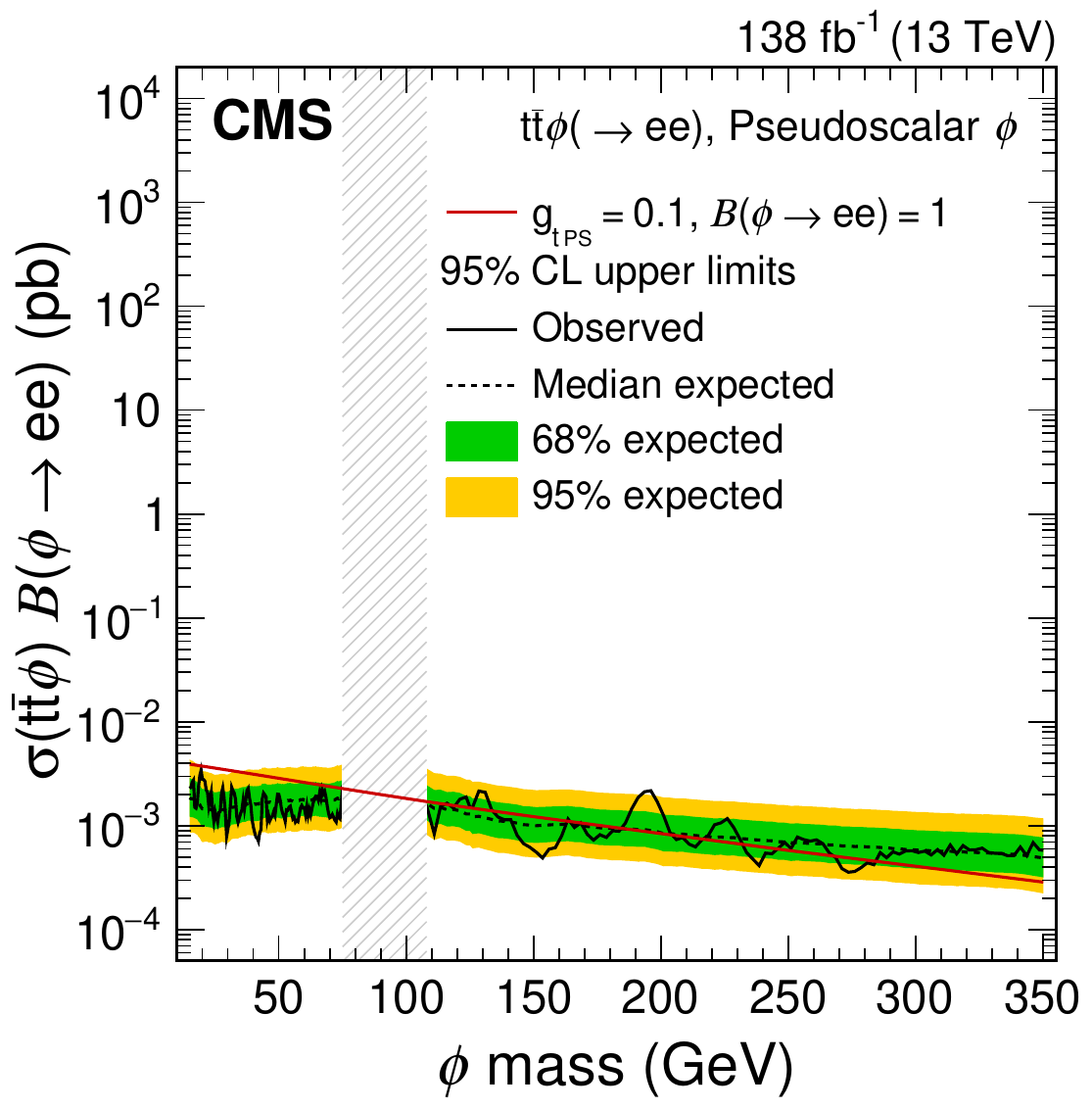}
\includegraphics[width=0.41\textwidth]{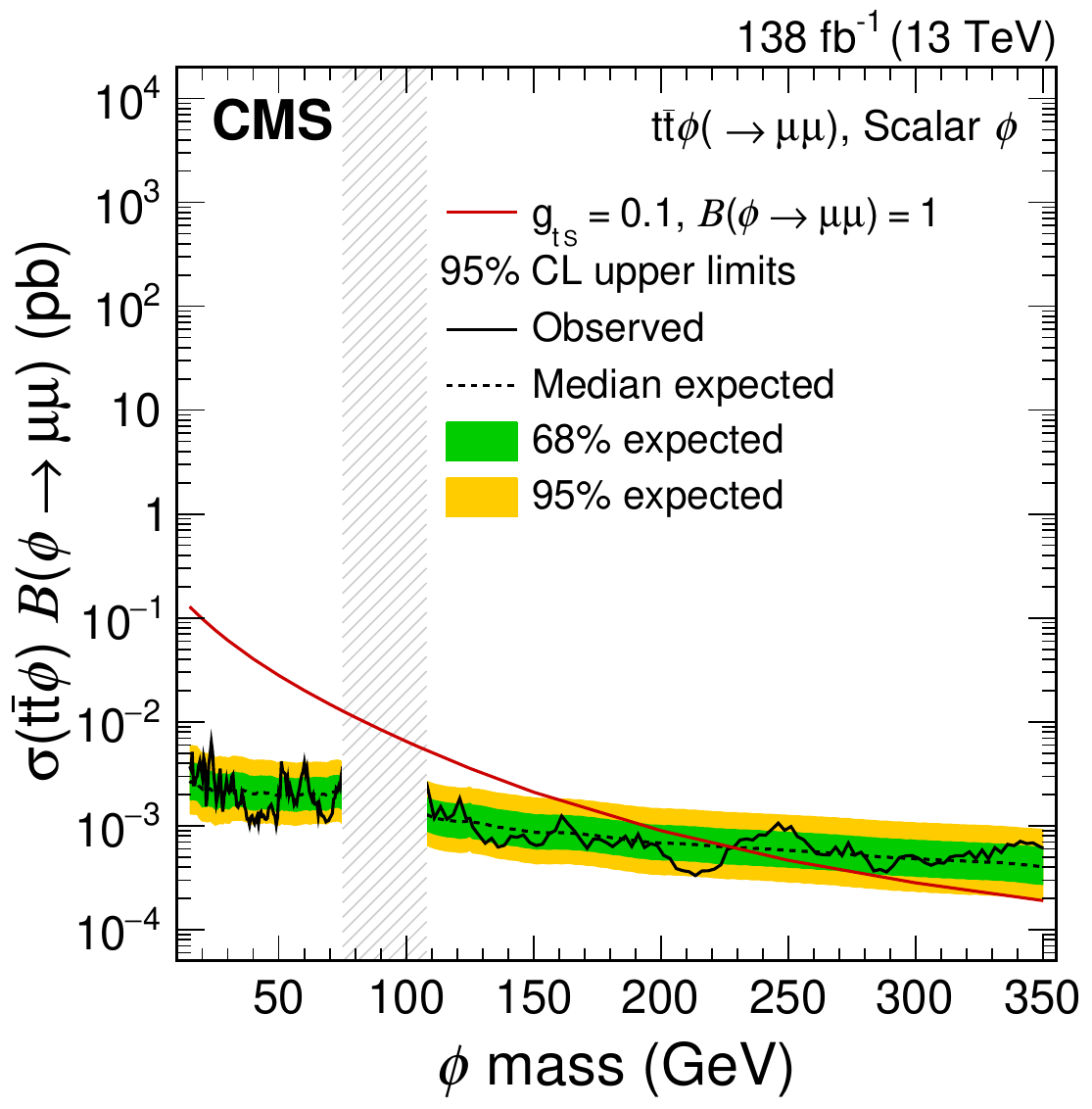}
\includegraphics[width=0.41\textwidth]{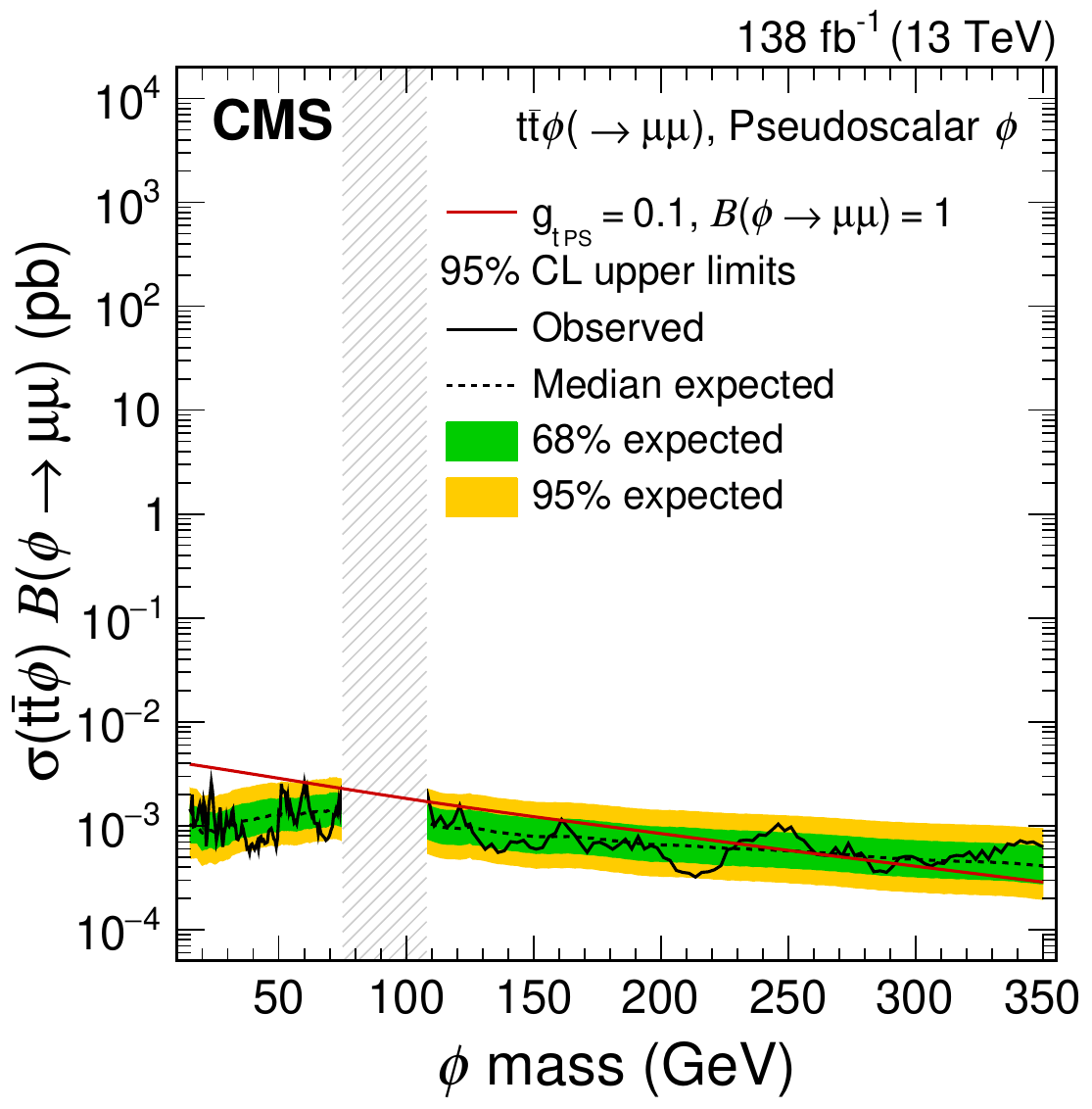}
\includegraphics[width=0.41\textwidth]{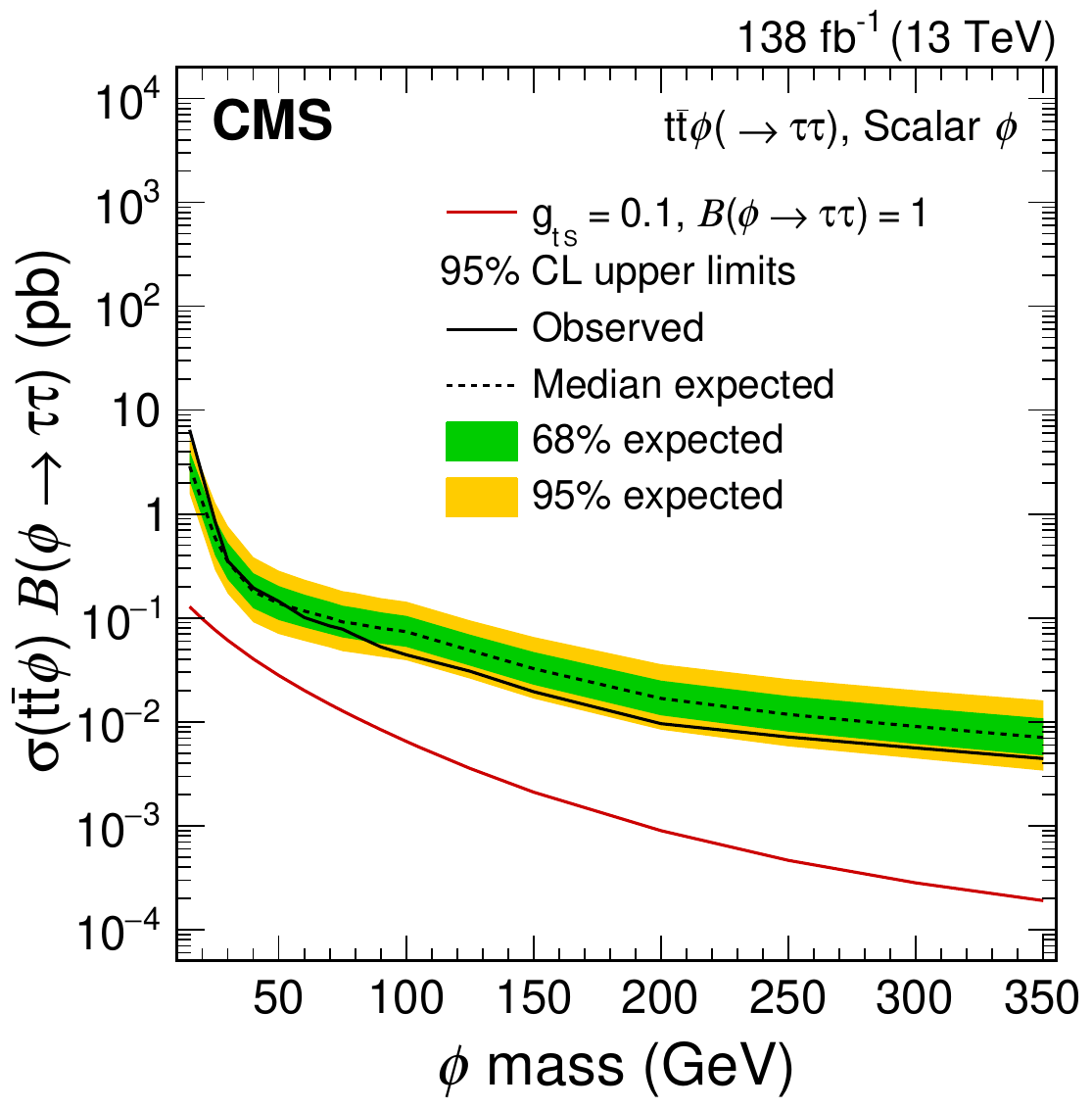}
\includegraphics[width=0.41\textwidth]{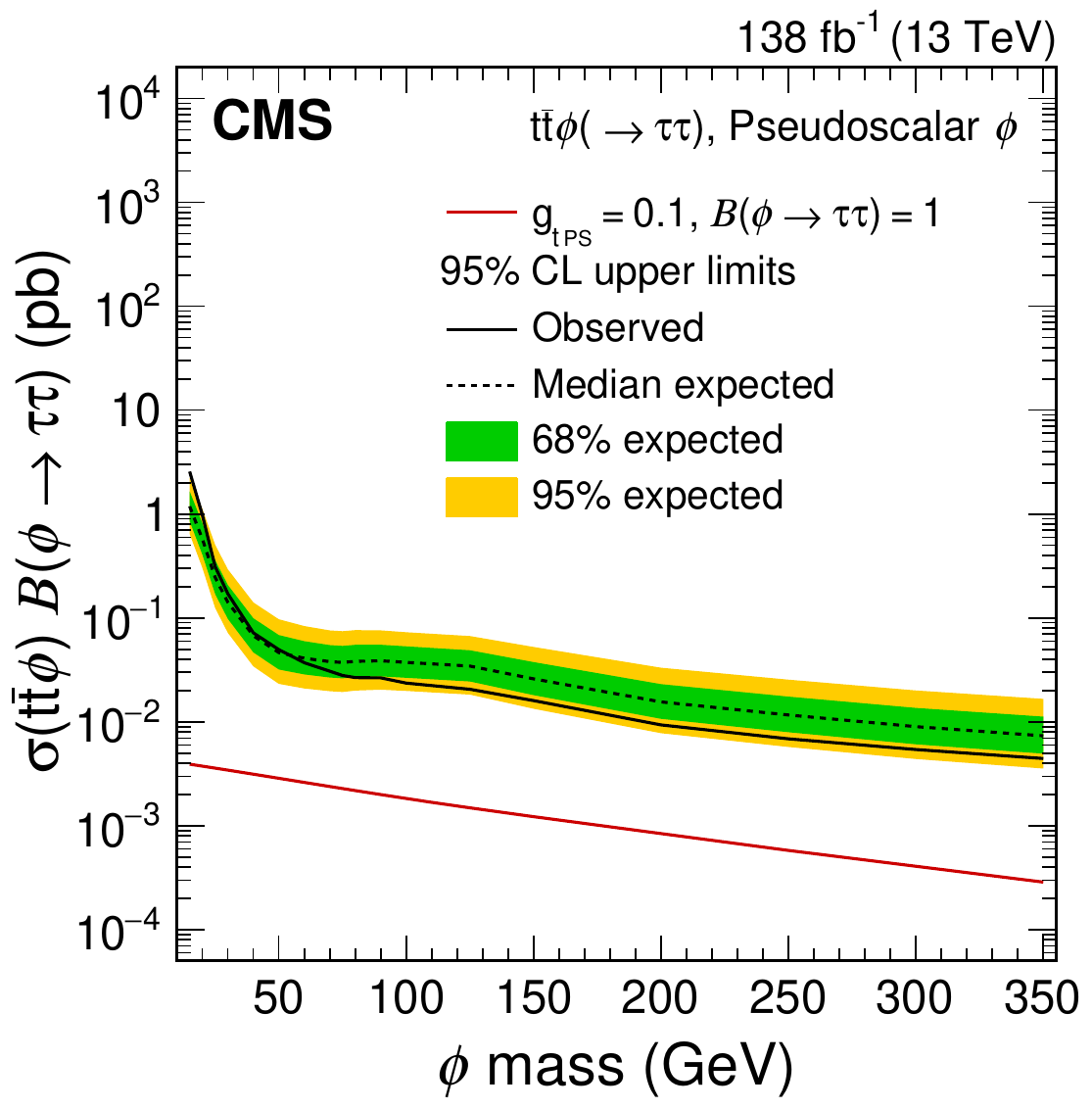}
\caption{\label{fig:ttPhiSPSLimitPlots} 
The 95\% confidence level upper limits on the product of the production cross section and branching fraction of the $\ttphi$ signal in the $\Pe\Pe$ (upper), $\PGm\PGm$ (middle) and $\PGt\PGt$ (lower) decay scenarios. 
The results for the scalar coupling are shown on the left and pseudoscalar on the right. 
The vertical gray band indicates the mass region not considered in the analysis.
The red line is the theoretical prediction for the product of the production cross section and branching fraction of the $\ttphi$ signal.
}
\end{figure*}

In total, 37 dilepton mass spectra are probed, corresponding to the 12 $\Xphi\to{\Pe\Pe}$, 12 $\Xphi\to{\PGm\PGm}$, and 13 $\Xphi\to{\PGt\PGt}$ SRs defined in Section~\ref{subsec:SignalRegions}; these are illustrated in Figs.~\ref{fig:WZPhiee}--\ref{fig:ttPhiee}, \ref{fig:WZPhimm}--\ref{fig:ttPhimm}, and \ref{fig:WZPhitautau}--\ref{fig:ttPhitautau2}, respectively.
To test each given $\Xphi$ production, decay, and mass scenario, a subset of these SRs is used, resulting in $(3+3+2)\,3=24$ model-independent bounds on $\Xphi$ signal hypotheses, as shown in Figs.~\ref{fig:WPhiSPSLimitPlots}--\ref{fig:ttPhiSPSLimitPlots}.
For example, considering a $\Wphi\to\Pe\Pe$ signal scenario with S, PS, or H-like couplings at a $\phi$ mass of 50\GeV, the $\mee$ and $\meemin$ distributions in the mass range of 12--76\GeV are used in the 3L($\Pe\Pe\PGm$) and 3L($\Pe\Pe\Pe$) channels, respectively.
Similarly, for a $\ttphi\to\PGm\PGm$ signal scenario with S or PS couplings at a $\phi$ mass of 200\GeV, the $\mmm$, $\mmmmax$, and $\mmmmax$ distributions in the mass range of 106--366\GeV are used in the 3L($\Pe\PGm\PGm$), 3L($\Pe\Pe\Pe$), and 4L+3L1T+2L2T channels, respectively.
In the case of a $\Zphi\to\PGt\PGt$ signal with S, PS, or H-like couplings of any $\phi$ mass, $\memmin$ and $\mtautaumin$ distributions are used in 4L+2L2T and 2L2T (OSSF1) channels, whereas $\mltmin$ distributions are used in both 3L1T and 2L2T (OSSF0) channels.
In all cases involving contributions from more than one channel, all channels are considered in combination, including correlations between systematic uncertainties, as described in Section~\ref{sec:Systematics}.

No statistically significant deviation from the SM expectations is observed in any of the probed mass distributions. 
The largest local deviation is observed in the high mass $\Zphi\to\Pe\Pe$ search, corresponding to the $\Zphi(\Pe\Pe)$ SRHigh mass spectrum in Fig.~\ref{fig:WZPhiee}, where an excess at a $\phi$ mass of 156\GeV corresponding to 2.9 standard deviations is observed, without considering the look-elsewhere effect (LEE)~\cite{Gross:2010qma}. 
The corresponding global significance, obtained taking into account LEE in an $\mee$ range of 106--366\GeV, is 1.4 standard deviations.

Upper limits at 95\% confidence level (\CL) are set on the product of the production cross sections and branching fractions, $\sigma(\Xphi)\,\mathcal{B}(\phi\to\Pell\Pell)$, using a modified frequentist approach based on the \CLs criterion~\cite{Junk:1999kv,Read:2002hq} in the asymptotic approximation~\cite{Cowan:2010js,ATLAS:2011tau}.
For each signal hypothesis, a binned maximum-likelihood fit is performed to discriminate between the potential signal and the SM background processes.
The systematic uncertainties and their correlations, described in Section~\ref{sec:Systematics}, are incorporated in the likelihood as nuisance parameters with log-normal probability density functions.  
The statistical uncertainties in the signal and background estimates are modeled with gamma functions.
The expected and observed upper limits on the probed signals are provided in Figs.~\ref{fig:WPhiSPSLimitPlots}-\ref{fig:ttPhiSPSLimitPlots}.
The $\Zphi\to\Pe\Pe/\PGm\PGm$ sensitivity is driven by the 4L channel, while the $\Wphi\to\Pe\Pe/\PGm\PGm$ and $\ttphi\to\Pe\Pe/\PGm\PGm$ sensitivities are driven by the 3L channels with 0 $\PQb$ jets and 1 or more $\PQb$ jets, respectively. 
For the $\Xphi\to\PGt\PGt$ signals, the sensitivity at high $\phi$ mass is driven by the channels with at least two $\tauh$s. 

For all $\Xphi$ signals, the expected upper limits are the most stringent for signals with the dimuon decay modes, with the dielectron modes less stringent by as much as a factor of two for low $\phi$ mass hypotheses because of the lower electron reconstruction and selection efficiencies. 
For high $\phi$ masses, the expected constraints are comparable for all signals with any coupling scenario with dielectron or dimuon coupling scenario.
Similarly, constraints on the $\PGt\PGt$ decay modes are less stringent than those on the light lepton decay modes throughout, limited by the $\PGt$ lepton energy resolution, the reconstruction and identification efficiencies, and the higher SM background contributions in these final states, especially for $\phi$ masses below 40\GeV.

For the $\Wphi$ signal, there are three coupling scenarios, among which the pseudoscalar coupling for a 15\GeV $\phi\to\PGm\PGm$ results in the most stringent limit of about $10$\unit{fb}.
For scalar and H-like couplings at the same mass and decay mode, the $\sigma\,\mathcal{B}$ values above 20 and 100\unit{fb} are excluded, respectively. 
For $\phi\to\Pe\Pe$, the limits are about two times less stringent, and for $\phi\to\PGt\PGt$, the upper limits are in the range 150--250\unit{fb} at a $\phi$ mass of 40\GeV. 
For a $\phi$ mass of 350\GeV, the upper limits are in the range 0.8--2.0\unit{fb} across all couplings with dielectron and dimuon decay modes, whereas for $\PGt\PGt$ decays, the limits are around 8\unit{fb}.

The observed constraints on the $\Zphi$ signal are similar for the dielectron and dimuon decay modes, where values of the $\sigma\,\mathcal{B}$ are excluded above 20--30, 20--30, and 50--60\unit{fb} at a $\phi$ mass of 15\GeV for  scalar, pseudoscalar, and H-like couplings, respectively; 
for the $\Zphi\to\PGt\PGt$ signal at a mass hypothesis of 40\GeV, the upper limits for the same couplings are about 300, 200, and 800\unit{fb}.
For a $\phi$ mass of 350\GeV, these constraints are about 1 (10)\unit{fb} for all coupling scenarios of dielectron and dimuon (of $\PGt\PGt$) decays.

For the $\ttphi\to\PGm\PGm/\Pe\Pe$ signal, values of the $\sigma\,\mathcal{B}$ above 4--7 and 1.5--2.5\unit{fb} are excluded for the $\phi$ mass of 15\GeV for the scalar and pseudoscalar coupling scenarios, respectively. 
For a 350\GeV $\phi$ boson, these constraints for both decay modes and coupling scenarios are about 0.6\unit{fb}. 
For the $\ttphi\to\PGt\PGt$ signal and the scalar (pseudoscalar) coupling scenario, the upper limit on the $\sigma\,\mathcal{B}$ varies from 200 (80)\unit{fb} for a $\phi$ mass of 40\GeV to 5\unit{fb} for a mass of 350\GeV. 

For all $\Xphi$ signal scenarios, the differences in the low-mass exclusion limits of scalar, pseudoscalar, and H-like signals result from different Lorentz structures of the interactions, which affect the signal acceptance.  

The exclusions are also reinterpreted as upper limits on the coupling parameters, and are provided in digital format in the \textsc{HEPData} record~\cite{hepdata}.
For the $\Wphi$ and $\Zphi$ signal scenarios with scalar and pseudoscalar couplings, limits on the product of the inverse square mass scale and branching fraction to leptons, $(1/\LambdaS )^2\,\mathcal{B}(\phi \to \ell \ell)$ and $(1/\LambdaPS )^2\,\mathcal{B}(\phi \to \ell \ell)$, are derived.
For the associated production of $\phi$ bosons with top quark pairs, the limits on the product of the coupling to top quarks and the branching fraction to leptons, $\gtS^2\,\mathcal{B}(\phi \to \ell \ell)$ and $\gtPS^2\,\mathcal{B}(\phi \to \ell \ell)$, are derived for a scalar and pseudoscalar $\phi$, respectively.

\begin{table*}[hbt!]
\centering
\topcaption{
A summary of model-dependent scenarios, and the corresponding subsets of SRs combined in the interpretations.
}\label{tab:CombinationsTable}
\cmsTable{
\renewcommand{\arraystretch}{1.1}
\begin{scotch}{l  c  l}
Signal                                                 & Parameter                                    & \multicolumn{1}{c}{SRs used in combination} \\ \hline 
Fermiophilic dilaton-like $\phi$ production and decay  & \multirow{2}{*}{$\gtS^2$}                    &  For $\phi$ masses less than 30\GeV:  \\
$\ttphi(\;\to\PGm\PGm,\;\to\PGt\PGt)$ combination      &                                              &  $\ttphi(\PGm\PGm)$ SR1-2 and $\ttphi(\PGt\PGt)$ SR1-3, 5-7 \\
                                                       &                                              &  For $\phi$ masses more than or equal to 30\GeV:  \\ 
                                                       &                                              &  $\ttphi(\PGt\PGt)$ all SRs      \\[\cmsTabSkip]
Fermiophilic axion-like $\phi$ production and decay    & \multirow{2}{*}{$\gtPS^2$}                   &  For $\phi$ masses less than 30\GeV:  \\
$\ttphi(\;\to\PGm\PGm,\;\to\PGt\PGt)$ combination      &                                              &  $\ttphi(\PGm\PGm)$ SR1-2 and $\ttphi(\PGt\PGt)$ SR1-3, 5-7 \\
                                                       &                                              &  For $\phi$ masses more than or equal to 30\GeV:  \\
                                                       &                                              &  $\ttphi(\PGt\PGt)$ all SRs      \\[\cmsTabSkip]
H-like production $\Xphi(\;\to\Pe\Pe)$ combination     & $\sin^2\theta\,\mathcal{B}(\phi\to\Pe\Pe)$   & $\Wphi(\Pe\Pe)$/$\Zphi(\Pe\Pe)$ all SRs, and $\ttphi(\Pe\Pe)$ SR1-2  \\[\cmsTabSkip]
H-like production $\Xphi(\;\to\PGm\PGm)$ combination   & $\sin^2\theta\,\mathcal{B}(\phi\to\PGm\PGm)$ &  $\Wphi(\PGm\PGm)$/$\Zphi(\PGm\PGm)$ all SRs, and $\ttphi(\PGm\PGm)$ SR1-2 \\[\cmsTabSkip]
H-like $\phi$ production and decay                     & \multirow{2}{*}{$\sin^2\theta$}              &  For $\phi$ masses less than 30\GeV:  \\
$\Xphi(\;\to\PGm\PGm,\;\to\PGt\PGt)$ combination       &                                              &  $\Xphi(\;\to\PGm\PGm)$ combination and $\ttphi(\PGt\PGt)$ SR2-3 \\
                                                       &                                              &  For $\phi$ masses more than or equal to 30\GeV:  \\    
                                                       &                                              &  $\ttphi(\PGt\PGt$)  all SRs     \\
\end{scotch}
}
\end{table*}

\subsection{Model-dependent results}\label{subsec:ModelDependentResults}

Several model-dependent exclusions are also presented. 
These are obtained from a weighted combination of nonoverlapping SRs, as summarized in Table~\ref{tab:CombinationsTable}, that target multiple decay modes, production modes, or both, relevant for the signal model under consideration.

\begin{figure*}[hbt!]
\centering
\includegraphics[width=0.41\textwidth]{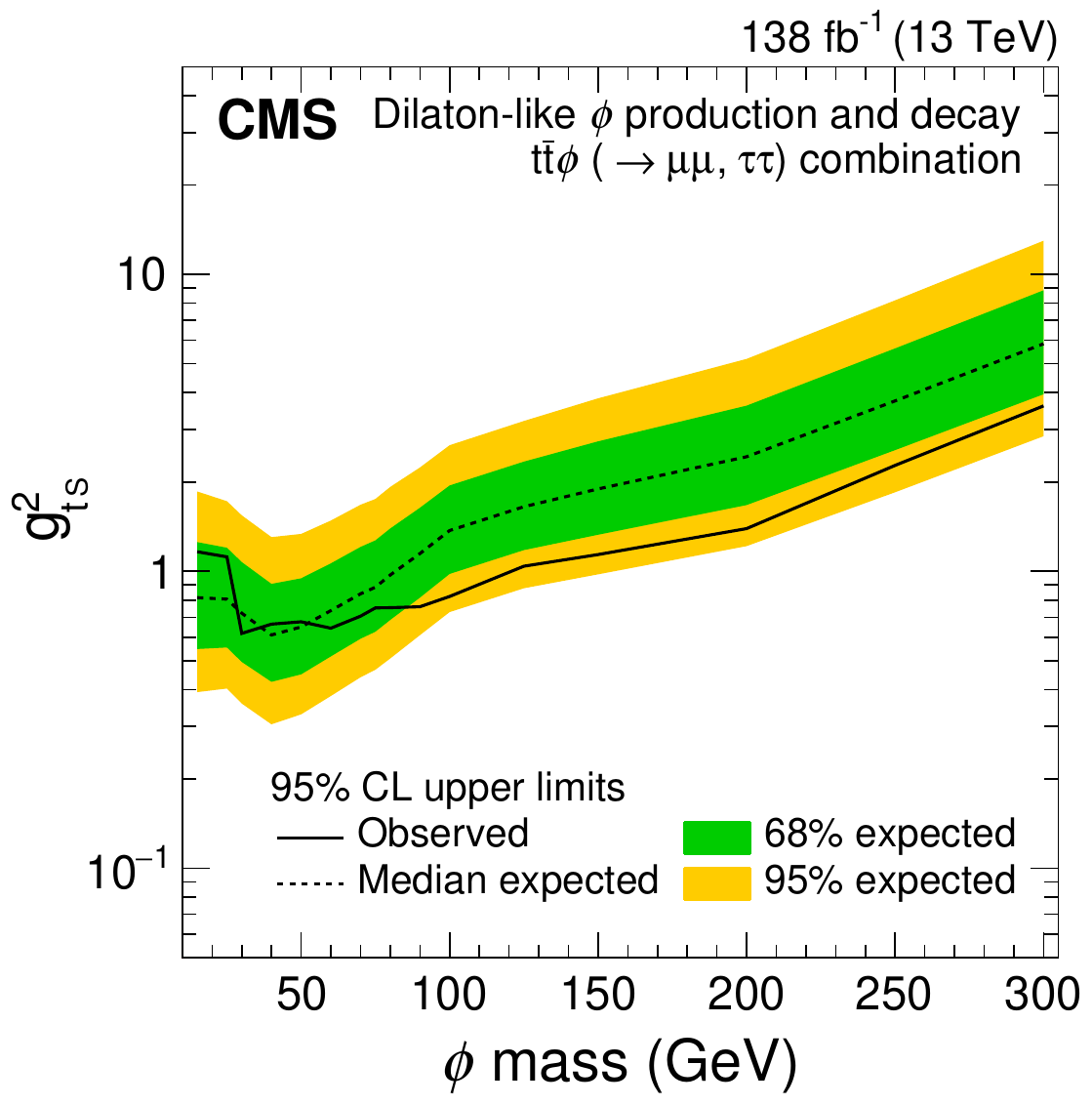}
\includegraphics[width=0.41\textwidth]{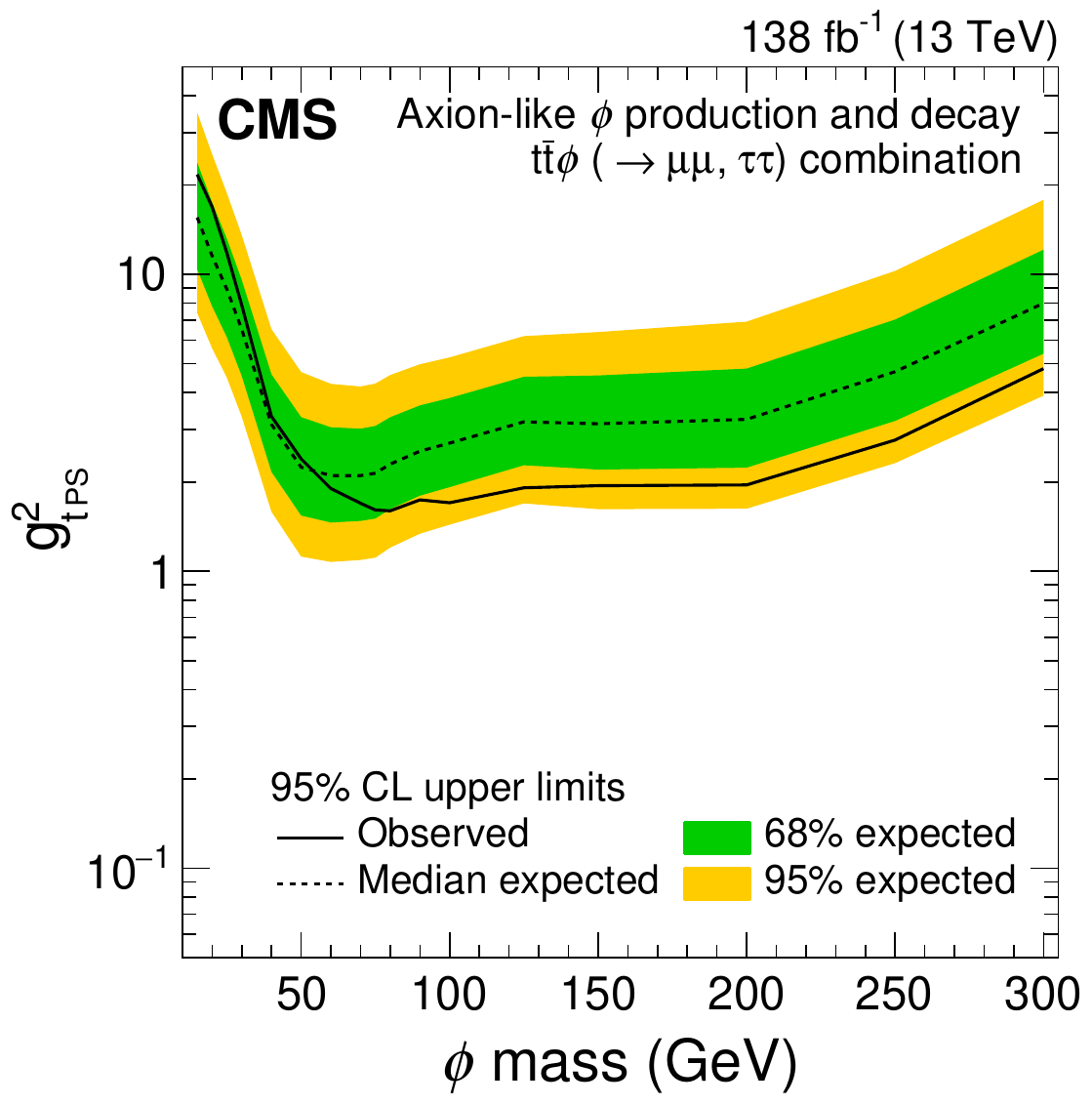}
\caption{\label{fig:AxionDilaton} 
The 95\% confidence level upper limits on $\gtS^2$ and $\gtPS^2$ for the dilaton- and axion-like $\ttphi$ signal model (left and right).
Masses of the $\phi$ boson above 300\GeV are not probed for the dilaton- and axion-like signal models as the $\phi$ branching fraction into top quark-antiquark pairs becomes nonnegligible.
}
\end{figure*}

Firstly, the $\ttphi$ mode is interpreted in the context of fermiophilic dilaton-like scalar boson and fermiophilic axion-like pseudoscalar boson signal models. 
These $\phi$ couplings are proportional to the fermion mass in both production and decay, so the $\PGm\PGm$ and $\PGt\PGt$ channels are combined to probe the coupling to top quarks for each of these scenarios. 
The combined 95\% \CL exclusions are shown in Fig.~\ref{fig:AxionDilaton}.

\begin{figure*}[hbt!]
\centering
\includegraphics[width=0.41\textwidth]{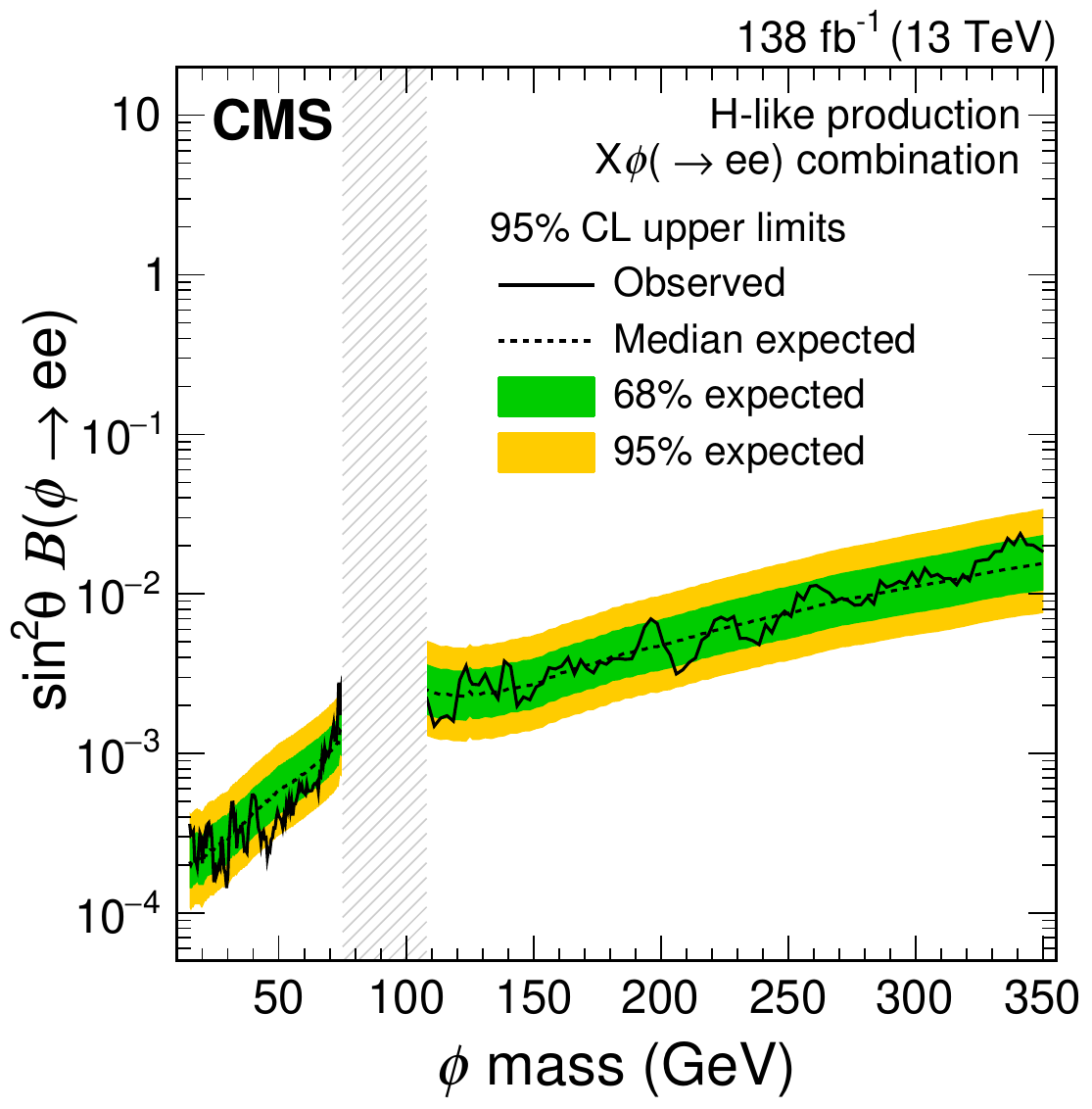}
\includegraphics[width=0.41\textwidth]{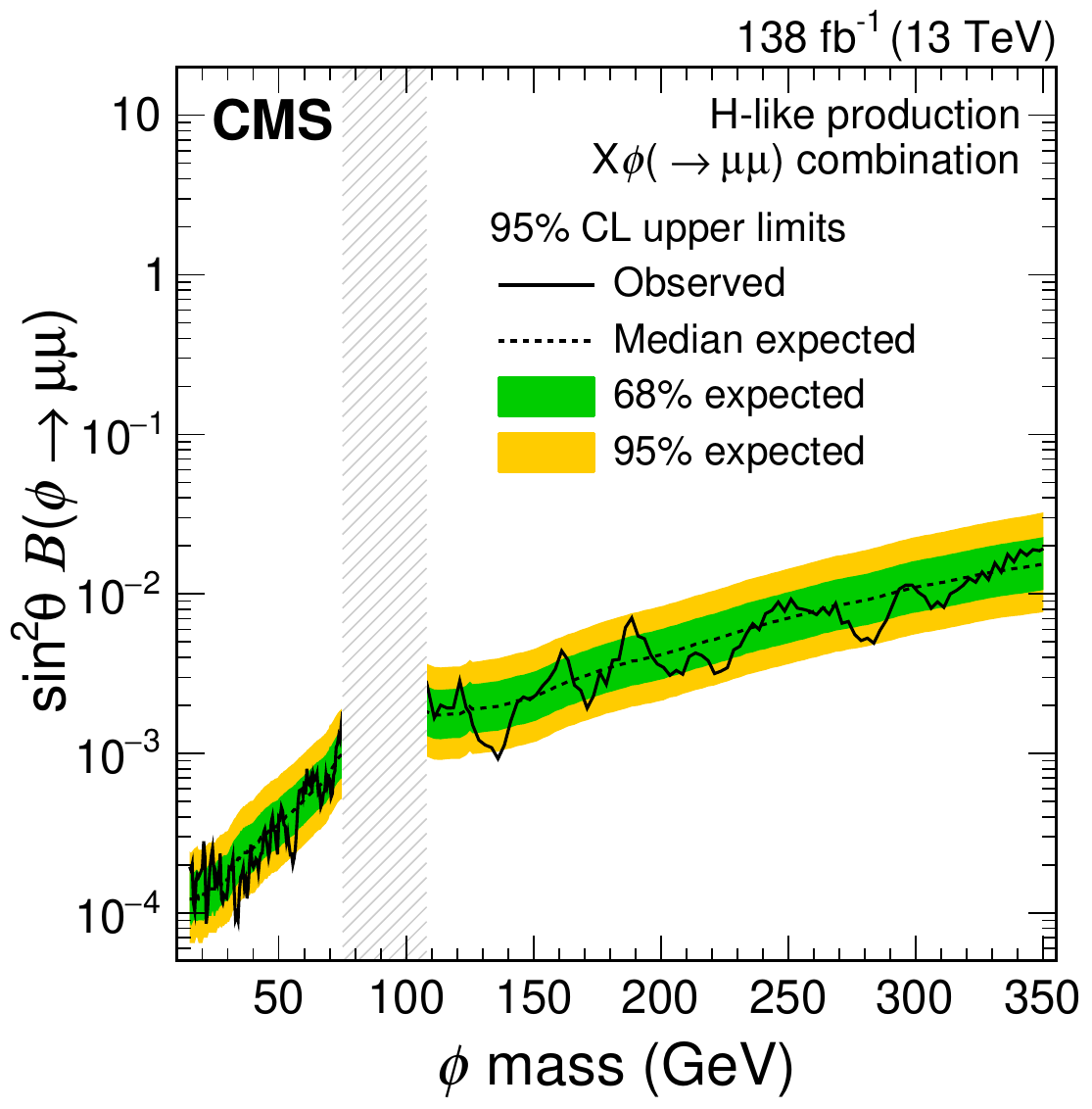}
\caption{\label{fig:HiggslikeProduction} 
The 95\% confidence level upper limits on the product of $\sin^2\theta$ and branching fraction for the H-like production of $\Xphi\to\Pe\Pe$ and $\Xphi\to\PGm\PGm$ (left and right).
The vertical gray band indicates the mass region not considered in the analysis.
}
\end{figure*}

For the H-like production scenario, the constraints from the $\Wphi$, $\Zphi$, and $\ttphi$ production modes are combined and labelled as the $\Xphi$ combination.
Upper limits at 95\% \CL are derived on the product of the mixing angle, $\sin^2\theta$, and the branching fractions to lepton pairs.
For $\phi$ bosons decaying to electrons or muons, the most stringent expected limit is obtained by combining all $\Wphi$ and $\Zphi$ signal regions with $\ttphi$ signal regions SR1 and SR2, which cover 3L events with one or more $\PQb$ jets and are independent of the $\Wphi$ and $\Zphi$ signal regions. 
The combined $\Xphi$ signal model limits on the product of the mixing angle and the branching fractions are shown in Fig.~\ref{fig:HiggslikeProduction}.
For a $\phi$ mass of 125\GeV, the $\Xphi$ combination excludes $\sin^2\theta\,\mathcal{B}(\phi \to \Pe \Pe)$ above $2.7 \times 10^{-3}$ and $\sin^2\theta\,\mathcal{B}(\phi \to \PGm \PGm)$ above $1.5 \times 10^{-3}$.
No combination is performed under the H-like production scenario for $\phi$ bosons decaying to tau leptons, as the most stringent expected limits over most of the mass range result from the $\ttphi$ signal regions alone, and therefore are proportional to the upper limits obtained for the scalar $\ttphi\to\PGt\PGt$ signal scenario. These are also provided in digital format in the \textsc{HEPData} record~\cite{hepdata}.

\begin{figure}[hbt!]
\centering
\includegraphics[width=0.41\textwidth]{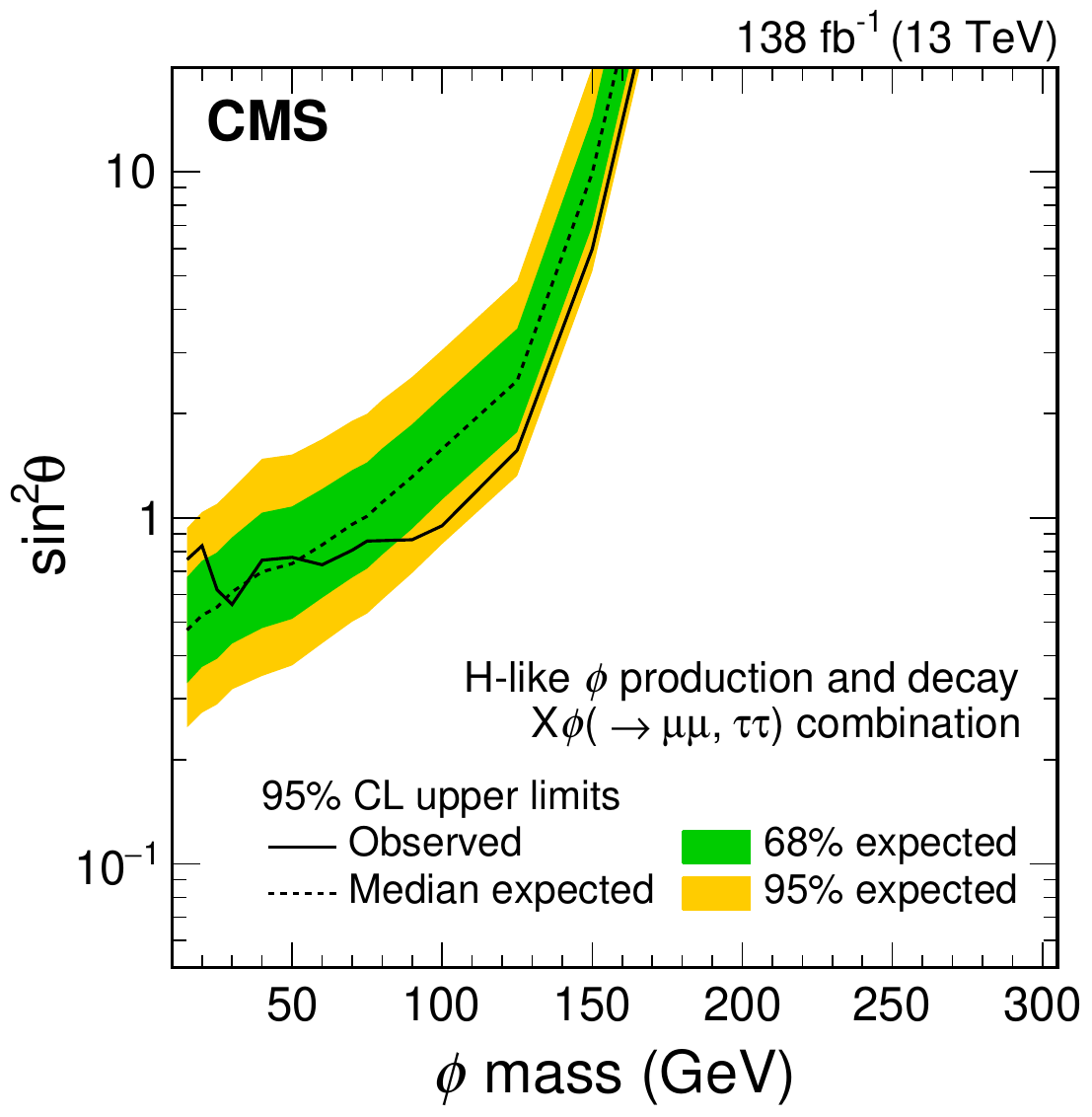}
\caption{\label{fig:HiggsLikeProductionDecay} 
The 95\% confidence level upper limits on $\sin^2\theta$ for the H-like production and decay of $\Xphi$ signal model, where the $\phi$ boson has the same branching fractions to $\PGm\PGm$ and $\PGt\PGt$ as an SM Higgs boson with the same mass.
}
\end{figure}

Assuming further that the $\phi$ branching fractions, particularly $\mathcal{B}(\phi \to \PGm \PGm)$ and $\mathcal{B}(\phi \to \PGt \PGt)$, are equal to those of the SM Higgs boson as a function of mass, the independent $\PGm\PGm$ and $\PGt\PGt$ channels are combined to derive a combined exclusion on $\sin^2\theta$ for an H-like $\phi$ model, as illustrated in Fig.~\ref{fig:HiggsLikeProductionDecay}.
The $\mathcal{B}(\phi \to \PGm \PGm)$ and $\mathcal{B}(\phi \to \PGt \PGt)$ values as functions of mass are obtained using the HDECAY program v.6.61~\cite{Djouadi:1997yw,Djouadi:2018xqq}.
This combination is dominated by the $\PGm\PGm$ mode for $\phi$ masses below 30\GeV, and by the $\PGt\PGt$ mode for higher $\phi$ masses.
Values of $\sin^2\theta \leq 1$ are excluded in the $\phi$ mass range 15--102\GeV.
Although values of $\sin^2\theta > 1$ are unphysical in this scenario, limits on $\sin^2\theta$ treated as a free unconstrained parameter can be considered in more general interpretations, as presented in Fig.~\ref{fig:HiggsLikeProductionDecay}.

In all model-dependent signal scenarios where $\phi$ is allowed to decay into pairs of leptons of all three flavors, the $\Xphi\to\PGt\PGt$ signal contributions in which both tau leptons decay leptonically into nonresonant $\Pe\Pe$ and $\PGm\PGm$ pairs are not considered in the $\mee$ and $\mmm$ spectra.
Such contributions are found to be negligible across the three model dependent interpretations carried out in this analysis.

\clearpage

\section{Summary}\label{sec:Summary}

A search for beyond-the-standard-model phenomena producing resonant dilepton signatures of any flavor in multilepton events has been performed using $\Pp\Pp$ collision data collected with the CMS detector at $\sqrt{s} = 13\TeV$, corresponding to an integrated luminosity of 138\fbinv.
The results provide direct and model independent constraints on the allowed parameter space for new spin-0 particles, $\phi$, with scalar, pseudoscalar, or H-like couplings. 
The $\phi$ bosons are assumed to be produced in association with a \PW or \PZ boson or a top quark-antiquark (\ttbar) pair, and decay into $\Pe\Pe$, $\PGm\PGm$, or $\PGt\PGt$ pairs.
Constraints are calculated at $95\%$ confidence level on the product of the production cross section and leptonic branching fraction of such bosons with masses in the range 15--350\GeV. No statistically significant excess is observed over the standard model background in the probed mass spectra. 
Over this mass range, the product of the cross section and branching fraction for the $\PGt\PGt$ ($\Pe\Pe$ and $\PGm\PGm$) final states is excluded above 0.004--35, 0.004--80, and 0.008--250\unit{pb} (0.5--50, 0.5--30, and 1--200\unit{fb}) as a function of $\phi$ mass for scalar, pseudoscalar, and H-like bosons, respectively.

Several model-dependent interpretations have also been considered. 
The $\ttphi$ mode provides the first direct bounds on the coupling of the $\phi$ boson to top quarks in the context of fermiophilic models. 
For a fermiophilic dilaton-like model with scalar couplings, the most stringent limit on the coupling is 0.63--0.66, obtained in the $\phi$ mass range 40--60\GeV. For a fermiophilic axion-like model with pseudoscalar couplings, the most stringent limit on the coupling is 1.59, obtained for a $\phi$ mass of 70\GeV.
To constrain the Higgs-$\phi$ mixing angle, $\sin^2\theta$, in the case where the $\phi$ is H-like, the independent $\Wphi$, $\Zphi$, and $\ttphi$ signal regions are combined.
The observed (expected) upper limit on $\sin^2\theta$ is 1.2 (1.9) for a $\phi$ mass of 125\GeV; the most stringent observed exclusion is obtained for a $\phi$ mass of 30\GeV, corresponding to an upper limit on $\sin^2\theta$ of 0.59 (0.64).

\begin{acknowledgments}
We congratulate our colleagues in the CERN accelerator departments for the excellent performance of the LHC and thank the technical and administrative staffs at CERN and at other CMS institutes for their contributions to the success of the CMS effort. In addition, we gratefully acknowledge the computing centers and personnel of the Worldwide LHC Computing Grid and other centers for delivering so effectively the computing infrastructure essential to our analyses. Finally, we acknowledge the enduring support for the construction and operation of the LHC, the CMS detector, and the supporting computing infrastructure provided by the following funding agencies: SC (Armenia), BMBWF and FWF (Austria); FNRS and FWO (Belgium); CNPq, CAPES, FAPERJ, FAPERGS, and FAPESP (Brazil); MES and BNSF (Bulgaria); CERN; CAS, MoST, and NSFC (China); MINCIENCIAS (Colombia); MSES and CSF (Croatia); RIF (Cyprus); SENESCYT (Ecuador); ERC PRG, RVTT3 and MoER TK202 (Estonia); Academy of Finland, MEC, and HIP (Finland); CEA and CNRS/IN2P3 (France); SRNSF (Georgia); BMBF, DFG, and HGF (Germany); GSRI (Greece); NKFIH (Hungary); DAE and DST (India); IPM (Iran); SFI (Ireland); INFN (Italy); MSIP and NRF (Republic of Korea); MES (Latvia); LMTLT (Lithuania); MOE and UM (Malaysia); BUAP, CINVESTAV, CONACYT, LNS, SEP, and UASLP-FAI (Mexico); MOS (Montenegro); MBIE (New Zealand); PAEC (Pakistan); MES and NSC (Poland); FCT (Portugal); MESTD (Serbia); MCIN/AEI and PCTI (Spain); MOSTR (Sri Lanka); Swiss Funding Agencies (Switzerland); MST (Taipei); MHESI and NSTDA (Thailand); TUBITAK and TENMAK (Turkey); NASU (Ukraine); STFC (United Kingdom); DOE and NSF (USA).
  
\hyphenation{Rachada-pisek} Individuals have received support from the Marie-Curie program and the European Research Council and Horizon 2020 Grant, contract Nos.\ 675440, 724704, 752730, 758316, 765710, 824093, 101115353, and COST Action CA16108 (European Union); the Leventis Foundation; the Alfred P.\ Sloan Foundation; the Alexander von Humboldt Foundation; the Science Committee, project no. 22rl-037 (Armenia); the Belgian Federal Science Policy Office; the Fonds pour la Formation \`a la Recherche dans l'Industrie et dans l'Agriculture (FRIA-Belgium); the Agentschap voor Innovatie door Wetenschap en Technologie (IWT-Belgium); the F.R.S.-FNRS and FWO (Belgium) under the ``Excellence of Science -- EOS" -- be.h project n.\ 30820817; the Beijing Municipal Science \& Technology Commission, No. Z191100007219010 and Fundamental Research Funds for the Central Universities (China); the Ministry of Education, Youth and Sports (MEYS) of the Czech Republic; the Shota Rustaveli National Science Foundation, grant FR-22-985 (Georgia); the Deutsche Forschungsgemeinschaft (DFG), under Germany's Excellence Strategy -- EXC 2121 ``Quantum Universe" -- 390833306, and under project number 400140256 - GRK2497; the Hellenic Foundation for Research and Innovation (HFRI), Project Number 2288 (Greece); the Hungarian Academy of Sciences, the New National Excellence Program - \'UNKP, the NKFIH research grants K 124845, K 124850, K 128713, K 128786, K 129058, K 131991, K 133046, K 138136, K 143460, K 143477, 2020-2.2.1-ED-2021-00181, and TKP2021-NKTA-64 (Hungary); the Council of Science and Industrial Research, India; ICSC -- National Research Center for High Performance Computing, Big Data and Quantum Computing, funded by the NextGenerationEU program (Italy); the Latvian Council of Science; the Ministry of Education and Science, project no. 2022/WK/14, and the National Science Center, contracts Opus 2021/41/B/ST2/01369 and 2021/43/B/ST2/01552 (Poland); the Funda\c{c}\~ao para a Ci\^encia e a Tecnologia, grant CEECIND/01334/2018 (Portugal); the National Priorities Research Program by Qatar National Research Fund; MCIN/AEI/10.13039/501100011033, ERDF ``a way of making Europe", and the Programa Estatal de Fomento de la Investigaci{\'o}n Cient{\'i}fica y T{\'e}cnica de Excelencia Mar\'{\i}a de Maeztu, grant MDM-2017-0765 and Programa Severo Ochoa del Principado de Asturias (Spain); the Chulalongkorn Academic into Its 2nd Century Project Advancement Project, and the National Science, Research and Innovation Fund via the Program Management Unit for Human Resources \& Institutional Development, Research and Innovation, grant B37G660013 (Thailand); the Kavli Foundation; the Nvidia Corporation; the SuperMicro Corporation; the Welch Foundation, contract C-1845; and the Weston Havens Foundation (USA).  
\end{acknowledgments}

\bibliography{auto_generated}

\providecommand{\href}[2]{#2}\begingroup\raggedright\begin{thebibliography}{10}%
\makeatletter
\providecommand{\hrefCMSnoop }[0]{\@secondoftwo}%
\makeatother
\providecommand{\doi}{\texttt{doi:}\begingroup \urlstyle{tt}\Url}

\bibitem{Cacciapaglia:2019bqz}
\hrefCMSnoop {}{G.~Cacciapaglia, G.~Ferretti, T.~Flacke, and H.~Serodio,
  ``Light scalars in composite {Higgs} models'',} \textit{ Front. Phys.}
  \textbf{ 7} (2019) 22,
  \href{http://dx.doi.org/10.3389/fphy.2019.00022}{\doi{10.3389/fphy.2019.00022}},
  \href{http://www.arXiv.org/abs/1902.06890}{\texttt{arXiv:1902.06890}}.

\bibitem{Ellwanger:2009dp}
\hrefCMSnoop {}{U.~Ellwanger, C.~Hugonie, and A.~M. Teixeira, ``The
  next-to-minimal supersymmetric standard model'',} \textit{ Phys. Rept.}
  \textbf{ 496} (2010) 1,
  \href{http://dx.doi.org/10.1016/j.physrep.2010.07.001}{\doi{10.1016/j.physrep.2010.07.001}},
  \href{http://www.arXiv.org/abs/0910.1785}{\texttt{arXiv:0910.1785}}.

\bibitem{Maniatis:2009re}
\hrefCMSnoop {}{M.~Maniatis, ``The next-to-minimal supersymmetric extension of
  the standard model reviewed'',} \textit{ Int. J. Mod. Phys. A} \textbf{ 25}
  (2010) 3505,
  \href{http://dx.doi.org/10.1142/S0217751X10049827}{\doi{10.1142/S0217751X10049827}},
  \href{http://www.arXiv.org/abs/0906.0777}{\texttt{arXiv:0906.0777}}.

\bibitem{Buckley:2014fba}
\hrefCMSnoop {}{M.~R. Buckley, D.~Feld, and D.~Goncalves, ``Scalar simplified
  models for dark matter'',} \textit{ Phys. Rev. D} \textbf{ 91} (2015) 015017,
  \href{http://dx.doi.org/10.1103/PhysRevD.91.015017}{\doi{10.1103/PhysRevD.91.015017}},
  \href{http://www.arXiv.org/abs/1410.6497}{\texttt{arXiv:1410.6497}}.

\bibitem{Casolino:2015cza}
M.~Casolino\hrefCMSnoop {}{ { et~al.}, ``Probing a light {CP-odd} scalar in
  di-top-associated production at the {LHC}'',} \textit{ Eur. Phys. J. C}
  \textbf{ 75} (2015) 498,
  \href{http://dx.doi.org/10.1140/epjc/s10052-015-3708-y}{\doi{10.1140/epjc/s10052-015-3708-y}},
  \href{http://www.arXiv.org/abs/1507.07004}{\texttt{arXiv:1507.07004}}.

\bibitem{Chang:2017ynj}
\hrefCMSnoop {}{W.-F. Chang, T.~Modak, and J.~N. Ng, ``Signal for a light
  singlet scalar at the {LHC}'',} \textit{ Phys. Rev. D} \textbf{ 97} (2018)
  055020,
  \href{http://dx.doi.org/10.1103/PhysRevD.97.055020}{\doi{10.1103/PhysRevD.97.055020}},
  \href{http://www.arXiv.org/abs/1711.05722}{\texttt{arXiv:1711.05722}}.

\bibitem{Artoisenet:2013puc}
\hrefCMSnoop {}{P.~Artoisenet { et~al.}, ``A framework for {Higgs}
  characterisation'',} \textit{ JHEP} \textbf{ 11} (2013) 043,
  \href{http://dx.doi.org/10.1007/JHEP11(2013)043}{\doi{10.1007/JHEP11(2013)043}},
  \href{http://www.arXiv.org/abs/1306.6464}{\texttt{arXiv:1306.6464}}.

\bibitem{Ghosh:2020ipy}
\hrefCMSnoop {}{T.~Ghosh, H.-K. Guo, T.~Han, and H.~Liu, ``Electroweak phase
  transition with an {SU(2)} dark sector'',} \textit{ JHEP} \textbf{ 07} (2021)
  045,
  \href{http://dx.doi.org/10.1007/JHEP07(2021)045}{\doi{10.1007/JHEP07(2021)045}},
  \href{http://www.arXiv.org/abs/2012.09758}{\texttt{arXiv:2012.09758}}.

\bibitem{Gildener:1976ih}
\hrefCMSnoop {}{E.~Gildener and S.~Weinberg, ``Symmetry breaking and scalar
  bosons'',} \textit{ Phys. Rev. D} \textbf{ 13} (1976) 3333,
  \href{http://dx.doi.org/10.1103/PhysRevD.13.3333}{\doi{10.1103/PhysRevD.13.3333}}.

\bibitem{Goldberger:2007zk}
\hrefCMSnoop {}{W.~D. Goldberger, B.~Grinstein, and W.~Skiba, ``Distinguishing
  the {Higgs} boson from the dilaton at the {Large Hadron Collider}'',}
  \textit{ Phys. Rev. Lett.} \textbf{ 100} (2008) 111802,
  \href{http://dx.doi.org/10.1103/PhysRevLett.100.111802}{\doi{10.1103/PhysRevLett.100.111802}},
  \href{http://www.arXiv.org/abs/0708.1463}{\texttt{arXiv:0708.1463}}.

\bibitem{Ahmed:2019csf}
\hrefCMSnoop {}{A.~Ahmed, A.~Mariotti, and S.~Najjari, ``A light dilaton at the
  {LHC}'',} \textit{ JHEP} \textbf{ 05} (2020) 093,
  \href{http://dx.doi.org/10.1007/JHEP05(2020)093}{\doi{10.1007/JHEP05(2020)093}},
  \href{http://www.arXiv.org/abs/1912.06645}{\texttt{arXiv:1912.06645}}.

\bibitem{Barger:2011nu}
\hrefCMSnoop {}{V.~Barger, M.~Ishida, and W.-Y. Keung, ``Dilaton at the
  {LHC}'',} \textit{ Phys. Rev. D} \textbf{ 85} (2012) 015024,
  \href{http://dx.doi.org/10.1103/PhysRevD.85.015024}{\doi{10.1103/PhysRevD.85.015024}},
  \href{http://www.arXiv.org/abs/1111.2580}{\texttt{arXiv:1111.2580}}.

\bibitem{Georgi:1986df}
\hrefCMSnoop {}{H.~Georgi, D.~B. Kaplan, and L.~Randall, ``Manifesting the
  invisible axion at low-energies'',} \textit{ Phys. Lett. B} \textbf{ 169}
  (1986) 73,
  \href{http://dx.doi.org/10.1016/0370-2693(86)90688-X}{\doi{10.1016/0370-2693(86)90688-X}}.

\bibitem{Mimasu:2014nea}
\hrefCMSnoop {}{K.~Mimasu and V.~Sanz, ``{ALP}s at colliders'',} \textit{ JHEP}
  \textbf{ 06} (2015) 173,
  \href{http://dx.doi.org/10.1007/JHEP06(2015)173}{\doi{10.1007/JHEP06(2015)173}},
  \href{http://www.arXiv.org/abs/1409.4792}{\texttt{arXiv:1409.4792}}.

\bibitem{Brivio:2017ije}
I.~Brivio\hrefCMSnoop {}{ { et~al.}, ``{ALP}s effective field theory and
  collider signatures'',} \textit{ Eur. Phys. J. C} \textbf{ 77} (2017) 572,
  \href{http://dx.doi.org/10.1140/epjc/s10052-017-5111-3}{\doi{10.1140/epjc/s10052-017-5111-3}},
  \href{http://www.arXiv.org/abs/1701.05379}{\texttt{arXiv:1701.05379}}.

\bibitem{Bauer:2018uxu}
\hrefCMSnoop {}{M.~Bauer, M.~Heiles, M.~Neubert, and A.~Thamm, ``Axion-like
  particles at future colliders'',} \textit{ Eur. Phys. J. C} \textbf{ 79}
  (2019) 74,
  \href{http://dx.doi.org/10.1140/epjc/s10052-019-6587-9}{\doi{10.1140/epjc/s10052-019-6587-9}},
  \href{http://www.arXiv.org/abs/1808.10323}{\texttt{arXiv:1808.10323}}.

\bibitem{Schabinger:2005ei}
\hrefCMSnoop {}{R.~M. Schabinger and J.~D. Wells, ``A minimal spontaneously
  broken hidden sector and its impact on {Higgs} boson physics at the large
  hadron collider'',} \textit{ Phys. Rev. D} \textbf{ 72} (2005) 093007,
  \href{http://dx.doi.org/10.1103/PhysRevD.72.093007}{\doi{10.1103/PhysRevD.72.093007}},
  \href{http://www.arXiv.org/abs/hep-ph/0509209}{\texttt{arXiv:hep-ph/0509209}}.

\bibitem{Barger:2007im}
V.~Barger\hrefCMSnoop {}{ { et~al.}, ``{LHC} phenomenology of an extended
  standard model with a real scalar singlet'',} \textit{ Phys. Rev. D} \textbf{
  77} (2008) 035005,
  \href{http://dx.doi.org/10.1103/PhysRevD.77.035005}{\doi{10.1103/PhysRevD.77.035005}},
  \href{http://www.arXiv.org/abs/0706.4311}{\texttt{arXiv:0706.4311}}.

\bibitem{hepdata}
\hrefCMSnoop {}{}{HEPD}ata record for this analysis, 2024.
\newblock
  \href{http://dx.doi.org/10.17182/hepdata.132367}{\doi{10.17182/hepdata.132367}}.

\bibitem{ALEPH:1993sjl}
\hrefCMSnoop {}{{ALEPH} Collaboration, ``Search for a nonminimal {Higgs} boson
  produced in the reaction $\mathrm{e}^{+} \mathrm{e}^{-} \to
  \mathrm{h}\mathrm{Z}^{*}$'',} \textit{ Phys. Lett. B} \textbf{ 313} (1993)
  312,
  \href{http://dx.doi.org/10.1016/0370-2693(93)91228-F}{\doi{10.1016/0370-2693(93)91228-F}}.

\bibitem{L3:1996ome}
\hrefCMSnoop {}{{L3} Collaboration, ``Search for neutral {Higgs} boson
  production through the process $\mathrm{e}^{+} \mathrm{e}^{-} \to
  \mathrm{Z}^{*}\mathrm{H}^0$'',} \textit{ Phys. Lett. B} \textbf{ 385} (1996)
  454,
  \href{http://dx.doi.org/10.1016/0370-2693(96)00987-2}{\doi{10.1016/0370-2693(96)00987-2}}.

\bibitem{LEPWG:2003ing}
\hrefCMSnoop {}{{LEP working group for Higgs boson searches, ALEPH, DELPHI, L3
  and OPAL Collaborations}, ``Search for the standard model {Higgs} boson at
  {LEP}'',} \textit{ Phys. Lett. B} \textbf{ 565} (2003) 61,
  \href{http://dx.doi.org/10.1016/S0370-2693(03)00614-2}{\doi{10.1016/S0370-2693(03)00614-2}},
  \href{http://www.arXiv.org/abs/hep-ex/0306033}{\texttt{arXiv:hep-ex/0306033}}.

\bibitem{D0:2013cej}
\hrefCMSnoop {}{{D0} Collaboration, ``Combined search for the {Higgs} boson
  with the {D0} experiment'',} \textit{ Phys. Rev. D} \textbf{ 88} (2013)
  052011,
  \href{http://dx.doi.org/10.1103/PhysRevD.88.052011}{\doi{10.1103/PhysRevD.88.052011}},
  \href{http://www.arXiv.org/abs/1303.0823}{\texttt{arXiv:1303.0823}}.

\bibitem{CDF:2013eju}
\hrefCMSnoop {}{{CDF} Collaboration, ``Combination of searches for the {Higgs}
  boson using the full {CDF} data set'',} \textit{ Phys. Rev. D} \textbf{ 88}
  (2013) 052013,
  \href{http://dx.doi.org/10.1103/PhysRevD.88.052013}{\doi{10.1103/PhysRevD.88.052013}},
  \href{http://www.arXiv.org/abs/1301.6668}{\texttt{arXiv:1301.6668}}.

\bibitem{CDF:2013kiv}
\hrefCMSnoop {}{{CDF and D0 Collaborations}, ``{Higgs} boson studies at the
  {Tevatron}'',} \textit{ Phys. Rev. D} \textbf{ 88} (2013) 052014,
  \href{http://dx.doi.org/10.1103/PhysRevD.88.052014}{\doi{10.1103/PhysRevD.88.052014}},
  \href{http://www.arXiv.org/abs/1303.6346}{\texttt{arXiv:1303.6346}}.

\bibitem{CMS:2012bfw}
\hrefCMSnoop {}{{CMS Collaboration}, ``Search for the standard model {Higgs}
  boson produced in association with {W} and {Z} bosons in pp collisions at
  $\sqrt{s}=7$ {TeV}'',} \textit{ JHEP} \textbf{ 11} (2012) 088,
  \href{http://dx.doi.org/10.1007/JHEP11(2012)088}{\doi{10.1007/JHEP11(2012)088}},
  \href{http://www.arXiv.org/abs/1209.3937}{\texttt{arXiv:1209.3937}}.

\bibitem{ATLAS:2019vrd}
\hrefCMSnoop {}{{ATLAS Collaboration}, ``Measurement of the production cross
  section for a {Higgs} boson in association with a vector boson in the
  $\mathrm{H} \to \mathrm{WW}^{\ast} \to \ell\nu\ell\nu$ channel in pp
  collisions at $\sqrt{s}=13$ {TeV} with the {ATLAS} detector'',} \textit{
  Phys. Lett. B} \textbf{ 798} (2019) 134949,
  \href{http://dx.doi.org/10.1016/j.physletb.2019.134949}{\doi{10.1016/j.physletb.2019.134949}},
  \href{http://www.arXiv.org/abs/1903.10052}{\texttt{arXiv:1903.10052}}.

\bibitem{CMS:2014wdm}
\hrefCMSnoop {}{{CMS Collaboration}, ``Evidence for the 125 {GeV} {Higgs} boson
  decaying to a pair of $\tau$ leptons'',} \textit{ JHEP} \textbf{ 05} (2014)
  104,
  \href{http://dx.doi.org/10.1007/JHEP05(2014)104}{\doi{10.1007/JHEP05(2014)104}},
  \href{http://www.arXiv.org/abs/1401.5041}{\texttt{arXiv:1401.5041}}.

\bibitem{ATLAS:2015xst}
\hrefCMSnoop {}{{ATLAS Collaboration}, ``Evidence for the {Higgs}-boson
  {Yukawa} coupling to tau leptons with the {ATLAS} detector'',} \textit{ JHEP}
  \textbf{ 04} (2015) 117,
  \href{http://dx.doi.org/10.1007/JHEP04(2015)117}{\doi{10.1007/JHEP04(2015)117}},
  \href{http://www.arXiv.org/abs/1501.04943}{\texttt{arXiv:1501.04943}}.

\bibitem{CMS:2014tll}
\hrefCMSnoop {}{{CMS Collaboration}, ``{Search for the associated production of
  the {Higgs} boson with a top-quark pair}'',} \textit{ JHEP} \textbf{ 09}
  (2014) 087,
  \href{http://dx.doi.org/10.1007/JHEP09(2014)087}{\doi{10.1007/JHEP09(2014)087}},
  \href{http://www.arXiv.org/abs/1408.1682}{\texttt{arXiv:1408.1682}}.
  [Erratum: JHEP 10, 106 (2014)].

\bibitem{ATLAS:2018alq}
\hrefCMSnoop {}{{ATLAS Collaboration}, ``Search for new phenomena in events
  with same-charge leptons and b jets in pp collisions at $\sqrt{s}= 13$ {TeV}
  with the {ATLAS} detector'',} \textit{ JHEP} \textbf{ 12} (2018) 039,
  \href{http://dx.doi.org/10.1007/JHEP12(2018)039}{\doi{10.1007/JHEP12(2018)039}},
  \href{http://www.arXiv.org/abs/1807.11883}{\texttt{arXiv:1807.11883}}.

\bibitem{CMS:2020xwi}
\hrefCMSnoop {}{{CMS Collaboration}, ``Evidence for {Higgs} boson decay to a
  pair of muons'',} \textit{ JHEP} \textbf{ 01} (2021) 148,
  \href{http://dx.doi.org/10.1007/JHEP01(2021)148}{\doi{10.1007/JHEP01(2021)148}},
  \href{http://www.arXiv.org/abs/2009.04363}{\texttt{arXiv:2009.04363}}.

\bibitem{ATLAS:2020fzp}
\hrefCMSnoop {}{{ATLAS Collaboration}, ``A search for the dimuon decay of the
  standard model {Higgs} boson with the {ATLAS} detector'',} \textit{ Phys.
  Lett. B} \textbf{ 812} (2021) 135980,
  \href{http://dx.doi.org/10.1016/j.physletb.2020.135980}{\doi{10.1016/j.physletb.2020.135980}},
  \href{http://www.arXiv.org/abs/2007.07830}{\texttt{arXiv:2007.07830}}.

\bibitem{CMS:2018nsn}
\hrefCMSnoop {}{{CMS Collaboration}, ``Observation of {Higgs} boson decay to
  bottom quarks'',} \textit{ Phys. Rev. Lett.} \textbf{ 121} (2018) 121801,
  \href{http://dx.doi.org/10.1103/PhysRevLett.121.121801}{\doi{10.1103/PhysRevLett.121.121801}},
  \href{http://www.arXiv.org/abs/1808.08242}{\texttt{arXiv:1808.08242}}.

\bibitem{ATLAS:2018kot}
\hrefCMSnoop {}{{ATLAS Collaboration}, ``Observation of $\mathrm{H} \to
  \mathrm{b}\bar{\mathrm{b}}$ decays and {VH} production with the {ATLAS}
  detector'',} \textit{ Phys. Lett. B} \textbf{ 786} (2018) 59,
  \href{http://dx.doi.org/10.1016/j.physletb.2018.09.013}{\doi{10.1016/j.physletb.2018.09.013}},
  \href{http://www.arXiv.org/abs/1808.08238}{\texttt{arXiv:1808.08238}}.

\bibitem{ATLAS:2023ofo}
\hrefCMSnoop {}{{ATLAS Collaboration}, ``Search for a new pseudoscalar decaying
  into a pair of muons in events with a top-quark pair at $\sqrt{s} = 13$ tev
  with the {ATLAS} detector'',} \textit{ Phys. Rev. D} \textbf{ 108} (2023),
  no.~9, 092007,
  \href{http://dx.doi.org/10.1103/PhysRevD.108.092007}{\doi{10.1103/PhysRevD.108.092007}},
  \href{http://www.arXiv.org/abs/2304.14247}{\texttt{arXiv:2304.14247}}.

\bibitem{CMS:2019lwf}
\hrefCMSnoop {}{{CMS Collaboration}, ``Search for physics beyond the standard
  model in multilepton final states in proton-proton collisions at $\sqrt{s} =
  13$ {TeV}'',} \textit{ JHEP} \textbf{ 03} (2020) 051,
  \href{http://dx.doi.org/10.1007/JHEP03(2020)051}{\doi{10.1007/JHEP03(2020)051}},
  \href{http://www.arXiv.org/abs/1911.04968}{\texttt{arXiv:1911.04968}}.

\bibitem{CMS:2008xjf}
\hrefCMSnoop {}{{CMS Collaboration}, ``The {CMS} experiment at the {CERN}
  {LHC}'',} \textit{ JINST} \textbf{ 3} (2008) S08004,
  \href{http://dx.doi.org/10.1088/1748-0221/3/08/S08004}{\doi{10.1088/1748-0221/3/08/S08004}}.

\bibitem{CMS:2023gfb}
\hrefCMSnoop {}{{CMS Collaboration}, ``Development of the {CMS} detector for
  the {CERN LHC Run 3}'',} 2023.
  \href{http://www.arXiv.org/abs/2309.05466}{\texttt{arXiv:2309.05466}}.
  Accepted by \textit{JINST}.

\bibitem{CMS:2020cmk}
\hrefCMSnoop {}{{CMS Collaboration}, ``Performance of the {CMS} level-1 trigger
  in proton-proton collisions at $\sqrt{s} = 13$ {TeV}'',} \textit{ JINST}
  \textbf{ 15} (2020) P10017,
  \href{http://dx.doi.org/10.1088/1748-0221/15/10/P10017}{\doi{10.1088/1748-0221/15/10/P10017}},
  \href{http://www.arXiv.org/abs/2006.10165}{\texttt{arXiv:2006.10165}}.

\bibitem{CMS:2016ngn}
\hrefCMSnoop {}{{CMS Collaboration}, ``The {CMS} trigger system'',} \textit{
  JINST} \textbf{ 12} (2017) P01020,
  \href{http://dx.doi.org/10.1088/1748-0221/12/01/P01020}{\doi{10.1088/1748-0221/12/01/P01020}},
\href{http://www.arXiv.org/abs/1609.02366}{\texttt{arXiv:1609.02366}}.
%%CITATION = ARXIV:1609.02366;%%.

\bibitem{CMS:2021xjt}
\hrefCMSnoop {}{{CMS Collaboration}, ``Precision luminosity measurement in
  proton-proton collisions at $\sqrt{s} = 13$ {TeV} in 2015 and 2016 at
  {CMS}'',} \textit{ Eur. Phys. J. C} \textbf{ 81} (2021) 800,
  \href{http://dx.doi.org/10.1140/epjc/s10052-021-09538-2}{\doi{10.1140/epjc/s10052-021-09538-2}},
  \href{http://www.arXiv.org/abs/2104.01927}{\texttt{arXiv:2104.01927}}.

\bibitem{CMS:2018elu}
\href {http://cds.cern.ch/record/2621960}{{CMS Collaboration}, ``{CMS}
  luminosity measurement for the 2017 data-taking period at $\sqrt{s} = 13$
  {TeV}'',} CMS Physics Analysis Summary CMS-PAS-LUM-17-004, 2018.

\bibitem{CMS:2019jhq}
\href {http://cds.cern.ch/record/2676164}{{CMS Collaboration}, ``{CMS}
  luminosity measurement for the 2018 data-taking period at $\sqrt{s} = 13$
  {TeV}'',} CMS Physics Analysis Summary CMS-PAS-LUM-18-002, 2019.

\bibitem{Alwall:2014hca}
J.~Alwall\hrefCMSnoop {}{ { et~al.}, ``The automated computation of tree-level
  and next-to-leading order differential cross sections, and their matching to
  parton shower simulations'',} \textit{ JHEP} \textbf{ 07} (2014) 079,
  \href{http://dx.doi.org/10.1007/JHEP07(2014)079}{\doi{10.1007/JHEP07(2014)079}},
  \href{http://www.arXiv.org/abs/1405.0301}{\texttt{arXiv:1405.0301}}.

\bibitem{Nason:2004rx}
\hrefCMSnoop {}{P.~Nason, ``A new method for combining {NLO QCD} with shower
  {Monte Carlo} algorithms'',} \textit{ JHEP} \textbf{ 11} (2004) 040,
  \href{http://dx.doi.org/10.1088/1126-6708/2004/11/040}{\doi{10.1088/1126-6708/2004/11/040}},
  \href{http://www.arXiv.org/abs/hep-ph/0409146}{\texttt{arXiv:hep-ph/0409146}}.

\bibitem{Frixione:2007vw}
\hrefCMSnoop {}{S.~Frixione, P.~Nason, and C.~Oleari, ``Matching {NLO QCD}
  computations with parton shower simulations: the {POWHEG} method'',} \textit{
  JHEP} \textbf{ 11} (2007) 070,
  \href{http://dx.doi.org/10.1088/1126-6708/2007/11/070}{\doi{10.1088/1126-6708/2007/11/070}},
  \href{http://www.arXiv.org/abs/0709.2092}{\texttt{arXiv:0709.2092}}.

\bibitem{Alioli:2010xd}
\hrefCMSnoop {}{S.~Alioli, P.~Nason, C.~Oleari, and E.~Re, ``A general
  framework for implementing {NLO} calculations in shower {Monte Carlo}
  programs: the {POWHEG BOX}'',} \textit{ JHEP} \textbf{ 06} (2010) 043,
  \href{http://dx.doi.org/10.1007/JHEP06(2010)043}{\doi{10.1007/JHEP06(2010)043}},
  \href{http://www.arXiv.org/abs/1002.2581}{\texttt{arXiv:1002.2581}}.

\bibitem{Campbell:2010ff}
\hrefCMSnoop {}{J.~M. Campbell and R.~K. Ellis, ``{MCFM} for the {Tevatron} and
  the {LHC}'',} \textit{ Nucl. Phys. Proc. Suppl.} \textbf{ 205--206} (2010)
  10,
  \href{http://dx.doi.org/10.1016/j.nuclphysbps.2010.08.011}{\doi{10.1016/j.nuclphysbps.2010.08.011}},
  \href{http://www.arXiv.org/abs/1007.3492}{\texttt{arXiv:1007.3492}}.

\bibitem{Gao:2010qx}
Y.~Gao\hrefCMSnoop {}{ { et~al.}, ``Spin determination of single-produced
  resonances at hadron colliders'',} \textit{ Phys. Rev. D} \textbf{ 81} (2010)
  075022,
  \href{http://dx.doi.org/10.1103/PhysRevD.81.075022}{\doi{10.1103/PhysRevD.81.075022}},
  \href{http://www.arXiv.org/abs/1001.3396}{\texttt{arXiv:1001.3396}}.

\bibitem{Bolognesi:2012mm}
S.~Bolognesi\hrefCMSnoop {}{ { et~al.}, ``On the spin and parity of a
  single-produced resonance at the {LHC}'',} \textit{ Phys. Rev. D} \textbf{
  86} (2012) 095031,
  \href{http://dx.doi.org/10.1103/PhysRevD.86.095031}{\doi{10.1103/PhysRevD.86.095031}},
  \href{http://www.arXiv.org/abs/1208.4018}{\texttt{arXiv:1208.4018}}.

\bibitem{Anderson:2013afp}
I.~Anderson\hrefCMSnoop {}{ { et~al.}, ``Constraining anomalous {HVV}
  interactions at proton and lepton colliders'',} \textit{ Phys. Rev. D}
  \textbf{ 89} (2014) 035007,
  \href{http://dx.doi.org/10.1103/PhysRevD.89.035007}{\doi{10.1103/PhysRevD.89.035007}},
  \href{http://www.arXiv.org/abs/1309.4819}{\texttt{arXiv:1309.4819}}.

\bibitem{Gritsan:2016hjl}
\hrefCMSnoop {}{A.~V. Gritsan, R.~R{\"o}ntsch, M.~Schulze, and M.~Xiao,
  ``Constraining anomalous {Higgs} boson couplings to the heavy flavor fermions
  using matrix element techniques'',} \textit{ Phys. Rev. D} \textbf{ 94}
  (2016) 055023,
  \href{http://dx.doi.org/10.1103/PhysRevD.94.055023}{\doi{10.1103/PhysRevD.94.055023}},
  \href{http://www.arXiv.org/abs/1606.03107}{\texttt{arXiv:1606.03107}}.

\bibitem{Ball:2014uwa}
\hrefCMSnoop {}{{NNPDF} Collaboration, ``Parton distributions for the {LHC Run
  II}'',} \textit{ JHEP} \textbf{ 04} (2015) 040,
  \href{http://dx.doi.org/10.1007/JHEP04(2015)040}{\doi{10.1007/JHEP04(2015)040}},
\href{http://www.arXiv.org/abs/1410.8849}{\texttt{arXiv:1410.8849}}.
%%CITATION = ARXIV:1410.8849;%%.

\bibitem{Ball:2017nwa}
\hrefCMSnoop {}{{NNPDF} Collaboration, ``Parton distributions from
  high-precision collider data'',} \textit{ Eur. Phys. J. C} \textbf{ 77}
  (2017) 663,
  \href{http://dx.doi.org/10.1140/epjc/s10052-017-5199-5}{\doi{10.1140/epjc/s10052-017-5199-5}},
\href{http://www.arXiv.org/abs/1706.00428}{\texttt{arXiv:1706.00428}}.
%%CITATION = ARXIV:1706.00428;%%.

\bibitem{Sjostrand:2014zea}
T.~Sj{\"o}strand\hrefCMSnoop {}{ { et~al.}, ``An introduction to {PYTHIA
  8.2}'',} \textit{ Comput. Phys. Commun.} \textbf{ 191} (2015) 159,
  \href{http://dx.doi.org/10.1016/j.cpc.2015.01.024}{\doi{10.1016/j.cpc.2015.01.024}},
\href{http://www.arXiv.org/abs/1410.3012}{\texttt{arXiv:1410.3012}}.
%%CITATION = ARXIV:1410.3012;%%.

\bibitem{Khachatryan:2015pea}
\hrefCMSnoop {}{{CMS Collaboration}, ``Event generator tunes obtained from
  underlying event and multiparton scattering measurements'',} \textit{ Eur.
  Phys. J. C} \textbf{ 76} (2016) 155,
  \href{http://dx.doi.org/10.1140/epjc/s10052-016-3988-x}{\doi{10.1140/epjc/s10052-016-3988-x}},
\href{http://www.arXiv.org/abs/1512.00815}{\texttt{arXiv:1512.00815}}.
%%CITATION = ARXIV:1512.00815;%%.

\bibitem{CMS:2019csb}
\hrefCMSnoop {}{{CMS Collaboration}, ``Extraction and validation of a new set
  of {CMS} {PYTHIA8} tunes from underlying-event measurements'',} \textit{ Eur.
  Phys. J. C} \textbf{ 80} (2020) 4,
  \href{http://dx.doi.org/10.1140/epjc/s10052-019-7499-4}{\doi{10.1140/epjc/s10052-019-7499-4}},
  \href{http://www.arXiv.org/abs/1903.12179}{\texttt{arXiv:1903.12179}}.

\bibitem{Alwall:2007fs}
\hrefCMSnoop {}{J.~Alwall { et~al.}, ``Comparative study of various algorithms
  for the merging of parton showers and matrix elements in hadronic
  collisions'',} \textit{ Eur. Phys. J. C} \textbf{ 53} (2008) 473,
  \href{http://dx.doi.org/10.1140/epjc/s10052-007-0490-5}{\doi{10.1140/epjc/s10052-007-0490-5}},
  \href{http://www.arXiv.org/abs/0706.2569}{\texttt{arXiv:0706.2569}}.

\bibitem{Frederix:2012ps}
\hrefCMSnoop {}{R.~Frederix and S.~Frixione, ``Merging meets matching in
  {MC@NLO}'',} \textit{ JHEP} \textbf{ 12} (2012) 061,
  \href{http://dx.doi.org/10.1007/JHEP12(2012)061}{\doi{10.1007/JHEP12(2012)061}},
\href{http://www.arXiv.org/abs/1209.6215}{\texttt{arXiv:1209.6215}}.
%%CITATION = ARXIV:1209.6215;%%.

\bibitem{Agostinelli:2002hh}
\hrefCMSnoop {}{{GEANT4} Collaboration, ``{\GEANTfour}---a simulation
  toolkit'',} \textit{ Nucl. Instrum. Meth. A} \textbf{ 506} (2003) 250,
\href{http://dx.doi.org/10.1016/S0168-9002(03)01368-8}{\doi{10.1016/S0168-9002(03)01368-8}}.
%%CITATION = NUIMA,A506,250;%%.

\bibitem{CMS-TDR-15-02}
\href {http://cds.cern.ch/record/2020886}{{CMS Collaboration}, ``Technical
  proposal for the {Phase-II} upgrade of the {Compact Muon Solenoid}'',} CMS
  Technical Proposal CERN-LHCC-2015-010, CMS-TDR-15-02, 2015.

\bibitem{Sirunyan:2017ulk}
\hrefCMSnoop {}{{CMS Collaboration}, ``Particle-flow reconstruction and global
  event description with the {CMS} detector'',} \textit{ JINST} \textbf{ 12}
  (2017) P10003,
  \href{http://dx.doi.org/10.1088/1748-0221/12/10/P10003}{\doi{10.1088/1748-0221/12/10/P10003}},
\href{http://www.arXiv.org/abs/1706.04965}{\texttt{arXiv:1706.04965}}.
%%CITATION = ARXIV:1706.04965;%%.

\bibitem{CMS:2020uim}
\hrefCMSnoop {}{{CMS Collaboration}, ``Electron and photon reconstruction and
  identification with the {CMS} experiment at the {CERN} {LHC}'',} \textit{
  JINST} \textbf{ 16} (2021) P05014,
  \href{http://dx.doi.org/10.1088/1748-0221/16/05/P05014}{\doi{10.1088/1748-0221/16/05/P05014}},
  \href{http://www.arXiv.org/abs/2012.06888}{\texttt{arXiv:2012.06888}}.

\bibitem{CMS-DP-2020-021}
\href {http://cds.cern.ch/record/2717925}{{CMS Collaboration}, ``{ECAL} 2016
  refined calibration and {Run2} summary plots'',} CMS Detector Performance
  Note CMS-DP-2020-021, 2020.

\bibitem{CMS:2018rym}
\hrefCMSnoop {}{{CMS Collaboration}, ``Performance of the {CMS} muon detector
  and muon reconstruction with proton-proton collisions at $\sqrt{s}=$ 13
  {TeV}'',} \textit{ JINST} \textbf{ 13} (2018) P06015,
  \href{http://dx.doi.org/10.1088/1748-0221/13/06/P06015}{\doi{10.1088/1748-0221/13/06/P06015}},
  \href{http://www.arXiv.org/abs/1804.04528}{\texttt{arXiv:1804.04528}}.

\bibitem{CMS:2018jrd}
\hrefCMSnoop {}{{CMS Collaboration}, ``Performance of reconstruction and
  identification of $\tau$ leptons decaying to hadrons and $\nu_{\tau}$ in pp
  collisions at $\sqrt{s}= 13$ {TeV}'',} \textit{ JINST} \textbf{ 13} (2018)
  \href{http://dx.doi.org/10.1088/1748-0221/13/10/P10005}{\doi{10.1088/1748-0221/13/10/P10005}},
  \href{http://www.arXiv.org/abs/1809.02816}{\texttt{arXiv:1809.02816}}.

\bibitem{Cacciari:2008gp}
\hrefCMSnoop {}{M.~Cacciari, G.~P. Salam, and G.~Soyez, ``The anti-\kt jet
  clustering algorithm'',} \textit{ JHEP} \textbf{ 04} (2008) 063,
  \href{http://dx.doi.org/10.1088/1126-6708/2008/04/063}{\doi{10.1088/1126-6708/2008/04/063}},
\href{http://www.arXiv.org/abs/0802.1189}{\texttt{arXiv:0802.1189}}.
%%CITATION = ARXIV:0802.1189;%%.

\bibitem{Cacciari:2011ma}
\hrefCMSnoop {}{M.~Cacciari, G.~P. Salam, and G.~Soyez, ``Fastjet user
  manual'',} \textit{ Eur. Phys. J. C} \textbf{ 72} (2012) 1896,
  \href{http://dx.doi.org/10.1140/epjc/s10052-012-1896-2}{\doi{10.1140/epjc/s10052-012-1896-2}},
\href{http://www.arXiv.org/abs/1111.6097}{\texttt{arXiv:1111.6097}}.
%%CITATION = ARXIV:1111.6097;%%.

\bibitem{CMS:2016lmd}
\hrefCMSnoop {}{{CMS Collaboration}, ``Jet energy scale and resolution in the
  {CMS} experiment in pp collisions at 8 {TeV}'',} \textit{ JINST} \textbf{ 12}
  (2017) P02014,
  \href{http://dx.doi.org/10.1088/1748-0221/12/02/P02014}{\doi{10.1088/1748-0221/12/02/P02014}},
\href{http://www.arXiv.org/abs/1607.03663}{\texttt{arXiv:1607.03663}}.
%%CITATION = ARXIV:1607.03663;%%.

\bibitem{Sirunyan:2019kia}
\hrefCMSnoop {}{{CMS Collaboration}, ``Performance of missing transverse
  momentum reconstruction in proton-proton collisions at $\sqrt{s} =$ 13 {TeV}
  using the {CMS} detector'',} \textit{ JINST} \textbf{ 14} (2019) P07004,
  \href{http://dx.doi.org/10.1088/1748-0221/14/07/P07004}{\doi{10.1088/1748-0221/14/07/P07004}},
  \href{http://www.arXiv.org/abs/1903.06078}{\texttt{arXiv:1903.06078}}.

\bibitem{Bertolini:2014bba}
\hrefCMSnoop {}{D.~Bertolini, P.~Harris, M.~Low, and N.~Tran, ``Pileup per
  particle identification'',} \textit{ JHEP} \textbf{ 10} (2014) 059,
  \href{http://dx.doi.org/10.1007/JHEP10(2014)059}{\doi{10.1007/JHEP10(2014)059}},
  \href{http://www.arXiv.org/abs/1407.6013}{\texttt{arXiv:1407.6013}}.

\bibitem{Sirunyan:2020foa}
\hrefCMSnoop {}{{CMS Collaboration}, ``Pileup mitigation at {CMS} in 13 {TeV}
  data'',} \textit{ JINST} \textbf{ 15} (2020) P09018,
  \href{http://dx.doi.org/10.1088/1748-0221/15/09/p09018}{\doi{10.1088/1748-0221/15/09/p09018}},
  \href{http://www.arXiv.org/abs/2003.00503}{\texttt{arXiv:2003.00503}}.

\bibitem{Sirunyan:2017ezt}
\hrefCMSnoop {}{{CMS Collaboration}, ``Identification of heavy-flavour jets
  with the {CMS} detector in pp collisions at 13 {TeV}'',} \textit{ JINST}
  \textbf{ 13} (2018) P05011,
  \href{http://dx.doi.org/10.1088/1748-0221/13/05/P05011}{\doi{10.1088/1748-0221/13/05/P05011}},
\href{http://www.arXiv.org/abs/1712.07158}{\texttt{arXiv:1712.07158}}.
%%CITATION = ARXIV:1712.07158;%%.

\bibitem{CMS:2022prd}
\hrefCMSnoop {}{{CMS Collaboration}, ``Identification of hadronic tau lepton
  decays using a deep neural network'',} \textit{ JINST} \textbf{ 17} (2022)
  P07023,
  \href{http://dx.doi.org/10.1088/1748-0221/17/07/P07023}{\doi{10.1088/1748-0221/17/07/P07023}},
  \href{http://www.arXiv.org/abs/2201.08458}{\texttt{arXiv:2201.08458}}.

\bibitem{CMS:2015nep}
\hrefCMSnoop {}{{CMS Collaboration}, ``Search for third-generation scalar
  leptoquarks in the t$\tau$ channel in proton-proton collisions at $\sqrt{s} =
  8$ {TeV}'',} \textit{ JHEP} \textbf{ 07} (2015) 042,
  \href{http://dx.doi.org/10.1007/JHEP11(2016)056}{\doi{10.1007/JHEP11(2016)056}},
  \href{http://www.arXiv.org/abs/1503.09049}{\texttt{arXiv:1503.09049}}.
  [Erratum: JHEP 11 (2016) 056].

\bibitem{CMS:2011aa}
\hrefCMSnoop {}{{CMS Collaboration}, ``Measurement of the inclusive {W} and {Z}
  production cross sections in pp collisions at $\sqrt{s}=7$ {TeV}'',} \textit{
  JHEP} \textbf{ 10} (2011) 132,
  \href{http://dx.doi.org/10.1007/JHEP10(2011)132}{\doi{10.1007/JHEP10(2011)132}},
  \href{http://www.arXiv.org/abs/1107.4789}{\texttt{arXiv:1107.4789}}.

\bibitem{Cranmer:2000du}
\hrefCMSnoop {}{K.~S. Cranmer, ``Kernel estimation in high-energy physics'',}
  \textit{ Comput. Phys. Commun.} \textbf{ 136} (2001) 198,
  \href{http://dx.doi.org/10.1016/S0010-4655(00)00243-5}{\doi{10.1016/S0010-4655(00)00243-5}},
  \href{http://www.arXiv.org/abs/hep-ex/0011057}{\texttt{arXiv:hep-ex/0011057}}.

\bibitem{Cacciari:2003fi}
M.~Cacciari\hrefCMSnoop {}{ { et~al.}, ``The $\mathrm{t}\bar{\mathrm{t}}$
  cross-section at 1.8 and 1.96 {TeV}: {A} study of the systematics due to
  parton densities and scale dependence'',} \textit{ JHEP} \textbf{ 04} (2004)
  068,
  \href{http://dx.doi.org/10.1088/1126-6708/2004/04/068}{\doi{10.1088/1126-6708/2004/04/068}},
\href{http://www.arXiv.org/abs/hep-ph/0303085}{\texttt{arXiv:hep-ph/0303085}}.
%%CITATION = HEP-PH/0303085;%%.

\bibitem{CMS:2018mlc}
\hrefCMSnoop {}{{CMS Collaboration}, ``Measurement of the inelastic
  proton-proton cross section at $ \sqrt{s}=13 $ {TeV}'',} \textit{ JHEP}
  \textbf{ 07} (2018) 161,
  \href{http://dx.doi.org/10.1007/JHEP07(2018)161}{\doi{10.1007/JHEP07(2018)161}},
  \href{http://www.arXiv.org/abs/1802.02613}{\texttt{arXiv:1802.02613}}.

\bibitem{ATLAS:2016ygv}
\hrefCMSnoop {}{{ATLAS Collaboration}, ``Measurement of the inelastic
  proton-proton cross section at $\sqrt{s} = 13$ {TeV} with the {ATLAS}
  detector at the {LHC}'',} \textit{ Phys. Rev. Lett.} \textbf{ 117} (2016)
  182002,
  \href{http://dx.doi.org/10.1103/PhysRevLett.117.182002}{\doi{10.1103/PhysRevLett.117.182002}},
  \href{http://www.arXiv.org/abs/1606.02625}{\texttt{arXiv:1606.02625}}.

\bibitem{Gross:2010qma}
\hrefCMSnoop {}{E.~Gross and O.~Vitells, ``Trial factors for the look elsewhere
  effect in high energy physics'',} \textit{ Eur. Phys. J. C} \textbf{ 70}
  (2010) 525,
  \href{http://dx.doi.org/10.1140/epjc/s10052-010-1470-8}{\doi{10.1140/epjc/s10052-010-1470-8}},
\href{http://www.arXiv.org/abs/1005.1891}{\texttt{arXiv:1005.1891}}.
%%CITATION = ARXIV:1005.1891;%%.

\bibitem{Junk:1999kv}
\hrefCMSnoop {}{T.~Junk, ``Confidence level computation for combining searches
  with small statistics'',} \textit{ Nucl. Instrum. Meth. A} \textbf{ 434}
  (1999) 435,
  \href{http://dx.doi.org/10.1016/S0168-9002(99)00498-2}{\doi{10.1016/S0168-9002(99)00498-2}},
\href{http://www.arXiv.org/abs/hep-ex/9902006}{\texttt{arXiv:hep-ex/9902006}}.
%%CITATION = HEP-EX/9902006;%%.

\bibitem{Read:2002hq}
\hrefCMSnoop {}{A.~L. Read, ``Presentation of search results: The {CL}$_{s}$
  technique'',} \textit{ J. Phys. G} \textbf{ 28} (2002) 2693,
\href{http://dx.doi.org/10.1088/0954-3899/28/10/313}{\doi{10.1088/0954-3899/28/10/313}}.
%%CITATION = JPAGA,G28,2693;%%.

\bibitem{Cowan:2010js}
\hrefCMSnoop {}{G.~Cowan, K.~Cranmer, E.~Gross, and O.~Vitells, ``Asymptotic
  formulae for likelihood-based tests of new physics'',} \textit{ Eur. Phys. J.
  C} \textbf{ 71} (2011) 1554,
  \href{http://dx.doi.org/10.1140/epjc/s10052-011-1554-0}{\doi{10.1140/epjc/s10052-011-1554-0}},
  \href{http://www.arXiv.org/abs/1007.1727}{\texttt{arXiv:1007.1727}}.
[Erratum: \DOI{10.1140/epjc/s10052-013-2501-z}].
%%CITATION = ARXIV:1007.1727;%%.

\bibitem{ATLAS:2011tau}
\href {http://cds.cern.ch/record/1379837}{{ATLAS and CMS Collaborations, and
  {LHC} Higgs Combination Group}, ``Procedure for the {LHC} {Higgs} boson
  search combination in summer 2011'',} Technical Report CMS-NOTE-2011-005,
  ATL-PHYS-PUB-2011-011, 2011.

\bibitem{Djouadi:1997yw}
\hrefCMSnoop {}{A.~Djouadi, J.~Kalinowski, and M.~Spira, ``{HDECAY}: {A}
  program for {Higgs} boson decays in the standard model and its supersymmetric
  extension'',} \textit{ Comput. Phys. Commun.} \textbf{ 108} (1998) 56,
  \href{http://dx.doi.org/10.1016/S0010-4655(97)00123-9}{\doi{10.1016/S0010-4655(97)00123-9}},
  \href{http://www.arXiv.org/abs/hep-ph/9704448}{\texttt{arXiv:hep-ph/9704448}}.

\bibitem{Djouadi:2018xqq}
\hrefCMSnoop {}{A.~Djouadi, J.~Kalinowski, M.~Muehlleitner, and M.~Spira,
  ``{HDECAY}: {Twenty}$++$ years after'',} \textit{ Comput. Phys. Commun.}
  \textbf{ 238} (2019) 214,
  \href{http://dx.doi.org/10.1016/j.cpc.2018.12.010}{\doi{10.1016/j.cpc.2018.12.010}},
  \href{http://www.arXiv.org/abs/1801.09506}{\texttt{arXiv:1801.09506}}.

\end{thebibliography}\endgroup

\cleardoublepage \appendix\section{The CMS Collaboration \label{app:collab}}\begin{sloppypar}\hyphenpenalty=5000\widowpenalty=500\clubpenalty=5000
\cmsinstitute{Yerevan Physics Institute, Yerevan, Armenia}
{\tolerance=6000
A.~Tumasyan\cmsAuthorMark{1}\cmsorcid{0009-0000-0684-6742}
\par}
\cmsinstitute{Institut f\"{u}r Hochenergiephysik, Vienna, Austria}
{\tolerance=6000
W.~Adam\cmsorcid{0000-0001-9099-4341}, J.W.~Andrejkovic, T.~Bergauer\cmsorcid{0000-0002-5786-0293}, S.~Chatterjee\cmsorcid{0000-0003-2660-0349}, K.~Damanakis\cmsorcid{0000-0001-5389-2872}, M.~Dragicevic\cmsorcid{0000-0003-1967-6783}, A.~Escalante~Del~Valle\cmsorcid{0000-0002-9702-6359}, P.S.~Hussain\cmsorcid{0000-0002-4825-5278}, M.~Jeitler\cmsAuthorMark{2}\cmsorcid{0000-0002-5141-9560}, N.~Krammer\cmsorcid{0000-0002-0548-0985}, L.~Lechner\cmsorcid{0000-0002-3065-1141}, D.~Liko\cmsorcid{0000-0002-3380-473X}, I.~Mikulec\cmsorcid{0000-0003-0385-2746}, P.~Paulitsch, J.~Schieck\cmsAuthorMark{2}\cmsorcid{0000-0002-1058-8093}, R.~Sch\"{o}fbeck\cmsorcid{0000-0002-2332-8784}, D.~Schwarz\cmsorcid{0000-0002-3821-7331}, M.~Sonawane\cmsorcid{0000-0003-0510-7010}, S.~Templ\cmsorcid{0000-0003-3137-5692}, W.~Waltenberger\cmsorcid{0000-0002-6215-7228}, C.-E.~Wulz\cmsAuthorMark{2}\cmsorcid{0000-0001-9226-5812}
% ...
\par}
\vskip\cmsinstskip
\dag:~Deceased\\
$^{1}$Also at Yerevan State University, Yerevan, Armenia\\
$^{2}$Also at TU Wien, Vienna, Austria\\
% ...
\end{sloppypar}
%%% END EDITABLE REGION %%%
% skeleton_end
\end{document}